\documentclass[a4paper,fleqn,usenatbib]{mnras}

\usepackage{newtxtext,newtxmath}

\usepackage[T1]{fontenc}
\usepackage{ae,aecompl}

\usepackage{graphicx}	
\usepackage{amsmath}	

\usepackage[dvipsnames]{xcolor}
\usepackage{multirow}

\title[Large-scale halo HI emission from intensity maps]{The neutral hydrogen distribution in large-scale haloes from 21-cm intensity maps}

\author[Tramonte \& Ma]{Denis Tramonte,$^{1,2,3}$\thanks{E-mail: tramonte@pmo.ac.cn}
Yin-Zhe Ma$^{2,3,1}$\thanks{E-mail: ma@ukzn.ac.za}
\\ 
$^{1}$Purple Mountain Observatory, No. 8 Yuanhua Road, Qixia District, Nanjing 210034, China \\
$^{2}$NAOC-UKZN Computational Astrophysics Center (NUCAC), University of Kwazulu-Natal, Durban, 4000, South Africa \\
$^{3}$School of Chemistry and Physics, University of KwaZulu-Natal, Westville Campus, Private Bag X54001, Durban, South Africa \\
}

\date{Accepted XXX. Received YYY; in original form ZZZ}

\pubyear{2020}

\begin{document}
\label{firstpage}
\pagerange{\pageref{firstpage}--\pageref{lastpage}}
\maketitle

\newcommand{\rs}{r_{\rm s}}
\newcommand{\kb}{k_{\text{B}} }
\newcommand{\de}{\text{d}}
\newcommand{\mvir}{M_{\rm v}}
\newcommand{\mvirt}{\tilde{M}_{\rm v}}
\newcommand{\mvirc}{\hat{M}_{\rm v}}

\newcommand{\logmvir}{\log_{10}{(M_{\rm v}/\text{M}_{\odot})}}
\newcommand{\logmvirt}{\log_{10}{(\tilde{M}_{\rm v}/\text{M}_{\odot})}}
\newcommand{\logmvirc}{\log_{10}{(\hat{M}_{\rm v}/\text{M}_{\odot})}}

\newcommand{\msun}{\text{M}_{\odot}}
\newcommand{\mhi}{M_{\rm HI}}
\newcommand{\nhi}{n_{\rm HI}}
\newcommand{\Nhi}{N_{\rm HI}}
\newcommand{\czhi}{c_{0,\rm HI}}
\newcommand{\czhit}{\tilde{c}_{0,\rm HI}}
\newcommand{\czhic}{\hat{c}_{0,\rm HI}}
\newcommand{\ch}{c_{\rm HI}}
\newcommand{\Rvir}{R_{\rm v}}
\newcommand{\thetavir}{\theta_{\rm v}}
\newcommand{\rhohi}{\rho_{\rm HI}}
\newcommand{\tb}{T_{\rm HI}}
\newcommand{\tbtheo}{T_{\rm HI}^{\rm (theo)}}
\newcommand{\tbobs}{T_{\rm HI}^{\rm (obs)}}
\newcommand{\ts}{T_{\rm s}}
\newcommand{\coldens}{N_{\rm HI}}
\newcommand{\DA}{D_{\rm A}}
\newcommand{\mr}{M_{\rm R}}
\newcommand{\tst}{T_{\rm stack}}
\newcommand{\wst}{W_{\rm stack}}

\begin{abstract}
We detect the neutral hydrogen (HI) radial brightness temperature profile in large-scale haloes 
by stacking 48,430 galaxies selected from the 2dFGRS catalogue onto a set of 21-cm 
intensity maps obtained with the Parkes radio telescope, spanning a total area of 
$\sim$1,300 $\text{deg}^2$ on the sky and covering the redshift range 
$0.06\lesssim z\lesssim 0.10$. Maps are obtained by removing both 10 and 20 foreground 
modes in the principal component analysis. We perform the stack at the map level and extract 
the profile from a circularly symmetrised version of the halo emission. We detect the HI 
halo emission with the significance $12.5\sigma$ for the 10-mode and 
$13.5\sigma$ for the 20-mode removed maps at the profile peak. We jointly fit for the 
observed halo mass $M_{\rm v}$ and the normalisation $c_{0,\rm HI}$ for the HI concentration 
parameter against the reconstructed profiles, using functional forms for the HI 
halo abundance proposed in the literature. 
We find $\log_{10}{(M_{\rm v}/\text{M}_{\odot})} = 16.1^{+0.1}_{-0.2}$, 
$c_{0,\rm HI}=3.5^{+0.7}_{-1.0}$ for the 10-mode and 
$\log_{10}{(M_{\rm v}/\text{M}_{\odot})} = 16.5^{+0.1}_{-0.2}$, 
$c_{0,\rm HI}=5.3^{+1.1}_{-1.7}$ for the 20-mode removed maps. 
These estimates show the detection of the integrated contribution from multiple galaxies 
located inside very massive haloes. We also consider sub-samples of 13,979 central and 
34,361 satellite 2dF galaxies separately, and obtain marginal differences suggesting satellite 
galaxies are HI-richer. This work shows for the first time the feasibility of testing 
theoretical models for the HI halo content directly on profiles extracted from 21-cm maps 
and opens future possibilities for exploiting upcoming HI intensity-mapping data.
\end{abstract}




\defcitealias{tramonte19}{T19}
\defcitealias{anderson18}{A18}

\section{Introduction}
\label{sec:intro}

In the most widely accepted $\Lambda$CDM cosmological paradigm, the evolution of the 
Universe witnesses the formation of structures in a bottom-up scenario in 
which smaller overdensities grow and merge driven by gravitational 
instability~\citep{peebles80,longair98,padmanabhan02}.
This process determines the transition from an early homogeneous Universe to a highly structured 
distribution of matter known as the cosmic web, made of an assembly of large voids, walls, filaments 
and nodes. Therefore, the mapping and reconstruction of the cosmic web is an important benchmark 
to test cosmological models.

Our primary source of information for mapping the large-scale structure (LSS) of the Universe is 
provided by optical/IR galaxy catalogues, which during the past decades have undergone a dramatic
improvement in terms of sensitivity and survey size. Among these, we can cite the the Two-degree 
Field Galaxy Redshift Survey ~\citep[2dFGRS]{colless01}, the Six-degree Field Galaxy 
Survey~\citep[][]{jones09}, 
the WiggleZ Dark Energy Survey~\citep{drinkwater10} 
the Baryon Oscillation Spectroscopic Survey~\citep[BOSS][]{schlegel09} 
and the Extended Baryon Oscillation Spectroscopic Survey~\citep[eBOSS][]{dawson16}.
In parallel, progress has also been made in the cosmic web modelling with numerical 
simulations, which provide details on the large-scale distribution of baryons 
to levels beyond observational constraints~\citep{cen99, dave99, cen06, haider16, cui19}.
Results from numerical simulations are very efficient in probing the distribution 
of different baryonic phases. A detailed study of how baryonic components with different
density and temperature are distributed across different 
cosmic web environments is described in~\citet{martizzi19}. These results show that 
the low-redshift diffuse baryonic gas in the LSS is expected to be mostly ionised.
Consequently, new observational techniques, 
tailored to map the hot gas distribution, have been exploited to map cosmic web 
structures besides galactic surveys. X-ray observations, which probe the hottest baryonic components,
are a relevant tool in this sense~\citep{zappacosta02, eckert15, bregman19, walker19, vazza19}. 
Observations of the Sunyaev-Zel'dovich effect
have also been largely exploited to study hot baryons in the cosmic 
web~\citep{van_waerbeke14, hernandez_monteagudo15, ma15, genova_santos15, tanimura19, degraaff19}.

In this effort of mapping baryons in the LSS it is interesting to consider the distribution of 
atomic hydrogen (HI) as the colder, neutral counterpart of the aforementioned probes. 
 In the post-reionisation Universe 
the majority of intergalactic medium (IGM) is ionised, and
almost all of the HI is believed to reside in dense clumps with column densities 
$\gtrsim 10^{20}\text{cm}^{-2}$, which are self-shielded against the ionising ultra-violet (UV)
 background, referred to as Damped Lyman-alpha sytems~\citep[DLAs,][]{wolfe05, prochaska09}.
DLAs are expected to trace the large-scale distribution of galaxies; the corresponding 
cross-correlation was measured, for example, in~\citet{font_ribera12}. Results from numerical 
simulations also prove that neutral hydrogen effectively traces the cosmic web at low 
redshift~\citep{popping09,duffy12,cunnama14,takeuchi14,horii17}. 
Mapping the large-scale neutral hydrogen distribution is therefore important not 
only for the reconstruction of the underlying cosmic web, but also to achieve a more 
complete characterisation of the different baryonic phases across cosmic structures. 
Perhaps the most interesting targets in this sense are the nodes of the cosmic web, i.e. massive 
dark matter haloes, where the interplay between atomic and molecular hydrogen phases is a key 
element in processes of galaxy formation and evolution~\citep{fu10, calette18}.

The total amount of HI in haloes hosting galaxy groups was investigated in~\citet{guo20}, 
where it was found to increase with the halo mass. The study  
separated the contribution of central and satellite galaxies, showing that the latter 
contribute for the majority of HI in haloes with masses above $\sim10^{12}\,h^{-1}\msun$.
Such mass represents the scale at which the HI content in central galaxies starts to be 
strongly suppressed by the feedback from active galactic nuclei~\citep{dimatteo05}, as 
already proposed by previous studies~\citep{kim11, kim17, zoldan17, baugh19}.  
Another conclusion is that the total HI content of a halo does not depend exclusively on 
its mass, but also on its richness and formation history~\citep{guo17}. Such a dependence 
is however less relevant for very massive haloes (above $\sim10^{13}\,h^{-1}\msun$) which 
are most likely assembled by mergers~\citep{genel10}.

These considerations highlight the importance of a proper characterisation of the abundance 
and distribution of HI in dark matter haloes. This subject has been the focus of several studies 
that adopt HI-oriented halo models and provide functional parametrisations for describing 
the HI halo content, fitted over numerical 
simulations~\citep{bagla10,villaescusa_navarro14,villaescusa_navarro18} or observational data, 
most commonly DLA-related quantities~\citep{barnes10,barnes14,padmanabhan17,padmanabhan17b}.  
Although numerical simulations can give valuable hints on this subject, ultimately the proposed 
models need to be tested and verified against observations.  Nonetheless, a universally accepted 
recipe to estimate the amount of HI residing in haloes of known mass, and its related spatial 
distribution, is still lacking. It is then worth considering how existing data can be used 
in new ways to provide alternative constraints.

A promising new probe for neutral gas in haloes is the 21-cm line emission due to the 
hyperfine-split transition in the ground state of neutral 
hydrogen~\citep{pritchard08}. This signal can be conveniently observed with ground-based 
facilities, and is therefore particularly useful at low redshift, where the Lyman-alpha line 
enters 
the ultra-violet range hindering observations of HI absorption. 
Radio 21-cm line observations can be used to detect HI-rich galaxies in large 
sky surveys. Among the resultant catalogues, we can cite the HI Parkes All Sky 
Survey~\citep[HIPASS,][]{meyer04,zwaan05} extending up to $z\sim0.04$
and the Arecibo Legacy Fast ALFA~\citep[ALFALFA,][]{martin10, hoppmann15, haynes18} 
extending up to $z\sim0.06$. HI sky surveys, however, become 
increasingly difficult at higher redshifts, as the faintness of the 21-cm emission requires long 
integration times, or reverting to stacking the spectra of 
galaxies~\citep{delhaize13, rhee13, rhee18}.  
An alternative approach is provided by the intensity mapping technique, which allows spanning large 
sky areas in 21-cm with low resolution to map out the integrated HI emission on large scales, 
foregoing the detection of individual objects. This type of surveys can be quickly carried out 
with ground-based facilities and by tuning the central frequency of receivers it is possible to 
probe the emitting gas at different redshifts, resulting in a tomography of HI in the Universe 
which is ideal for reconstructing the three-dimensional cosmic web.
The  feasibility of tracing cosmic structures using 21-cm maps was first shown by~\citet{pen09}, 
by measuring the cross-correlation between the HIPASS maps and the 6dFGRS catalogue. 
Subsequent works continued on this line of exploiting the correlation between 21-cm maps and galaxy 
catalogues (under the assumption of full correlation between HI fluctuations and galaxy 
overdensities at large scales), to provide constraints on the cosmic HI density and bias 
parameters.~\citet{chang10} combined 21-cm maps obtained with the Green Bank Telescope (GBT) 
with the DEEP2 survey catalogue~\citep{davis01}, while~\citet{masui13} combined GBT maps with 
the WiggleZ galaxy catalogue. Particularly relevant for the current work is the study described 
in~\citet{anderson18}, where the 21-cm maps are the result of observations performed with the 
Parkes radio telescope over the region spanned by the 2dFGRS catalogue. The main difference 
with respect to previous works is the much larger sky coverage, around 1300 deg$^2$. 

A different strategy has been explored in~\citet[][hereafter T19]{tramonte19}, where
the same combination of Parkes maps and 2dF galaxies has been employed to search for the HI 
emission of filamentary structures in the cosmic web. The signal was searched for directly at 
the map level, by coherently stacking the contribution from the region in between selected pairs 
of 2dF galaxies. The results set upper limits on the expected amount of HI in large-scale 
filaments, and for the first time showed the feasibility of this type of search directly on 
21-cm maps. In this work, we want to conduct a similar analysis, but focussing this time on 
the HI found within haloes. Once again, the 2dF galaxies are used to trace the location of the 
cosmic web nodes, and, by taking advantage of the large sky coverage of these maps, a blind 
stacking technique is employed to enhance the halo signal over the background. The first, main 
goal of this work is to detect the halo 21-cm emission and to measure the associated HI radial 
profile. Secondly, we want to choose suitable functional forms discussed in the literature, 
describing the HI abundance in haloes, and test their prediction on our measurements. This 
procedure will allow us to fit for the mean mass of the objects we are detecting and for the 
concentration of the HI gas. For the first time, models are tested directly on profiles 
reconstructed from 21-cm maps.   

We stress that the resolution of the available 21-cm intensity maps does 
not allow to resolve galactic-scale haloes in our study. This work targets instead 
the large-scale massive haloes in the 
cosmic web, at the scale of galaxy clusters and above ($M \gtrsim 10^{13}\,\msun$).
In this context the 2dF objects serve as tracers of the underlying large-scale cosmic web,
assuming full correlation between HI fluctuations and galaxy overdensity as in the aforementioned
cross-correlation studies. Any detectable continuous radial HI emission at these scales 
is not to be ascribed to a diffuse neutral component in the IGM, which is highly ionised, but rather
to the integrated contribution of the HI hosted in the halo member galaxies.
Nonetheless, the very methodology described in this work can be applied to characterise the 
HI amount and distribution in galactic-scale haloes, provided intensity maps with the required 
resolution are available. It represents therefore a promising tool to analyse the  
 next generation of HI survey data.

This paper is organised as follows. The description of the data set we employ is 
presented in Section~\ref{sec:data}. The methodology adopted for processing the 21-cm maps, 
and the corresponding results, are presented in Section~\ref{sec:stacking}
and further discussed in Section~\ref{sec:datadiscussion}. The theoretical model we employ
 for computing the HI halo profiles 
is detailed in Section~\ref{sec:theomodel}; in Section~\ref{sec:estimation} we describe instead
the corresponding parameter estimation. Finally, Section~\ref{sec:conclusions} lists the conclusions. 
Throughout this paper we adopt a flat-$\Lambda$CDM cosmological model with $\Omega_{\rm m}=0.3$, 
$\Omega_{\rm b}=0.049$ and $H_0=70\,\text{km}\,\text{s}^{-1}\text{Mpc}^{-1}$.


\begin{figure*}
\includegraphics[trim= 10mm 0mm 0mm 0mm, scale=0.36]{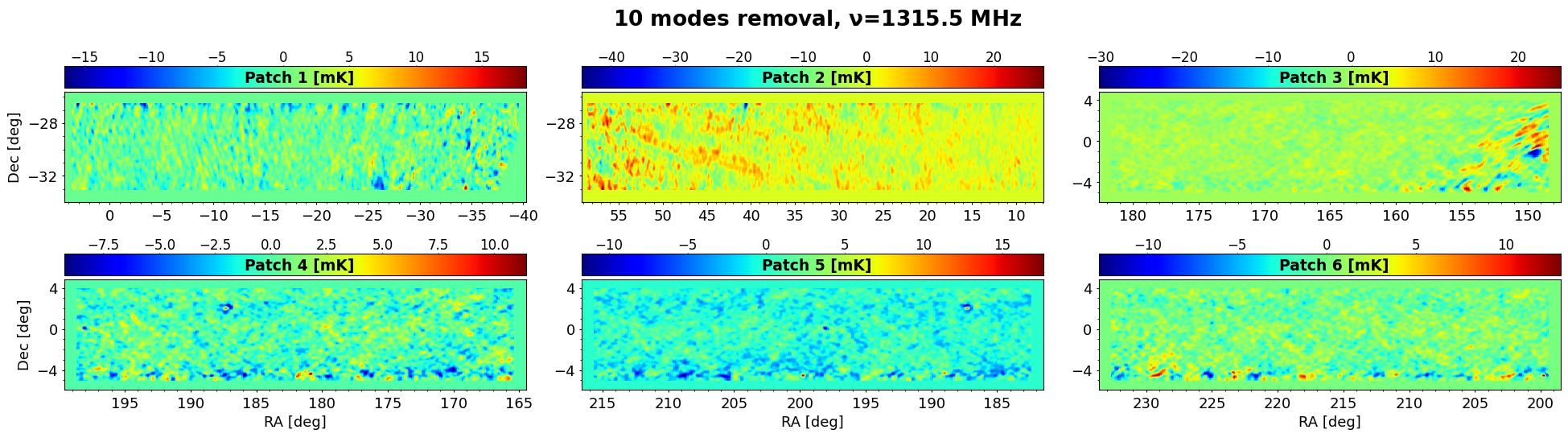}
\newline
\includegraphics[trim= 10mm 0mm 0mm 0mm, scale=0.36]{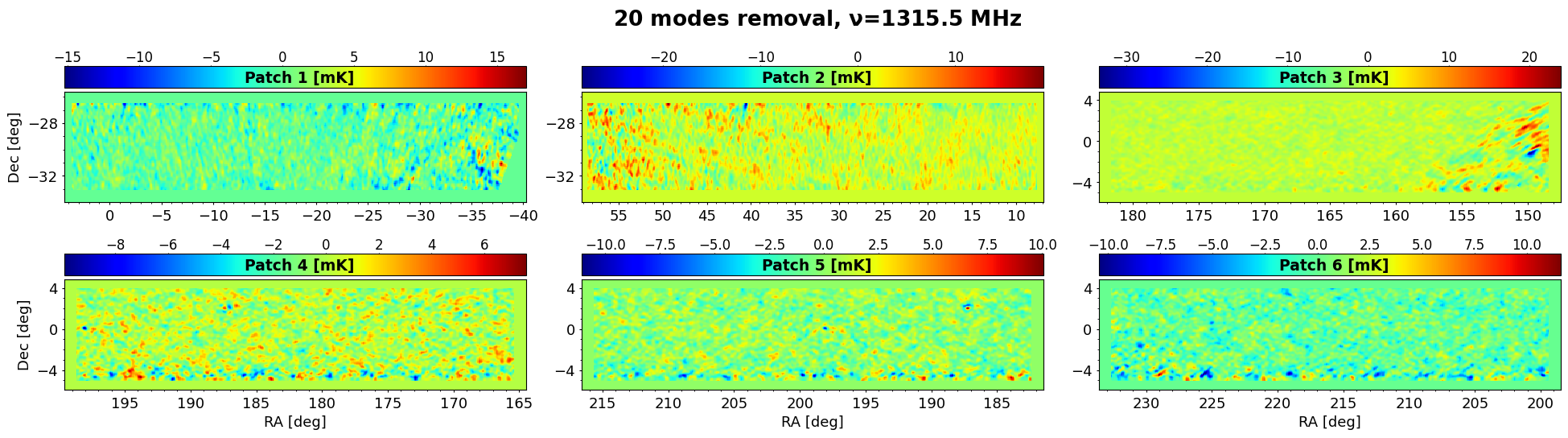}
\newline
\includegraphics[trim= 10mm 10mm 0mm 0mm, scale=0.36]{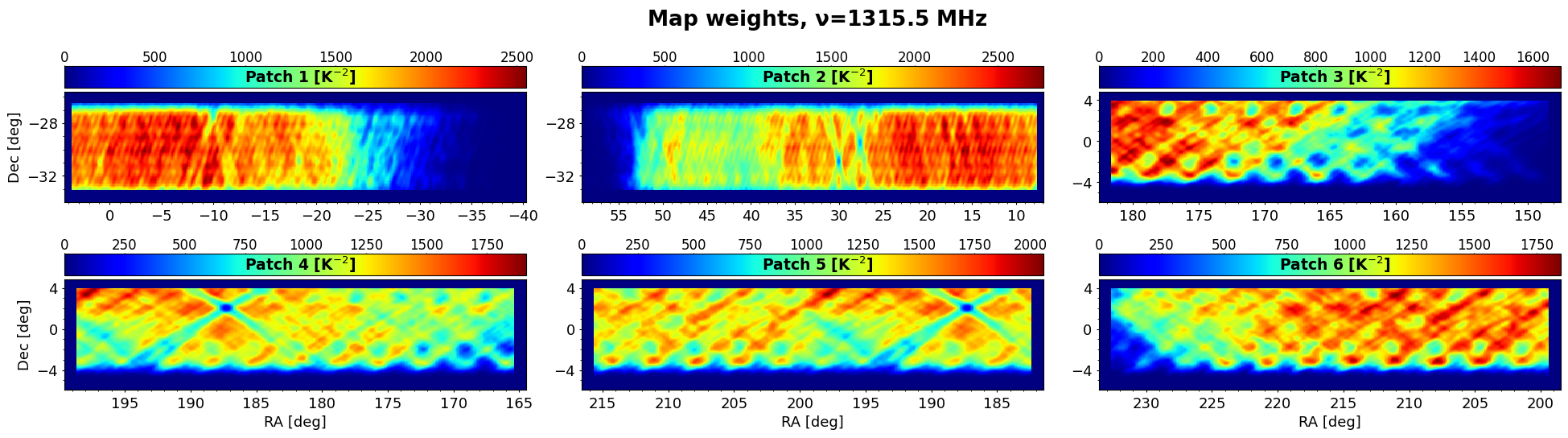}
	\caption{
	Top: Parkes 21-cm intensity maps of all six patches for the 10-mode removal case, 
	sliced at the Multibeam Receiver central observing frequency and plotted in mK units 
	in (RA, Dec) coordinates. Middle: same as in the top panel, but for the 20-mode 
	foreground removal case. Bottom: Parkes weight maps for the same frequency slice, 
	plotted in $\textrm{K}^{-2}$ units. The crossed stripes result from different telescope
	raster scans at constant elevation; the locations where bright radio sources have been 
	masked are also visible as dark spots.}
\label{fig:parkes}	
\end{figure*}

\begin{table*}
\centering
\caption{Table summarising the main properties of the data set used in this work. For each 
	Parkes patch we quote its central position and resolution (pixel size) in right 
	ascension (RA, 486 pixels), declination (Dec, 106 pixels) and frequency ($\nu$, 64 pixels). 
	The number of pixels in the three axes is the same for all patches. The RA pixel size 
	in the low declination patches is larger because it is measured in degrees of polar rotation. 
	Finally, the table reports the total number of 2dFGRS galaxies located within each patch, 
	for the three different galaxy samples employed in this work.}
\label{tab:dataset}	
\setlength{\tabcolsep}{1.2em}
\begin{tabular}{p{0.2cm}p{0.01cm}p{.8cm}p{1.1cm}p{0.01cm}p{0.8cm}p{1.1cm}p{0.01cm}p{.8cm}p{1.1cm}p{0.01cm}p{.9cm}p{.9cm}p{.9cm}}
\hline
	Patch & & \multicolumn{2}{c}{RA [deg]} & & \multicolumn{2}{c}{Dec [deg]} & & \multicolumn{2}{c}{$\nu$ [MHz]} & & $N_{\rm gal}$ & $N_{\rm gal}$ & $N_{\rm gal}$ \\ 
\\[-1em]
\cline{3-4}\cline{6-7}\cline{9-10}
\\[-1em]
	& & Centre &Pixel size& & Centre &Pixel size& & Centre &Pixel size& & [Full] & [Central] & [Satellite]\\
\hline
	1 & & -18.0 & 0.092 & & -29.75 & 0.080 & & 1315.5 & 1.0 & & 7343 & 2209 & 5134 \\ 
	2 & & 33.0 & 0.092 & & -29.75 & 0.080 & & 1315.5 & 1.0 & & 7708 & 2438 & 5270 \\ 
	3 & & 165.0 & 0.080 & & -0.05 & 0.080 & & 1315.5 & 1.0 & & 9394 & 2658 & 6736 \\ 
	4 & & 182.0 & 0.080 & & -0.05 & 0.080 & & 1315.5 & 1.0 & & 9960 & 2688 & 7272 \\ 
	5 & & 199.0 & 0.080 & & -0.05 & 0.080 & & 1315.5 & 1.0 & & 9625 & 2542 & 7083 \\ 
	6 & & 216.0 & 0.080 & & -0.05 & 0.080 & & 1315.5 & 1.0 & & 4310 & 1444 & 2866 \\ 
\hline
	Total & &  &  & &  &  & & & & &  48430  &  13979 & 34361 \\ 
\hline
\end{tabular}
\end{table*}
\section{Data set}
\label{sec:data}
The analysis addressed in this work requires the combined use of a galaxy catalogue, complete 
with angular and redshift information, and of foreground cleaned 21-cm maps, structured as 
three-dimensional data cubes. Clearly, the angular coverage and redshift span of these two 
data sets must have a significant overlap, in order to 
allow the association of the objects we are targeting with their HI emission.
The choice for this work consists of 21-cm maps from the Parkes 
radio telescope over the 2dF galaxy catalogue region.
This very data set was also employed in~\citetalias{tramonte19} and proved very efficient 
in the search for HI signal in intergalactic filaments. We review it in the following.

\subsection{21-cm maps}
\label{ssec:parkes}
The maps used in this work are the result of observations conducted using the Parkes radio 
telescope, equipped with the 21-cm Multibeam Receiver~\citep{staveley_smith96}, during a single 
week between April and May 2014; these observations specifically covered the volume spanned by 
the 2dF galaxy catalogue. In total, 152 hours of survey were conducted with the Multibeam 
Correlator, with a central observing frequency of 1315.5 GHz and a 64 MHz full bandwidth split 
into 1024 frequency channels with a 62.5 kHz resolution each. However, the frequency resolution 
was subsequently degraded to $1\,\text{MHz}$ after removing local radio frequency interference. 
The corresponding redshift range for the 21-cm emission spanned by the map is 
$0.06\lesssim z \lesssim 0.10$. The total sky coverage is approximately 1,300 square degrees, 
divided into a northern-galactic and a southern-galactic stripe. The map resolution is set by 
the Parkes beam, with a 14' full width at half maximum (FWHM).

The full processing of the Multibeam Receiver data to yield the final 21-cm maps is detailed 
in~\citet{anderson18}. In order to improve the efficiency 
of the map-making code, the northern stripe was split into four different patches, and the 
southern one into two. These patches, 
although partially overlapping, underwent bandpass calibration and foreground removal 
independently. The current study takes it into account by initially performing separate 
analysis on different patches (see Section~\ref{ssec:galstack}). Table~\ref{tab:dataset}
summarises the location, size and pixel resolution for all the patches, in both the 
angular and frequency space. 

One of the most critical steps in yielding the final HI maps is the removal 
of the other radio foregrounds entering this frequency range, particularly Galactic 
synchrotron, that are 2 to 3 orders of magnitude higher than the underlying HI 
signal. By taking advantage of the smooth spectral distribution of the foregrounds, 
compared to the clumpy nature of the HI emission~\citep{liu12}, the cleaning is done with a 
principal component analysis (PCA) which discards modes with higher frequency 
correlation~\citep{switzer15}. Although the first modes are expected to be dominated by 
foregrounds, they still carry a marginal HI signal. As a result, a higher number 
of removed modes yields cleaner maps but produces also a higher loss of 
the targeted signal. The work in~\citet{anderson18} established the 10-mode 
removal case as the optimal choice. In the current work we will use both the 
10- and 20-mode removed maps to assess the effect of the foreground removal 
on our results. The same strategy, indeed, was adopted in~\citetalias{tramonte19} 
and showed some difference between the two cases when considering individual 
patches. 

The 10 and 20-mode removed maps are therefore the starting point of our analysis, 
and are shown in Fig.~\ref{fig:parkes} sliced at the central observational frequency;
the value stored in each pixel is the observed HI brightness temperature. 
The same figure also shows the corresponding weight maps, which are necessary to 
account for the inhomogeneous sky coverage resulting from the telescope 
azimuthal drift scans. Each pixel stores the inverse squared noise weight, 
which is roughly proportional to the observing time with pointings within that 
pixel.

\subsection{Galaxy catalogue}
\label{ssec:2dfgalaxies}
We employ the galaxy catalogue from the 2dF Galaxy Redshift 
Survey~\citep{colless01}, which was carried out with the 
 Anglo-Australian Observatory 3.9-meter telescope 
over the years 1997 to 2002. The majority of the objects targeted by the 2dF 
spectroscopic observations lie in the fields covered by the Parkes HI maps. The 
spectroscopic catalogue, which is publicly available\footnote{``Catalogue of best 
spectroscopic observations'' at \url{http://www.2dfgrs.net/}.}, provides spectroscopic 
redshift information for 245,591 objects, and has to be appropriately queried to extract 
the most suitable galactic targets for our analysis. 

First of all, we consider only sources with a reliable positive redshift estimate 
(according to a specific quality flag in the data), lowering the total number of sources 
to 227,190. At this point, we split the catalogue into six sub-catalogues contained in 
the 21-cm map patches. We discard sources located within $1\,\text{deg}$ in declination 
and $3\,\text{deg}$ in right ascension from the Parkes patch edges (patches are indeed
more extended in right ascension than in declination by a factor $\sim 3$) as these regions are 
more likely affected by strong residual contaminations. 
For the same reason we excise all sources whose redshift falls within the lowest 10 MHz
or the highest 4 MHz in the frequency range covered by Parkes, 
following the criterion adopted by~\citet{anderson18}. Notice that if a galaxy falls in 
the intersection area of two patches, it is included into both of the corresponding 
sub-catalogues. This query reduces the total galaxy number to 48,430 objects, counting the 
repetitions in neighbouring patches. Hereafter we shall refer to this selected ensemble of 
galaxies as the full 2dF galaxy sample.

\begin{figure}
\includegraphics[trim= 20mm 20mm 0mm 0mm, scale=0.26]{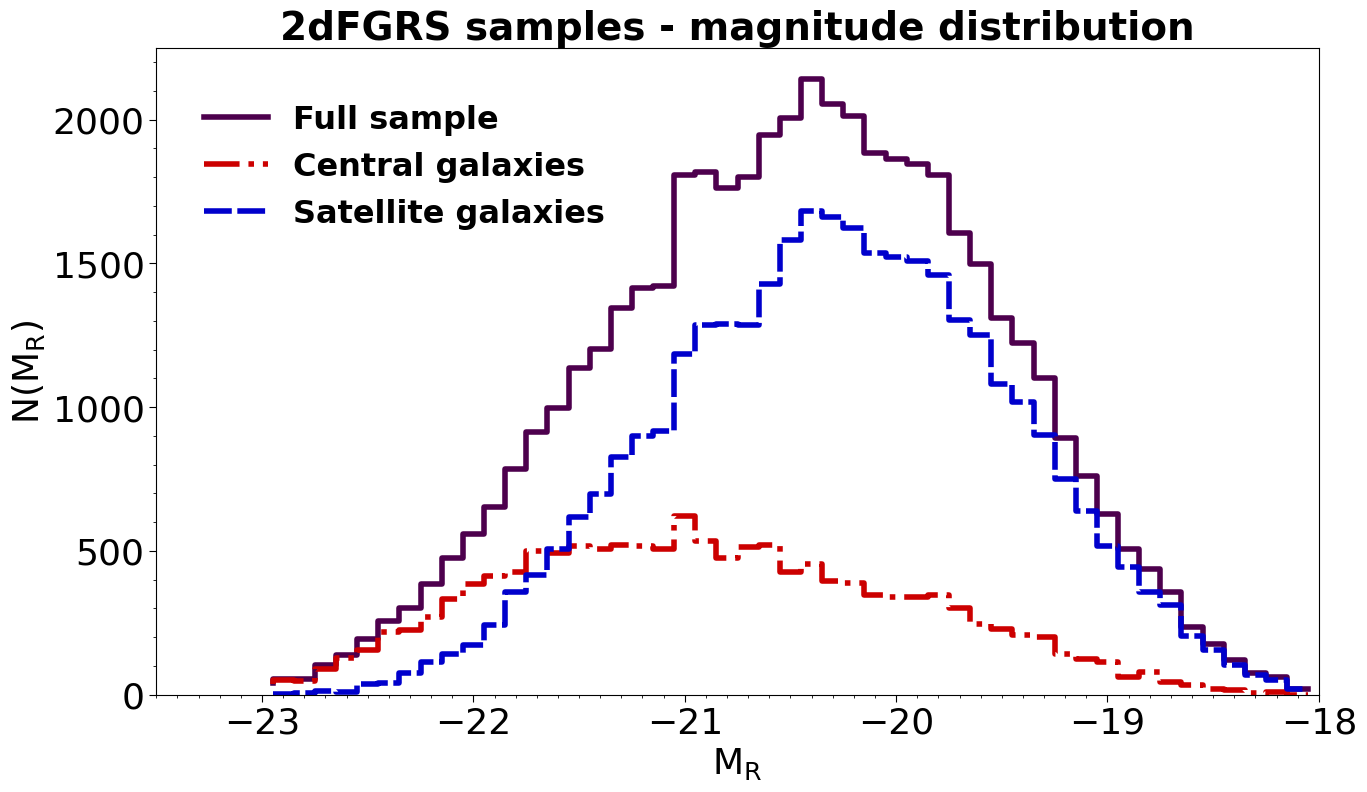}
\vspace{0.1cm}
\caption{R-band absolute magnitude distribution, computed over bins with size 0.1, 
for the three samples of galaxies considered in this work. These three samples are 
the full 2dF galaxy sample bounded by the Parkes data volume, the selected 
locally brighter galaxies (the central galaxy sample) and the complementary sample 
between the two, made by satellite galaxies.}
\label{fig:2df}	
\end{figure}

The selected galaxies serve to identify the position on the maps
of their hosting dark matter haloes, whose HI radial profile is the object of the present study. 
It is clear that the 2dF sample selected with the procedure we just described contains galaxies with 
different masses and morphological types, and belonging to different environments in the 
cosmic web. Although the results of our analysis are mostly insensitive to the properties of 
individual galaxies, we can still apply a further selection by  
exploiting their magnitude information, following the criterion adopted 
in~\citetalias{tramonte19}. In that case, the galaxy catalogue served to locate the 
endpoints of large-scale filaments, corresponding to massive haloes likely hosting 
galaxy clusters. The 2dF catalogue was then queried to extract the locally brighter 
galaxies, that are the most likely to be located in the centres of those haloes. We shall 
repeat the same selection here, and consider a galaxy ``central'' if no brighter galaxy is 
found within a projected separation of $1.0\,h^{-1}\text{Mpc}$ and a line-of-sight (LoS) 
separation of $|c\Delta z| =1000\,\text{km}\,\text{s}^{-1}$~\citep[this isolation criteria 
was initially proposed in][]{planck13}. We find in total 13,979 central galaxies. For the 
current work, however, it is also interesting to consider the complementary sample of 34,361 
satellite galaxies. As a result, we are left with three galaxy samples that will be used for 
our study: the full 2dF sample, the one made by the central galaxies and the one made by the 
satellite galaxies. In Fig.~\ref{fig:2df} we show the magnitude distributions of these three 
2dF sub-catalogues: as expected, the brighter, central galaxies have a distribution centred 
on lower values of the absolute magnitude, while the satellite galaxies are more numerous and 
fainter on average. Although it is not possible to study the profile for galaxies with similar 
magnitude, this strategy still provides a magnitude-based separation of the 2dF catalogue, 
which can be useful to assess the dependence of the halo HI content on the local galaxy 
environment. The way the full, the central and the satellite catalogues are distributed across 
the different patches is reported in Table~\ref{tab:dataset}.


\begin{figure*}
\includegraphics[trim= 0mm 0mm 0mm 0mm, scale=0.25]{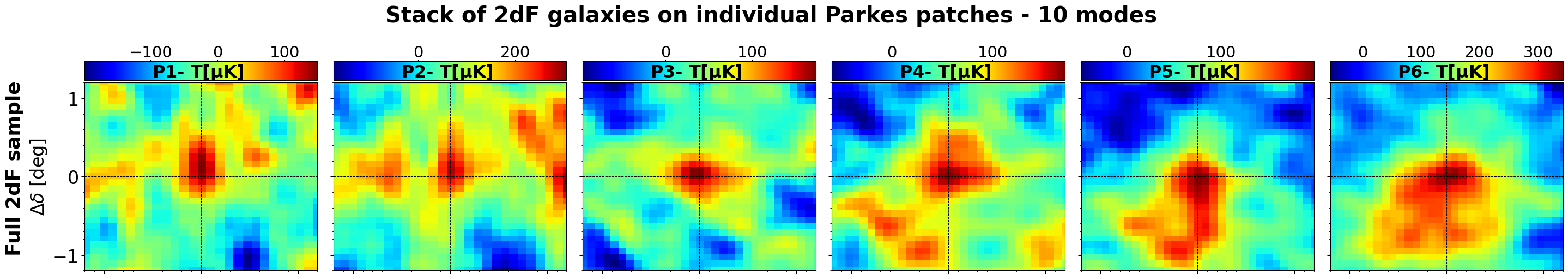}
\newline
\includegraphics[trim= 0mm 0mm 0mm 0mm, scale=0.25]{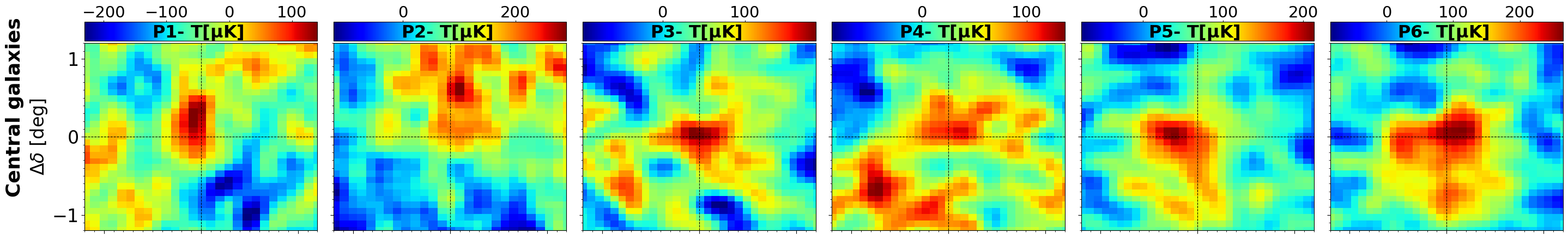}
\newline
\includegraphics[trim= 0mm 0mm 0mm 0mm, scale=0.25]{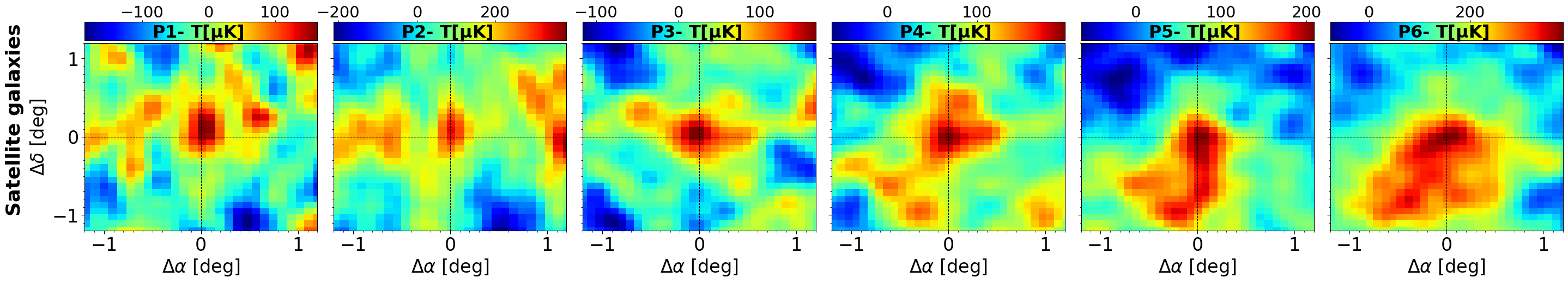}
\caption{Results of the galaxy stacking procedure described in Section~\ref{ssec:galstack}, 
	shown separately for the six Parkes patches in the 10-mode removed maps case. 
	Different rows show the results of stacking the full 2dF galaxy sample (top row), 
	the central galaxy sample (middle row) and the satellite galaxy sample (bottom row).}
\label{fig:patch10modes}	
\end{figure*}

\begin{figure*}
\includegraphics[trim= 0mm 0mm 0mm 0mm, scale=0.25]{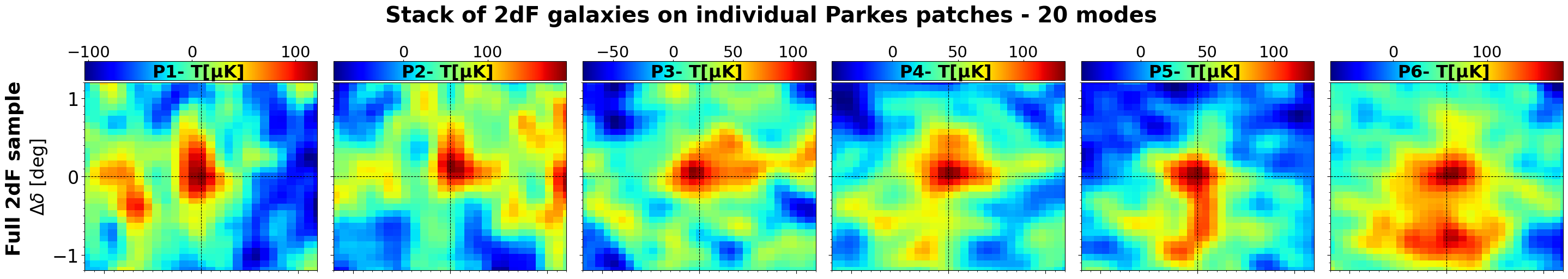}
\newline
\includegraphics[trim= 0mm 0mm 0mm 0mm, scale=0.25]{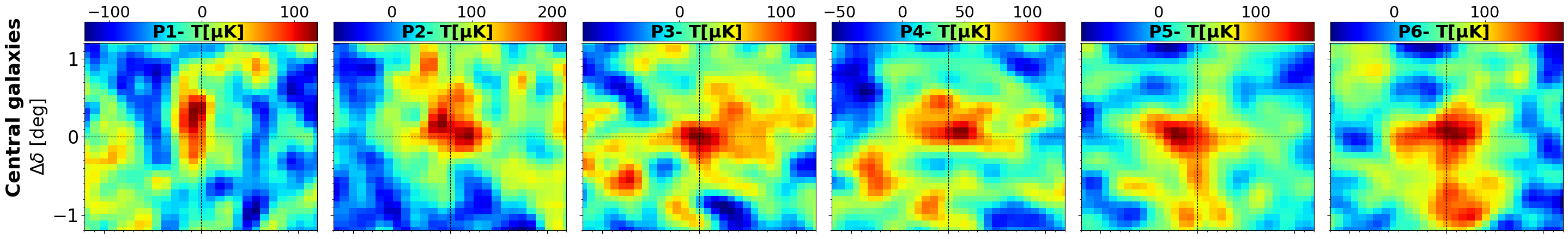}
\newline
\includegraphics[trim= 0mm 0mm 0mm 0mm, scale=0.25]{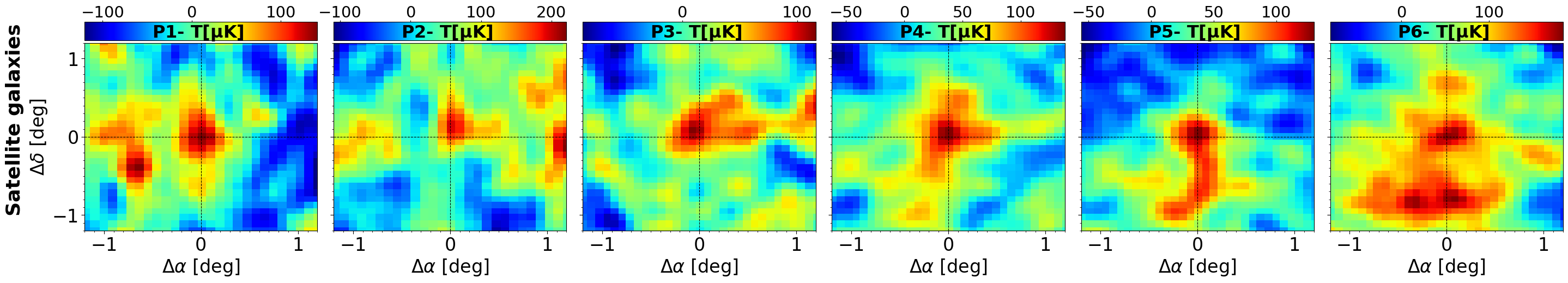}
\caption{Same as in Fig.~\ref{fig:patch10modes}, but showing the stacks on the 20-mode removed maps.}
\label{fig:patch20modes}	
\end{figure*}

\section{HI profiles from intensity maps}
\label{sec:stacking}

In this section, we describe the methodology we employ to combine the Parkes HI maps and the 
2dF galaxies for extracting the observed HI halo profile. The low resolution of intensity mapping 
does not allow the detection of individual galaxies at the map level, requiring further data 
processing to enhance the signal we are interested in. In our case, we revert to stacking the map 
area centred at the selected galaxy locations, taking advantage of the full statistics available 
for the three galaxy samples defined in Section~\ref{ssec:2dfgalaxies}. The HI profile is then 
extracted from the resultant stacked maps, after symmetrizing the signal around the halo centre. 
The uncertainty associated with this reconstructed profile is obtained with a bootstrap method 
by repeating the stacks on a set of randomized galaxy samples. Each of these steps is detailed 
in the following. A discussion of the results is instead provided in 
Section~\ref{sec:datadiscussion}. 

\begin{figure*}
\includegraphics[trim= -14mm 0mm 0mm 0mm, scale=0.19]{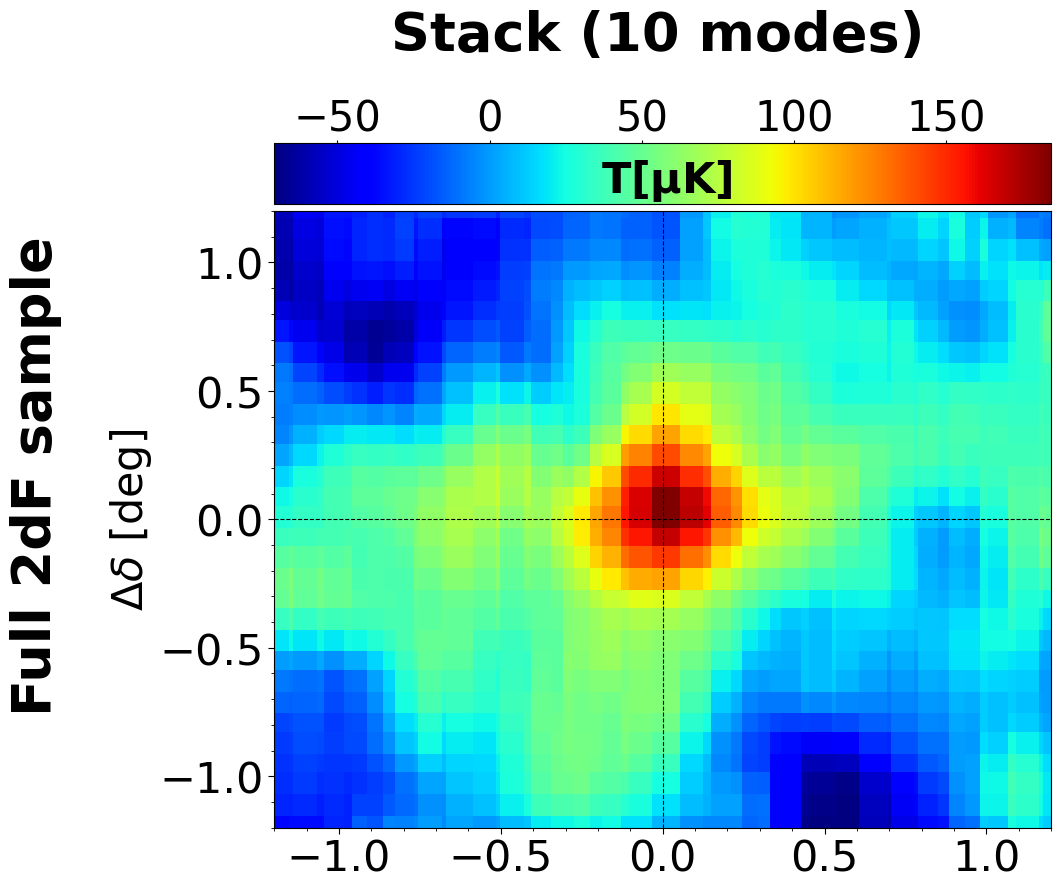}\quad\quad
\includegraphics[trim= 0mm 0mm 0mm 0mm, scale=0.19]{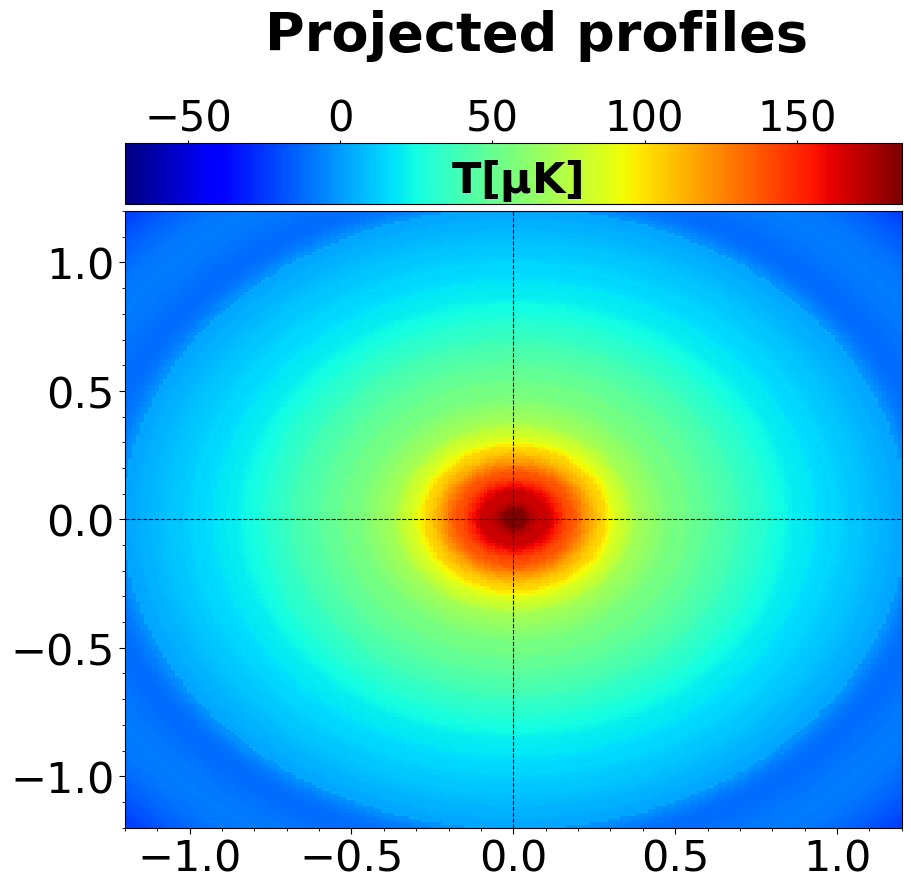}\quad\quad
\includegraphics[trim= 0mm 0mm 0mm 0mm, scale=0.19]{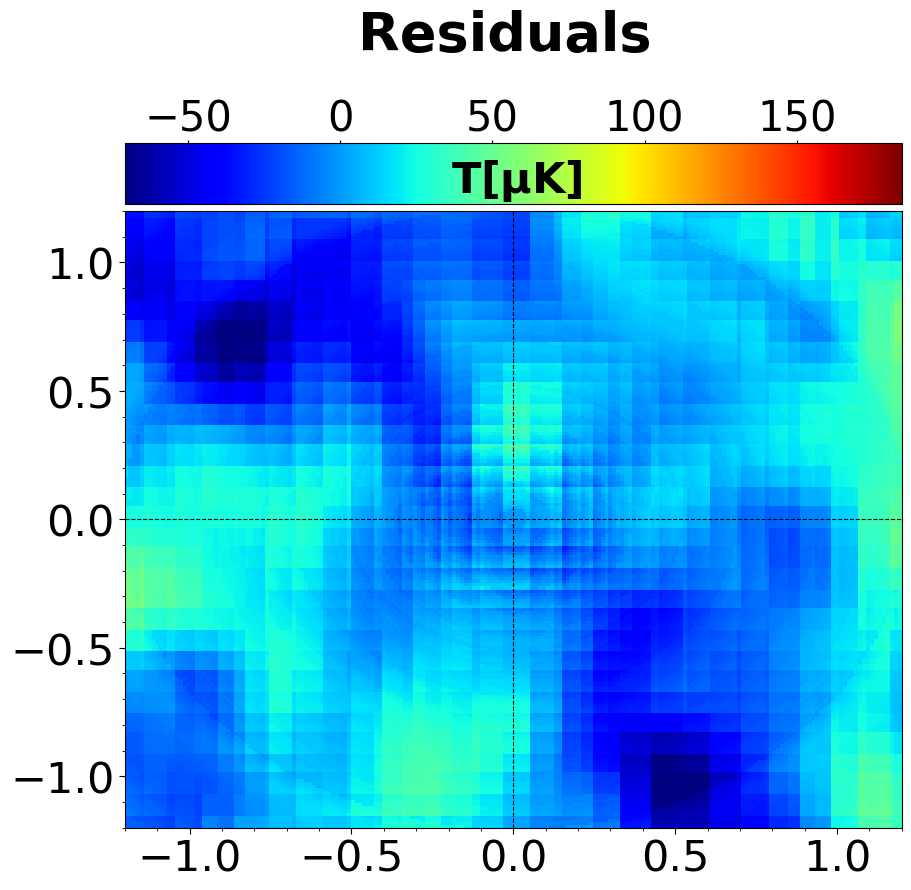}\quad
\newline
\includegraphics[trim= -25mm 0mm 0mm 0mm, scale=0.19]{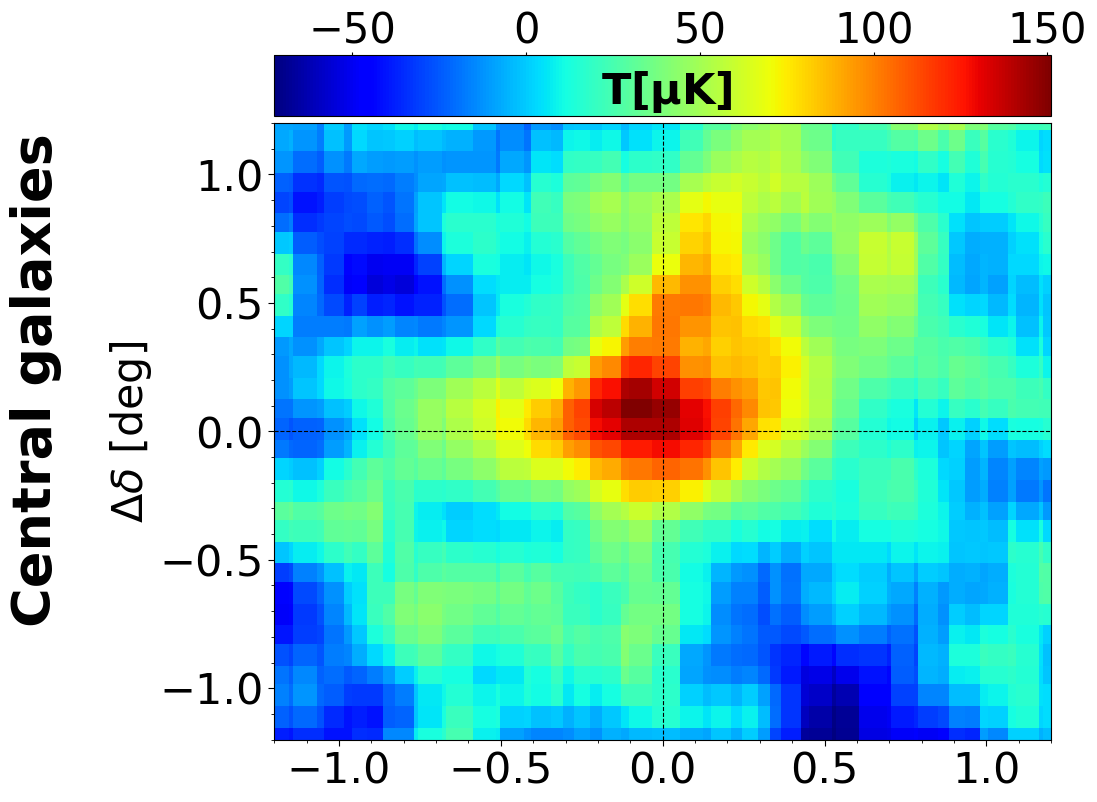}\quad
\includegraphics[trim= 0mm 0mm 0mm 0mm, scale=0.19]{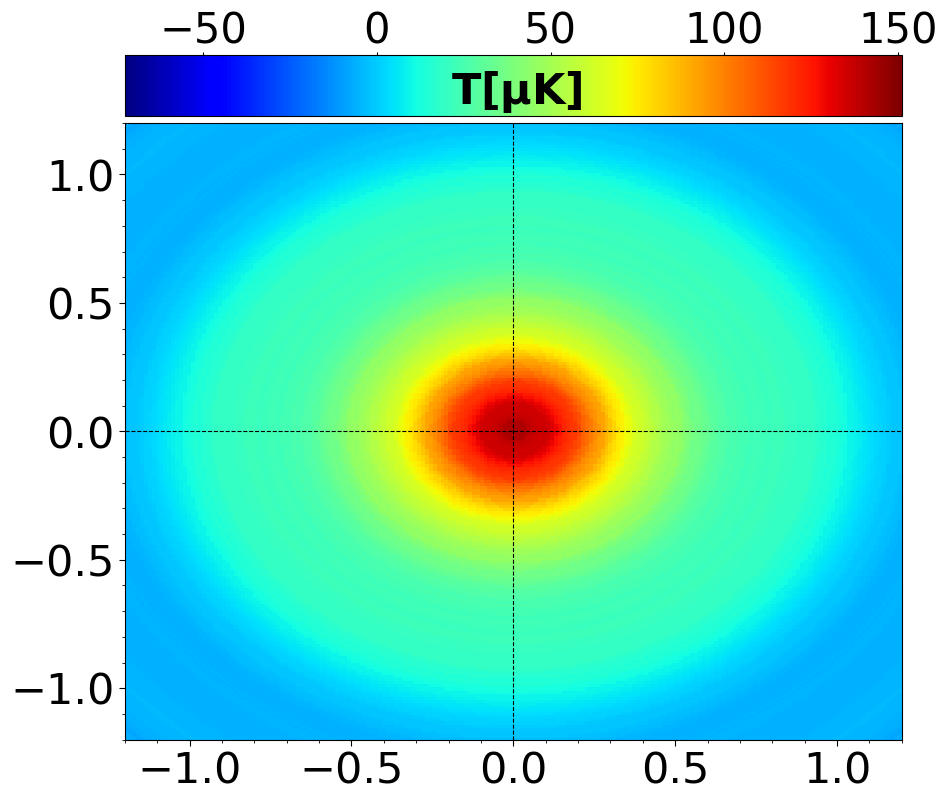}\quad
\includegraphics[trim= 0mm 0mm 0mm 0mm, scale=0.19]{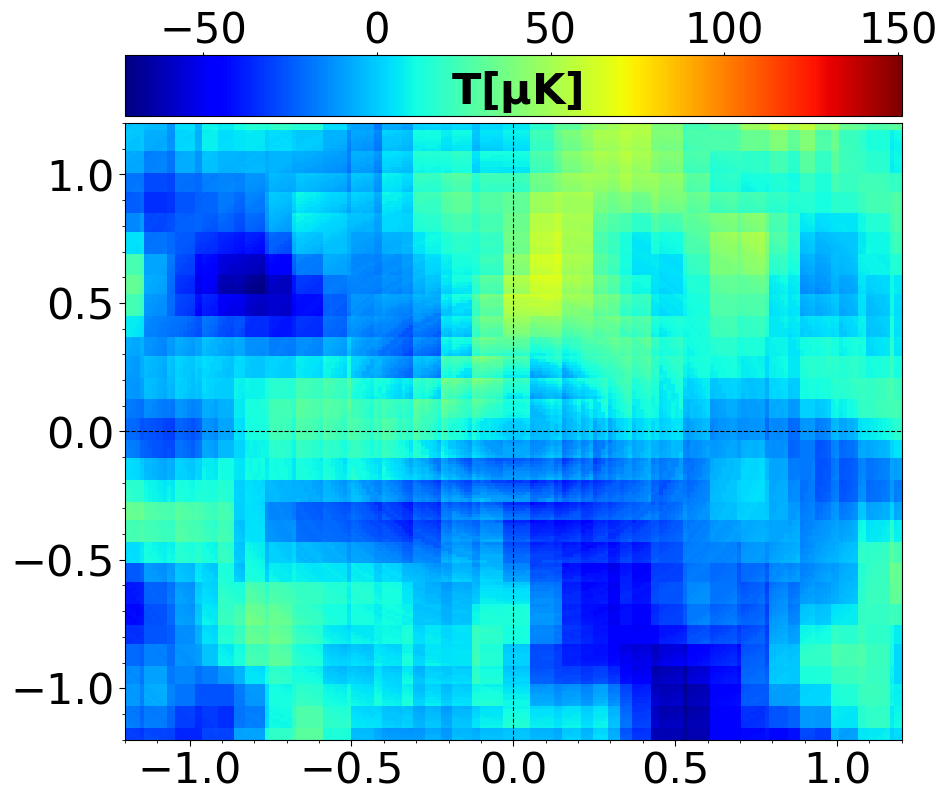}\quad
\newline
\includegraphics[trim= 0mm 0mm 0mm 0mm, scale=0.19]{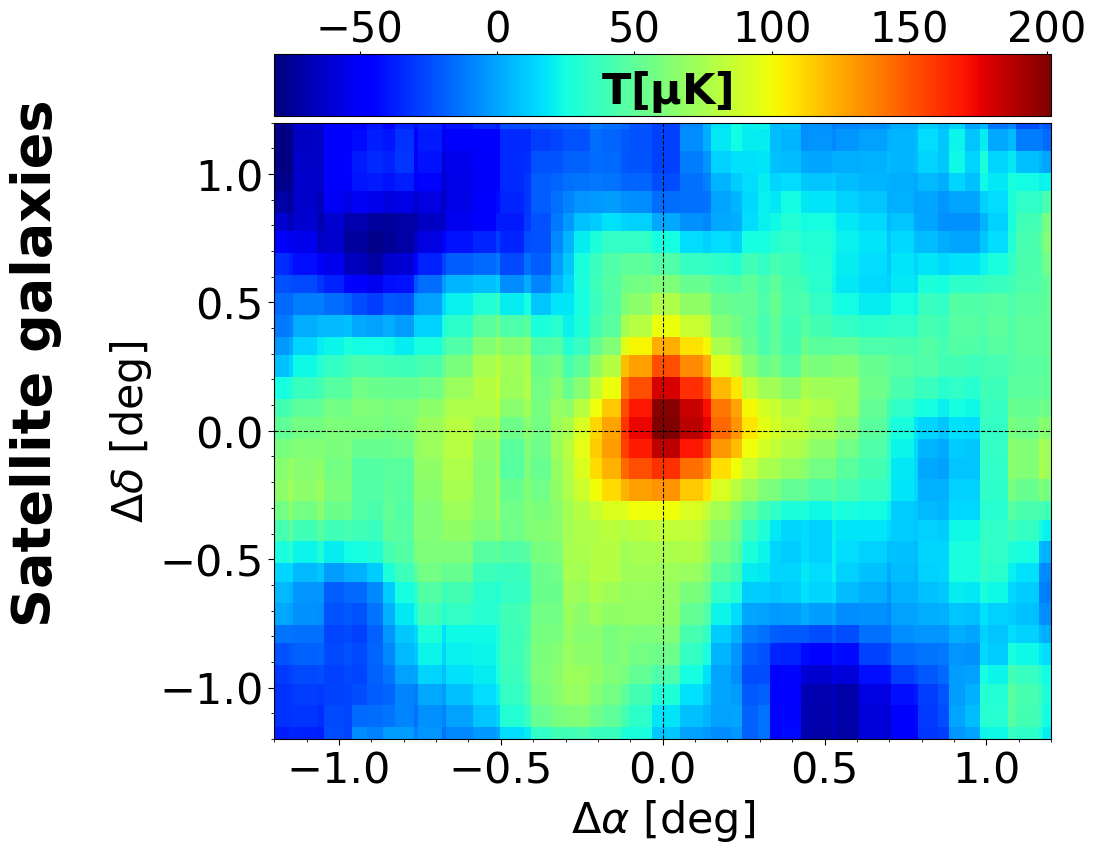}\quad
\includegraphics[trim= 0mm 0mm 0mm 0mm, scale=0.19]{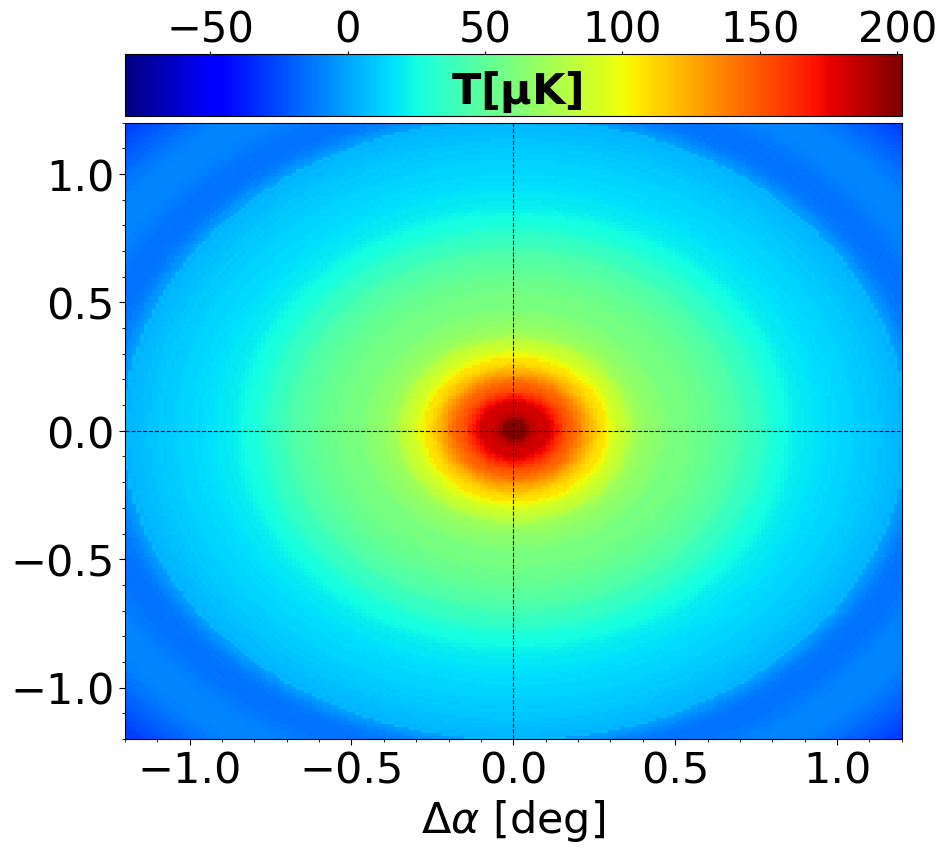}\quad
\includegraphics[trim= 0mm 0mm 0mm 0mm, scale=0.19]{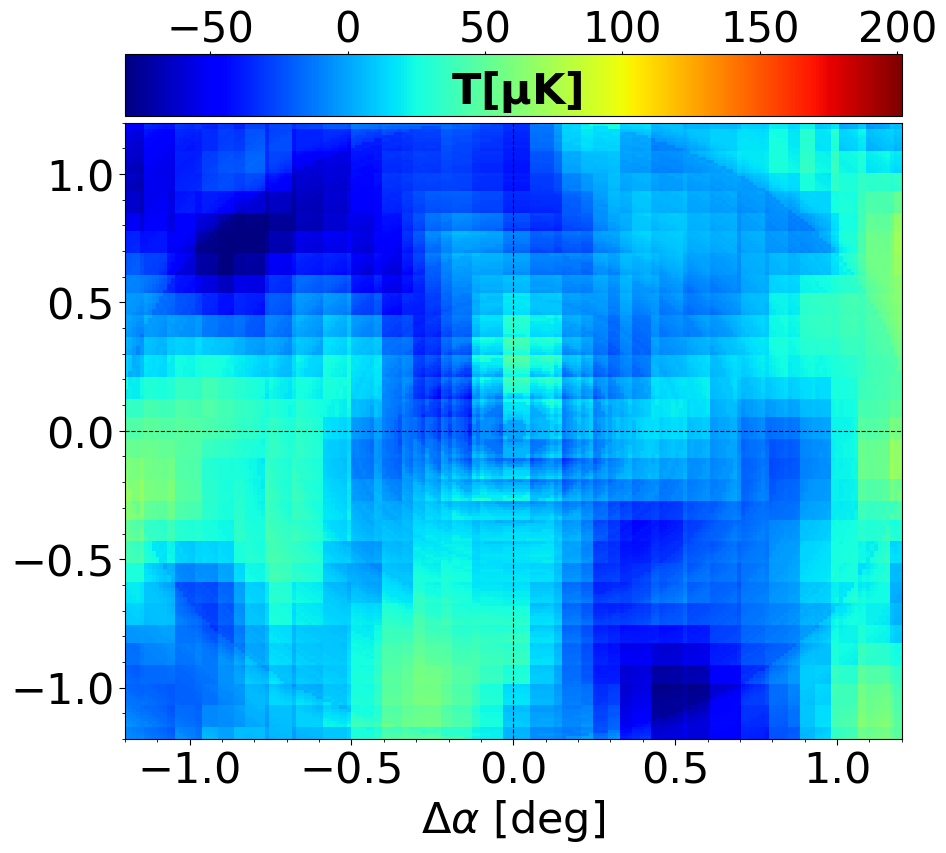}\quad
\caption{Results from the data analysis described in this section. Rows from top to bottom show, 
	in order, the results from stacking the full selected 2dF galaxy sample, the subsample 
	of locally brighter galaxies and the subsample of satellite galaxies. For each case, we 
	show the final stack obtained combined the six Parkes patches (left column), 
	the circular-symmetrised modelled emission obtained from the stack (central column) 
	and the residual map obtained by subtracting the model from the stack (right column).}
\label{fig:stack10modes}	
\end{figure*}

\begin{figure*}
\includegraphics[trim= -14mm 0mm 0mm 0mm, scale=0.19]{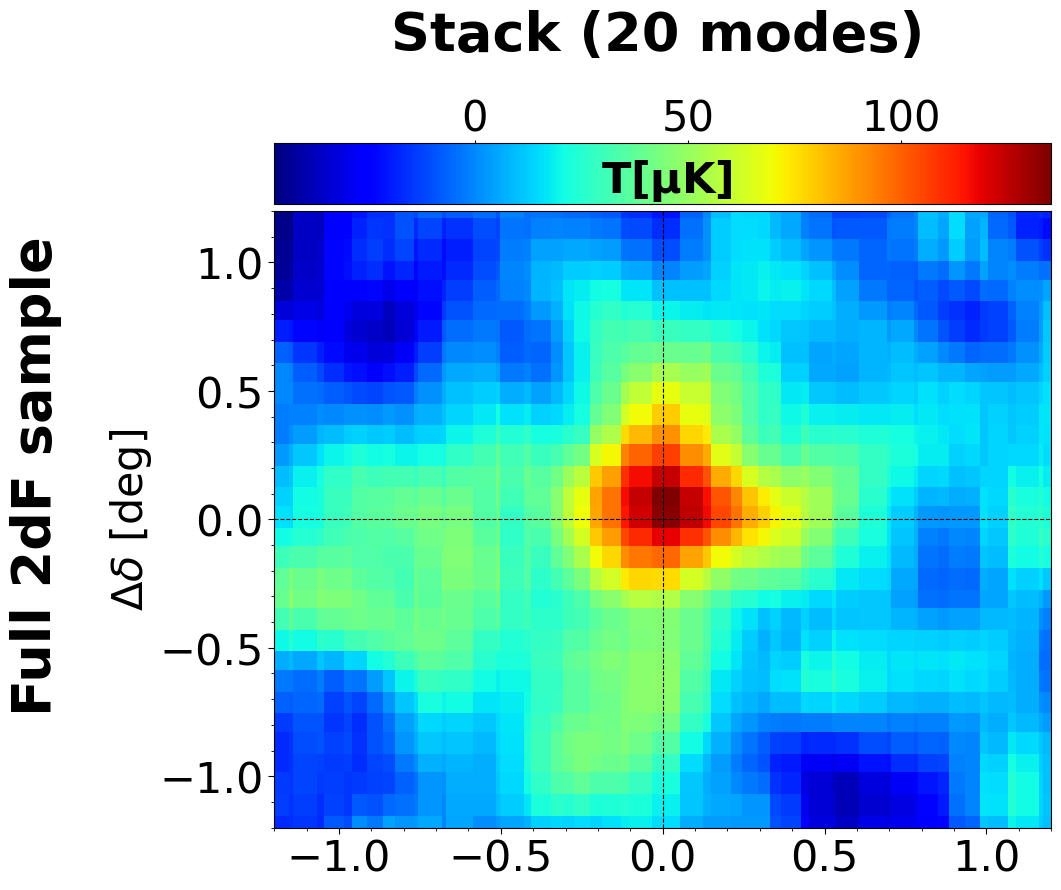}\quad
\includegraphics[trim= 0mm 0mm 0mm 0mm, scale=0.19]{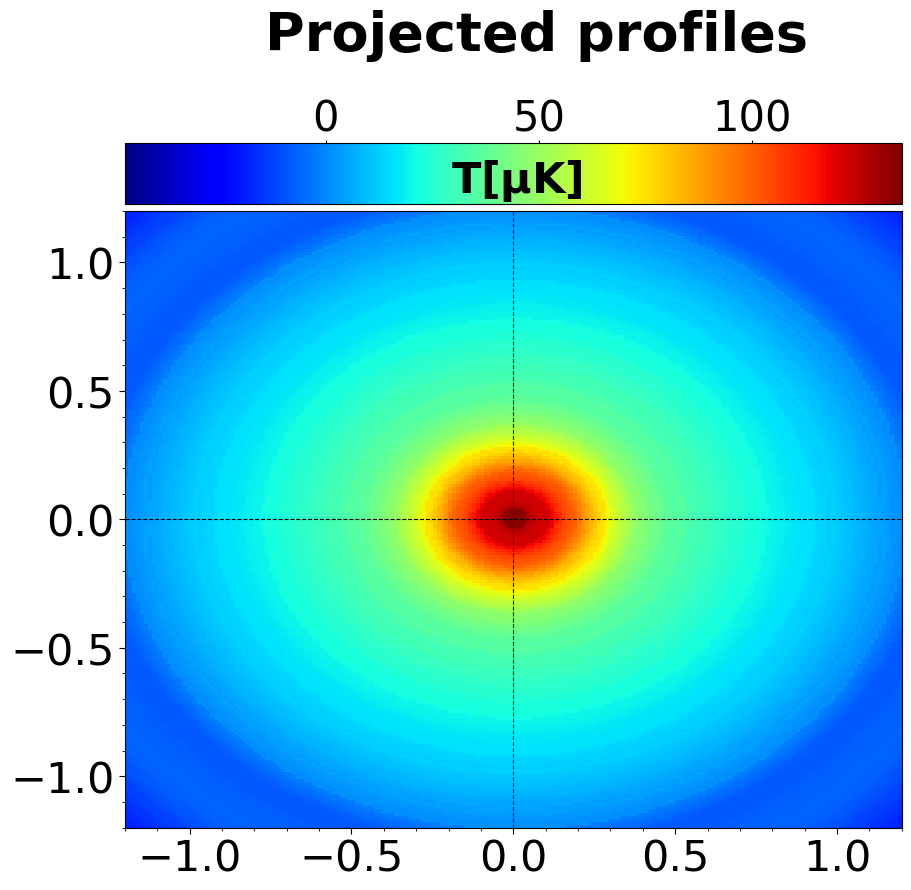}\quad
\includegraphics[trim= 0mm 0mm 0mm 0mm, scale=0.19]{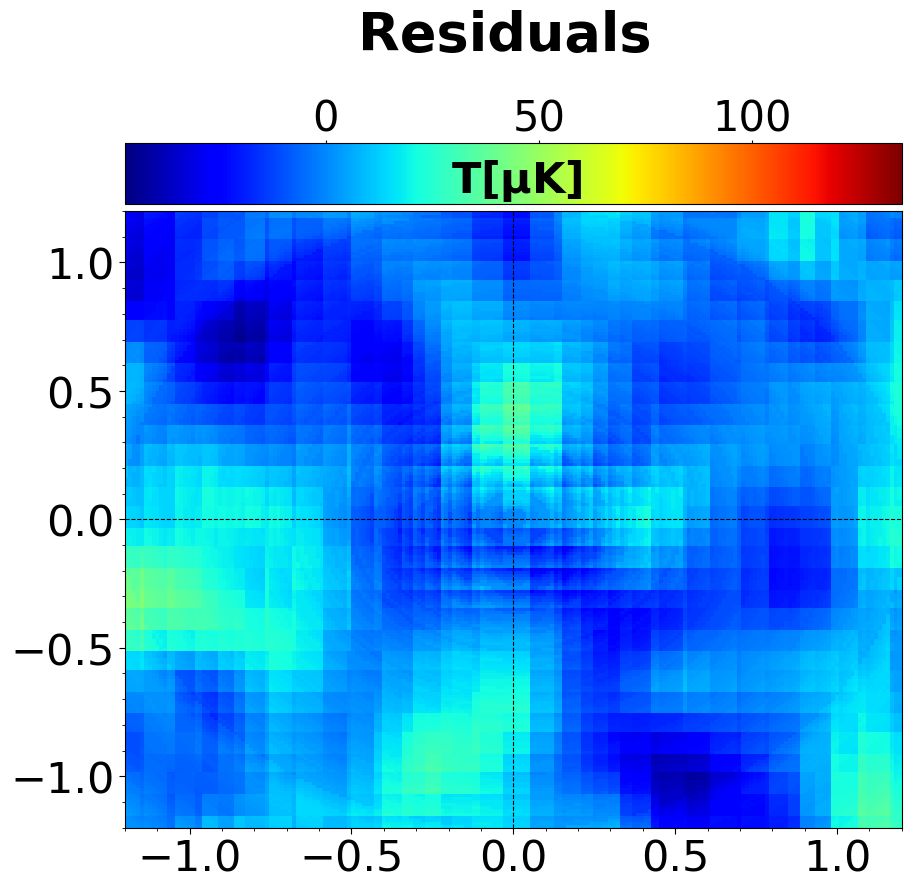}\quad
\newline
\includegraphics[trim= -25mm 0mm 0mm 0mm, scale=0.19]{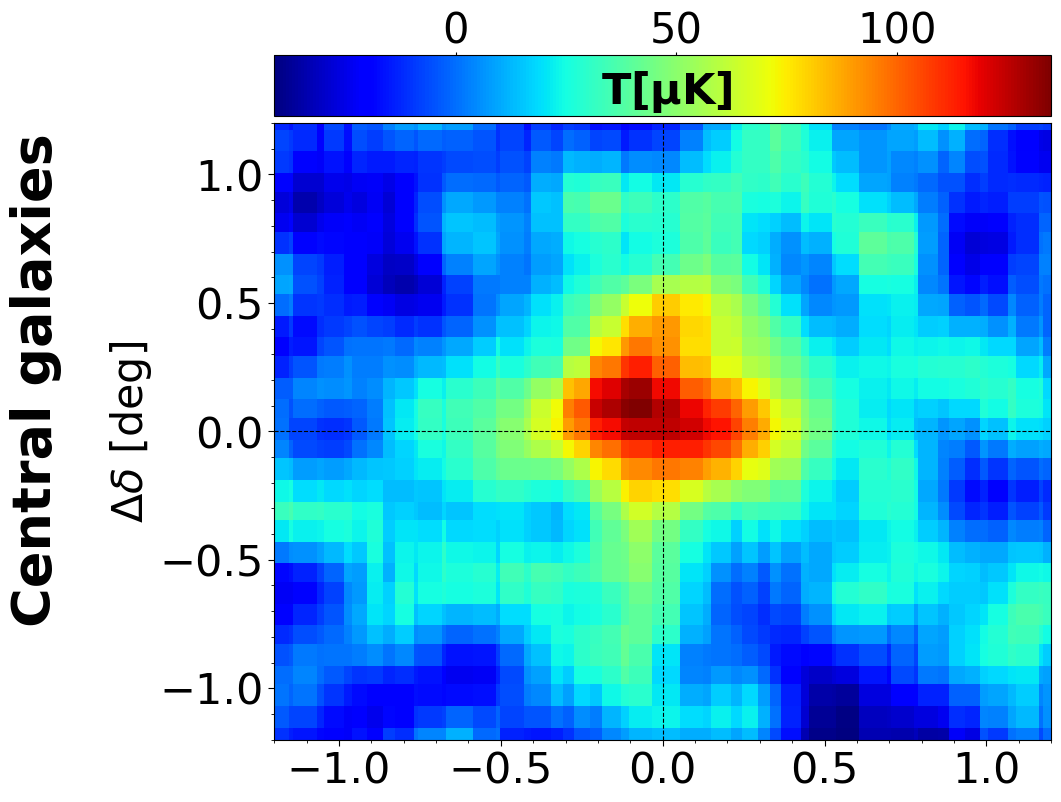}\quad
\includegraphics[trim= 0mm 0mm 0mm 0mm, scale=0.19]{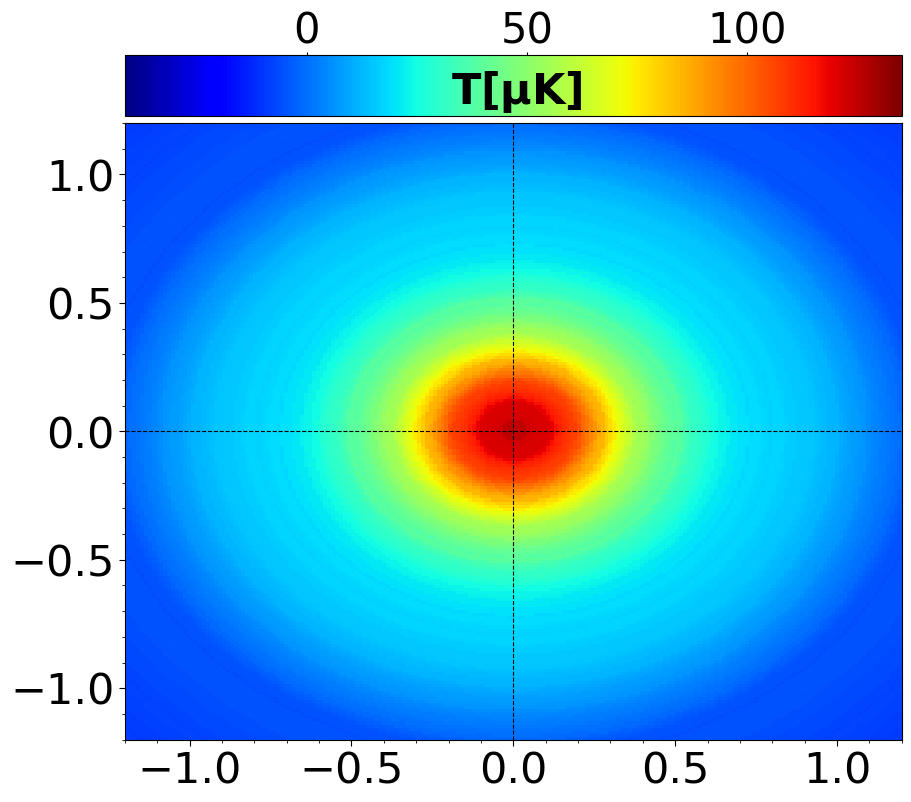}\quad
\includegraphics[trim= 0mm 0mm 0mm 0mm, scale=0.19]{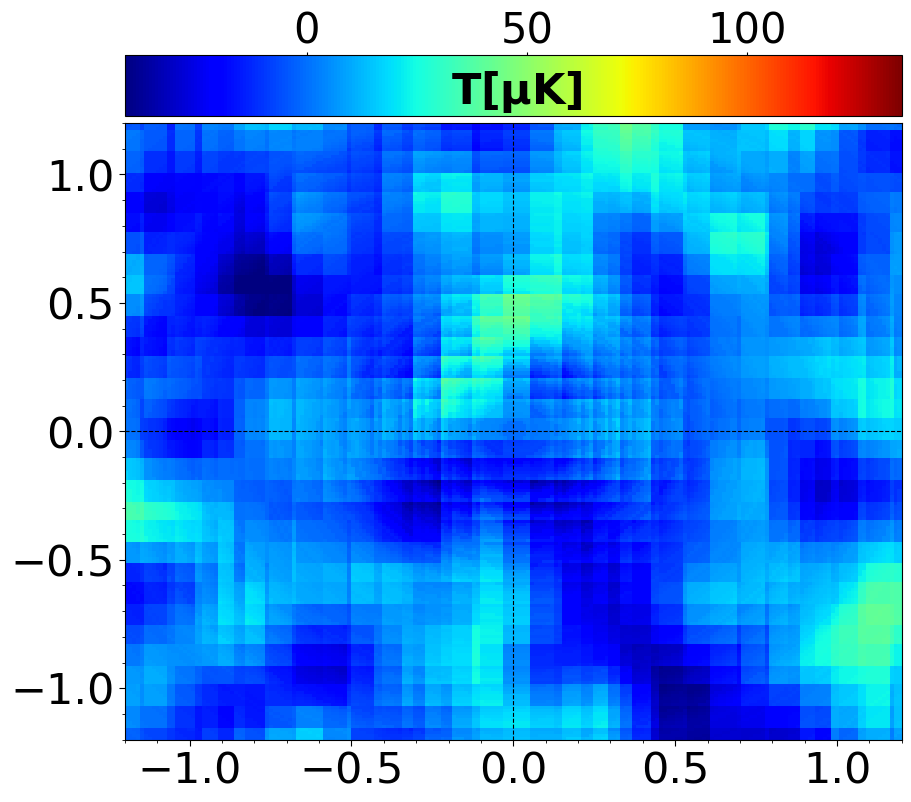}\quad
\newline
\includegraphics[trim= 0mm 0mm 0mm 0mm, scale=0.19]{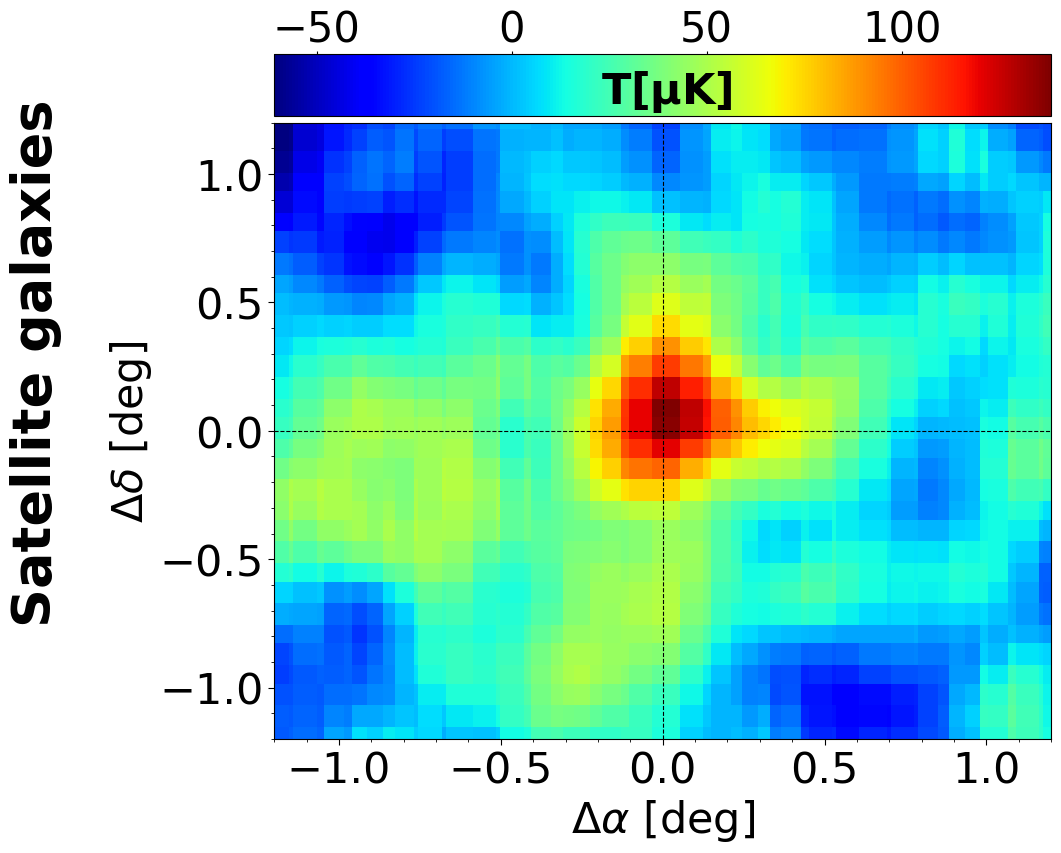}\quad
\includegraphics[trim= 0mm 0mm 0mm 0mm, scale=0.19]{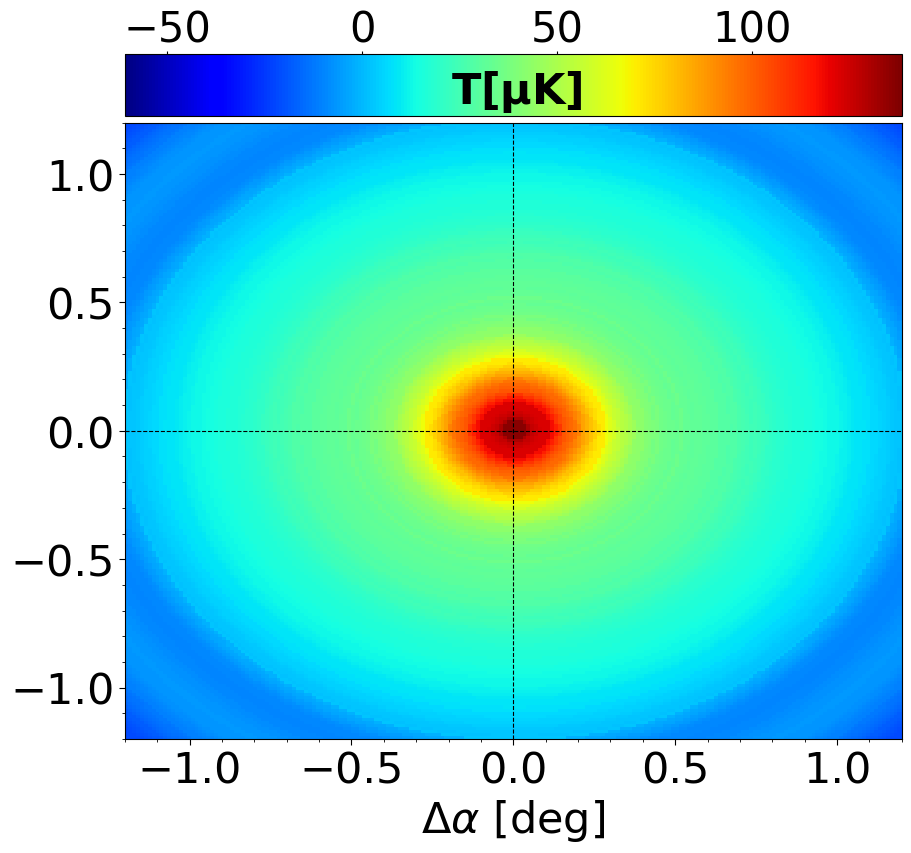}\quad
\includegraphics[trim= 0mm 0mm 0mm 0mm, scale=0.19]{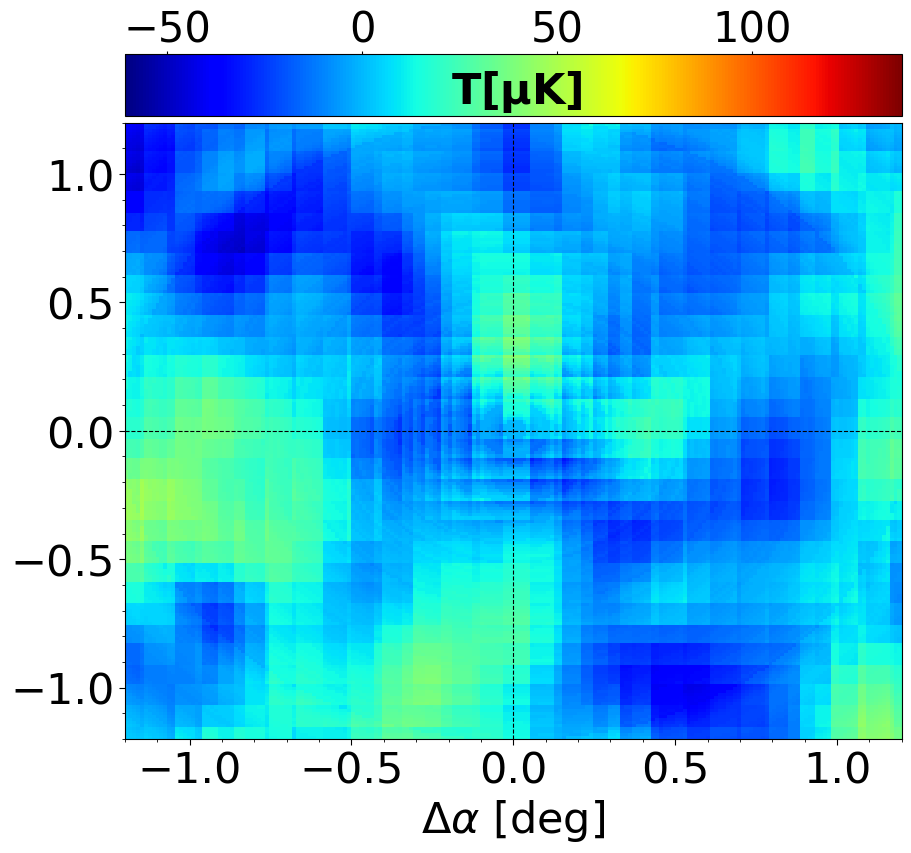}\quad
	\caption{Same as in Fig.~\ref{fig:stack10modes} but showing the results for the stacks on the 20-mode removed maps.}
\label{fig:stack20modes}	
\end{figure*}

\subsection{Galaxy stacking}
\label{ssec:galstack}

Initially, we consider each of the six Parkes patches separately, in combination with the 
2dF sub-sample overlapping with it. For each galaxy the 2dF catalogue reports its equatorial 
coordinates and its redshift; this information allows us to locate the galaxy in the 
corresponding three-dimensional Parkes HI patch. First of all, a two-dimensional HI map 
in (RA, Dec) is extracted by slicing the HI cube at the frequency corresponding to the 
galaxy redshift. From this slice, we then trim a sub-map centred on the galaxy position, 
with total extension $(2.4 \times 2.4)\,\text{deg}^2$. For a generic galaxy $i$ located 
in the $j$th patch, we call $T_i^{(j)}$ the resultant sub-map, with $T$ standing for 
brightness temperature, which is the observable plotted in the maps. The same procedure 
is applied to the Parkes weight cubes, yielding a weight map $W_i^{(j)}$ for this 
particular galaxy and patch. Let $N_j$ be the total number of galaxies in the $j$th patch. 
The final stack map $T^{(j)}$ for this patch is then computed as:
\begin{equation}
	T^{(j)} = \frac{1}{W^{(j)}}\sum_{i=1}^{N_j} T_i^{(j)} W_i^{(j)},
\end{equation}
with $W^{(j)}$ the total weight:
\begin{equation}
	W^{(j)} = \sum_{i=1}^{N_j} W_i^{(j)}.
\end{equation}

The above procedure is applied to both the 10 and 20-mode removed maps and using all 
our three 2dF sub-samples. For every choice of the foreground removal and the galaxy 
sample, we end up with six stack maps, one for each patch. All these stacks are shown 
in Fig.~\ref{fig:patch10modes}
for the 10-mode removal case and in Fig.~\ref{fig:patch20modes} for the 20-mode removal 
case. The stacking procedure described above does not imply any rotation of the field 
of view since it only produces a superposition of different regions extracted from the 
data cubes. Therefore, the final map axes retain the direction along the right ascension 
and declination axes of the Parkes maps, as it is made explicit
in our plots. The stack axes values are quoted as angular separations $\Delta\alpha$ and 
$\Delta\delta$ from the centre, corresponding to the expected position of the stacked 
galaxies. We also comment that the original Parkes maps pixelisation is still visible in 
the stacks, corresponding to the pixel size of 0.08 deg (0.092 deg for the right ascension 
in patches 1 and 2) quoted in Table~\ref{tab:dataset}.
 
The stack results in individual Parkes patches provide a clear detection of the HI halo 
emission, although the region surrounding the central peak is often affected by spurious 
structures. We will discuss this issue more in detail in Section~\ref{sec:datadiscussion}. 
For the moment it suffices to know that the (low) available statistics of the 
stacked galaxy samples is a major factor determining the appearance of a noisy background. 
To mitigate this effect, we choose to combine the contributions from all six patches together.
Formally, this results in a final stack map $\tst$ defined as:
\begin{equation}
	\tst = \frac{1}{\wst}\sum_{j=1}^{N_{\rm p}} T^{(j)} W^{(j)},
\end{equation}
where $N_{\rm p}=6$ is the total number of available patches and 
\begin{equation}
	\wst = \sum_{j=1}^{N_{\rm p}} W^{(j)}
\end{equation}
is the overall weight. Once again, this computation is repeated for the three 2dF 
sub-catalogues and for the two foreground removal schemes. The resultant stacks are 
shown in the first column of Fig.~\ref{fig:stack10modes}
for the 10-mode maps and in the first column of Fig.~\ref{fig:stack20modes} for the 
20-mode maps. The improvement of the available statistics is visible in these stacks, 
yielding more homogeneous backgrounds and more circularly symmetric halo peaks. We conclude 
that the quality of the individual patch stacks is too poor for any further analysis, and 
consider the overall stacks exclusively for the rest of this work. 

This procedure merges together the contribution from 
haloes of different sizes, the final result being an average of the stacked sample. 
This effect could be mitigated by scaling each sub-map radial units by the corresponding 
halo virial radius. 
Neverthless, the latter is not know beforehand in the blind stack we are performing, 
and any indirect estimate would not be accurate enough to be used as a 
scaling unit. The change in apparent size with redshift could be taken care of 
by performing the stack in linear coordinates; this, however, would imply to lose 
information on the angular scale of the signal we are observing in relation to the Parkes beam extension. 
We therefore perform all the stacks in angular coordinates, being aware that the resulting halo 
emission is to be interpreted as an average measurement. 
This procedure preserves the beam dilution effect consistently and enables a final 
direct comparison of the recovered signal with the beam size, as it is detailed in 
Section~\ref{ssec:profiles}.

\subsection{Profile extraction}
\label{ssec:profiles}

The final goal of our stacking analysis is to obtain the HI brightness temperature 
profile $T(\theta)$ for the large-scale haloes, expressed as a function of the 
observed angular separation 
$\theta$ from the halo centre. We do not make any assumption on the shape of such profile, 
and we extract it directly from the stack maps; the only condition we impose is that the halo 
profile is circularly symmetric. First, for each stack, we compute the average background temperature 
as the mean value of all pixels whose separation from the centre is higher than 1 degree. 
Because we are interested in the local excess temperature due to the HI emission, this background 
value is subtracted from the corresponding map. We then build a set of concentric annuli around 
the centre of each stack, corresponding to a set of bins in the angular separation from the centre 
of the halo. We choose a radial bin angular size of 0.015 deg, and for each annulus, we take the 
mean of the map pixel values found within its boundaries\footnote{Although the original 
Parkes maps pixel angular size is 0.08 deg, the stack maps shown in Figs.~\ref{fig:stack10modes}
and~\ref{fig:stack20modes} are interpolated to a finer resolution of $\sim 0.012\,\textrm{deg}$ 
pixel size. This ensures that none of the annuli used for building the radial profile is empty, 
and the coarse resolution of the original maps results in an oversampling of the radial profiles.} 
as the value of the HI brightness temperature profile at that radial separation. 

Before considering the resultant profiles, we show the results of this methodology at the 
map level. The second column in Figs.~\ref{fig:stack10modes} and~\ref{fig:stack20modes} shows 
the modelled halo HI brightness maps for each of the foreground removal and 2dF sample cases, 
built by assigning to each circular annulus the average found as just described. These maps are 
two-dimensional projections of the reconstructed halo profile. Thus by construction, they 
represent a circularly symmetrised version of the stacks shown in the first column. To assess 
how well the reconstructed halo profiles capture the features in the original stacks, the last 
column in the same figures reports the residual maps obtained by subtracting the profile maps 
from the stacks. We see that the halo peak is consistently removed, the residuals being usually 
below $\sim 50\,\mu\text{K}$ and typically located far away from the halo centre.  

\begin{figure*}
\includegraphics[trim= 0mm 0mm 0mm 0mm, scale=0.29]{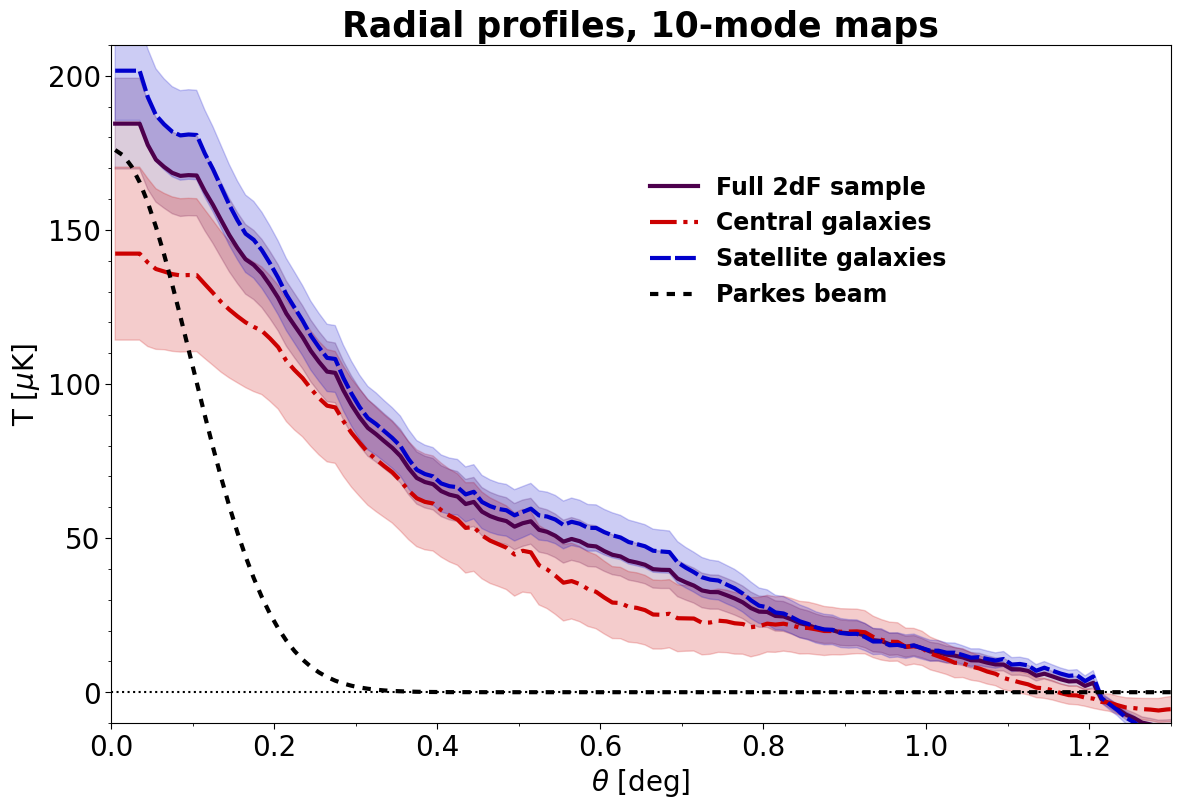}
\includegraphics[trim= 0mm 0mm 0mm 0mm, scale=0.29]{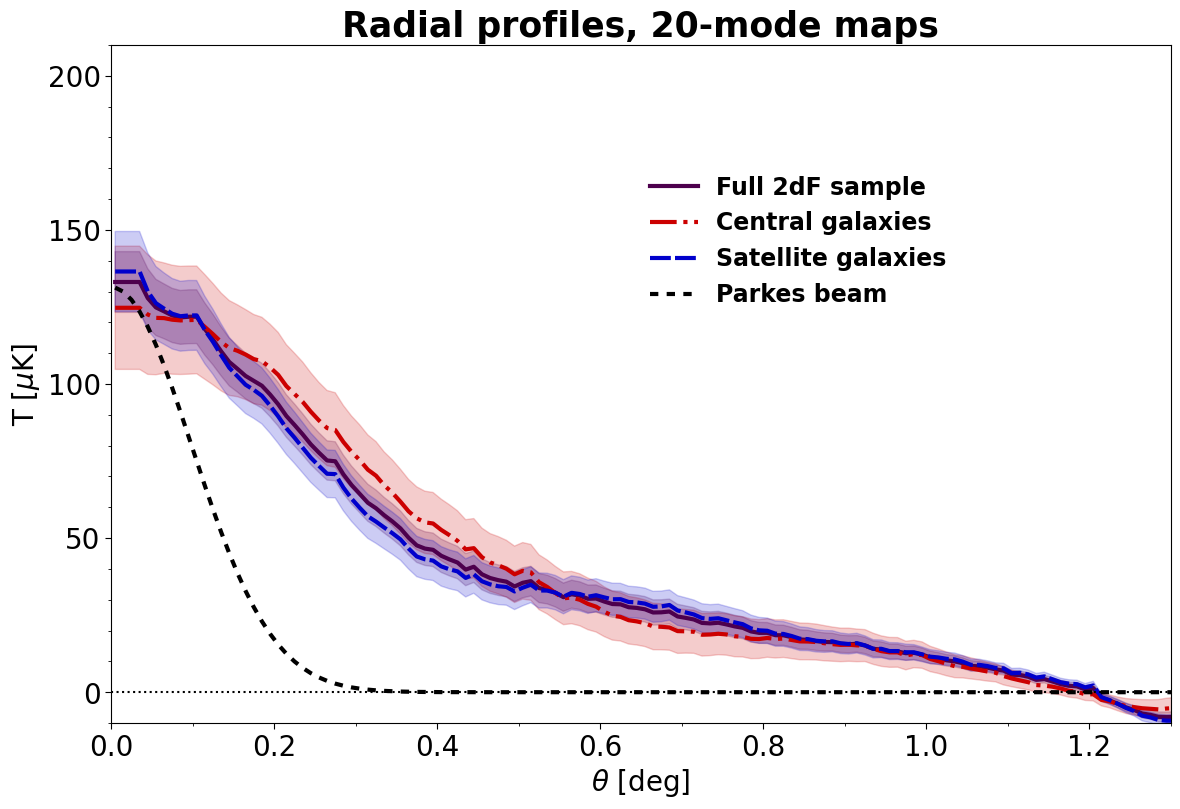}
	\caption{Radial profiles extracted from the stacks of the different galaxy samples used in this work showed for the case of the 10-mode removed maps (left panel) and the 20-mode removed maps (right panel). The shaded region around each profile shows the size of the error bar computed following the procedure described in section~\ref{ssec:errest}. For comparison, the 14' FWHM Parkes beam is also plotted, normalised at its maximum to the mean of the profile amplitudes.}
\label{fig:modeprofs}	
\end{figure*}

\begin{figure}
\includegraphics[trim= 0mm 0mm 0mm 0mm, scale=0.29]{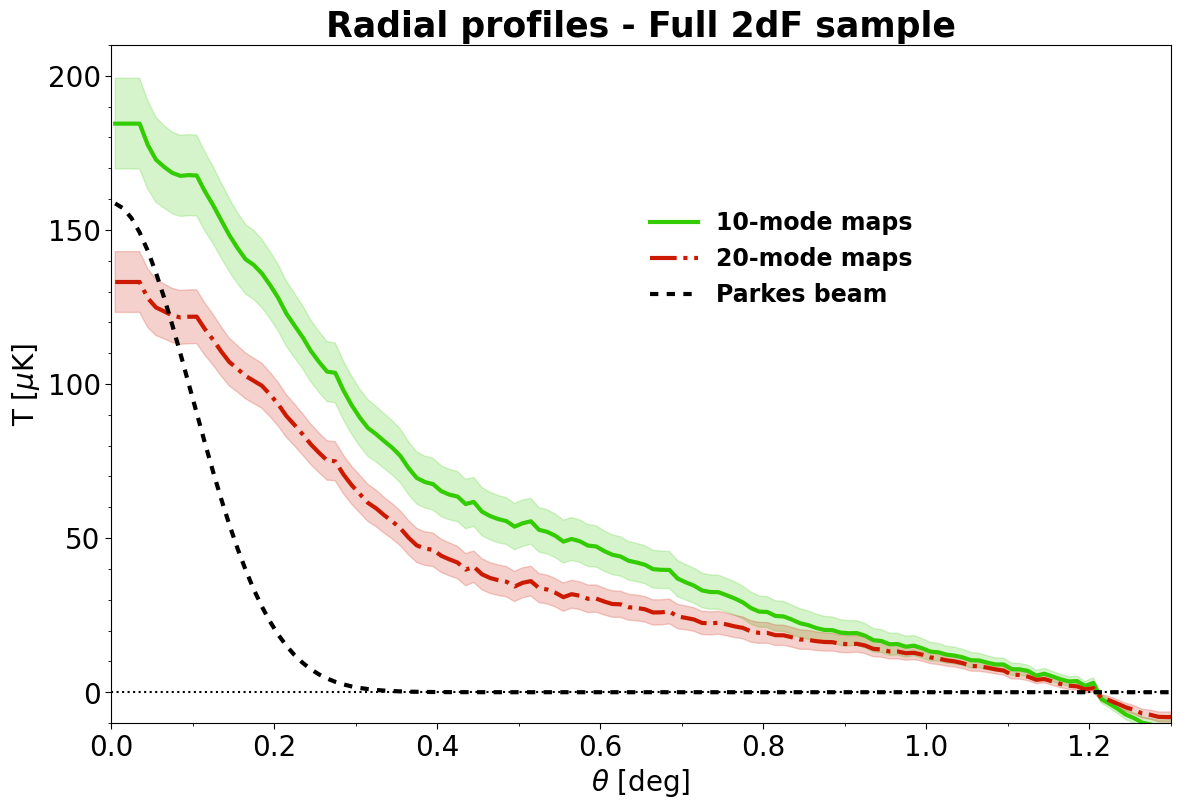}
\includegraphics[trim= 0mm 0mm 0mm 0mm, scale=0.29]{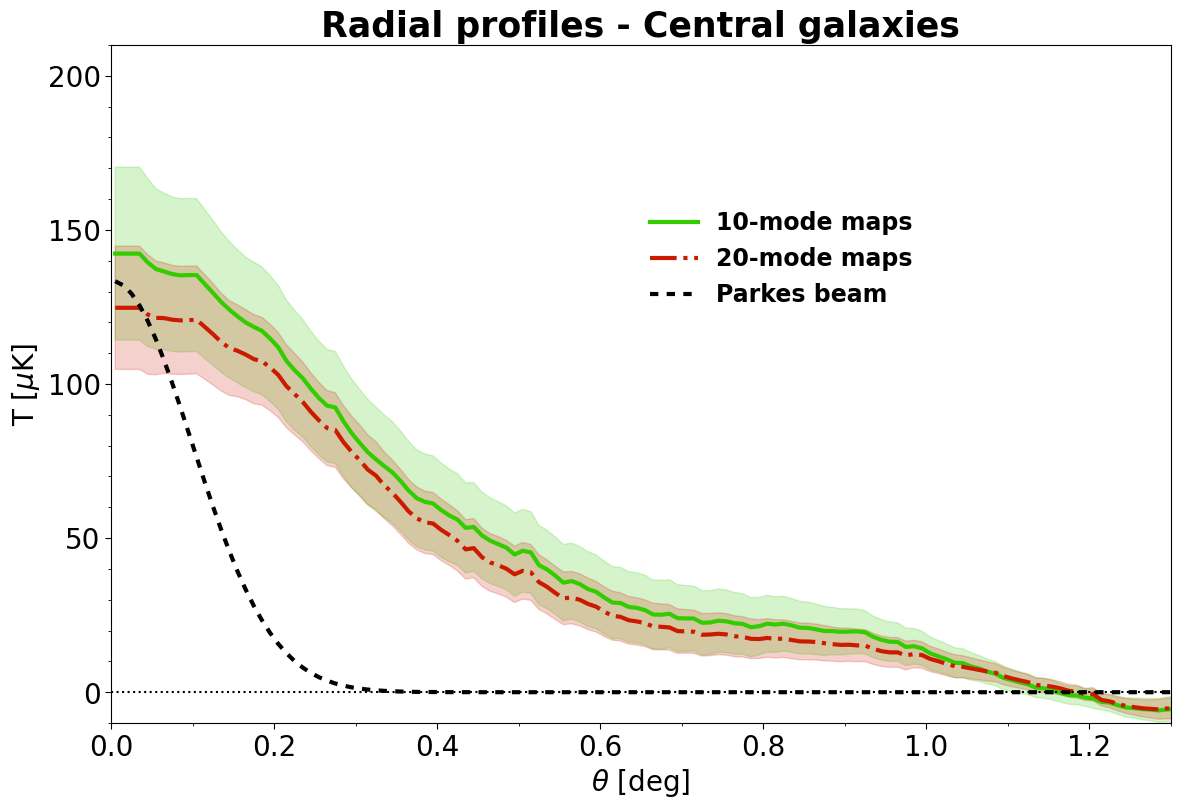}
\includegraphics[trim= 0mm 0mm 0mm 0mm, scale=0.29]{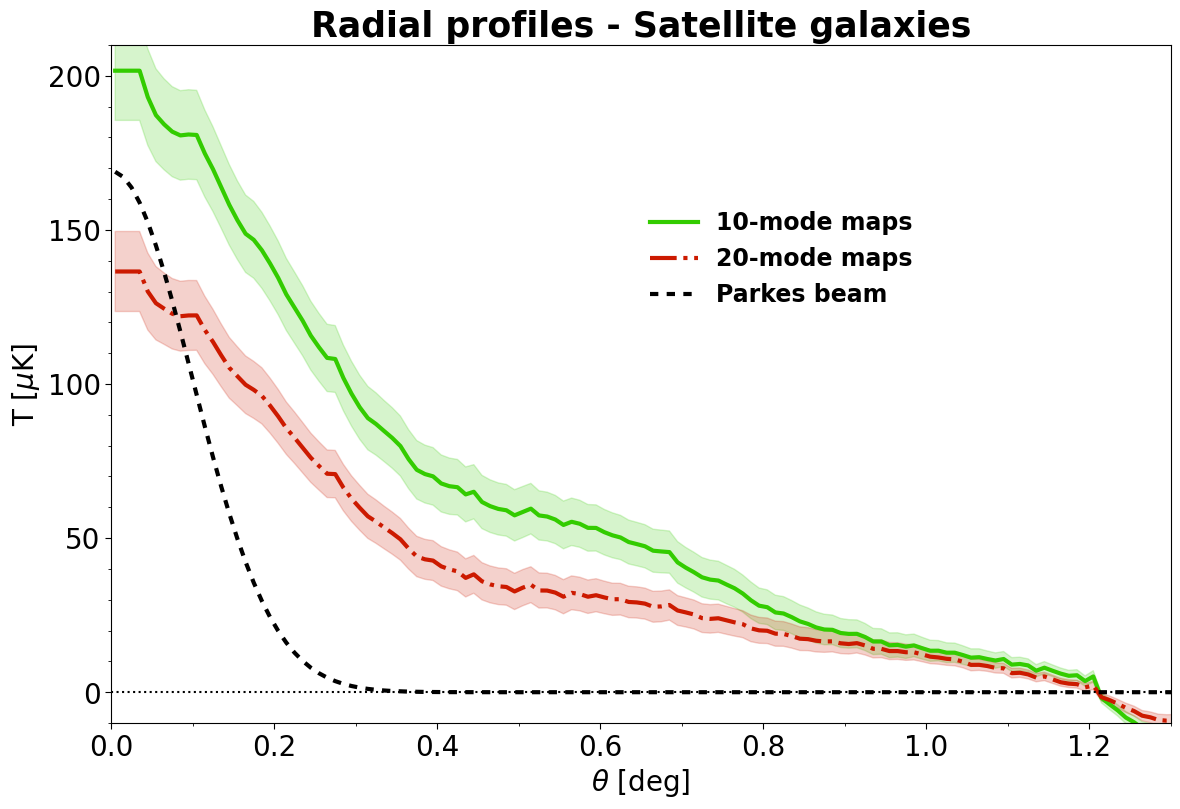}
	\caption{A complementary version of Fig.~\ref{fig:modeprofs}, in which the comparison is between profiles extracted from different foreground cleaned maps. From top to bottom, we show such a comparison for the full 2dF sample, the central galaxy sample and the satellite galaxy sample.}
	\label{fig:magprofs}	
\end{figure}

The resultant radial profiles are plotted in different combinations 
in Figs.~\ref{fig:modeprofs} and~\ref{fig:magprofs}. In Fig.~\ref{fig:modeprofs} we show, 
for each foreground removal case, the comparison of the profiles obtained stacking the three
2dF samples considered in this work. In Fig.~\ref{fig:magprofs}, instead, we show for each 
individual 2dF sample the comparison between the profiles extracted from different foreground 
removed maps. In these figures, the profiles are superimposed to a shaded region which 
quantifies the profile 1-$\sigma$ uncertainty, computed following the procedure described 
in Section~\ref{ssec:errest}.
In all of the panels, we show as well the radial profile for the Parkes beam, assumed for 
simplicity as perfectly Gaussian with $\theta_{\rm FWHM}=14'$, with an amplitude arbitrarily 
set to the mean between the peaks of the profiles plotted in the same panel. In this context, 
the beam profile serves to prove that the halo HI emission is actually resolved in these 
observations, and it is then meaningful to proceed with a study of the reconstructed profile.
The relationship between the profiles shown in Figs.~\ref{fig:modeprofs} and~\ref{fig:magprofs} 
is discussed in Section~\ref{sec:datadiscussion}.

\begin{table}
\centering
\caption{Peak brightness temperature value and uncertainty for the profiles extracted using different foreground removal maps and 2dF samples. The corresponding detection significance is also quoted for each case.}
\label{tab:significance}
\setlength{\tabcolsep}{1.2em}
\begin{tabular}{ c c c }
\hline
 & 10-modes & 20-modes \\
\hline
\multirow{2}{4em}{Full 2dF sample} & $T = (184.5 \pm 14.7)\,\mu\text{K} $ & $T = (133.1 \pm 9.9)\,\mu\text{K} $ \\
 & $(12.5 \sigma) $ & $(13.5 \sigma) $ \\
\hline
\multirow{2}{4em}{Central galaxies} & $T = (142.3 \pm 28.1)\,\mu\text{K} $ & $T = (124.7 \pm 20.0)\,\mu\text{K} $ \\
 & $(5.1 \sigma) $ & $(6.2 \sigma) $ \\
\hline
\multirow{2}{4em}{Satellite galaxies} & $T = (201.7 \pm 16.1)\,\mu\text{K} $ & $T = (136.5 \pm 13.0)\,\mu\text{K} $ \\
 & $(12.5 \sigma) $ & $(10.5 \sigma) $ \\
\hline
\end{tabular}
\end{table}

\begin{figure*}
\includegraphics[trim= -35mm 0mm 0mm 0mm, scale=0.19]{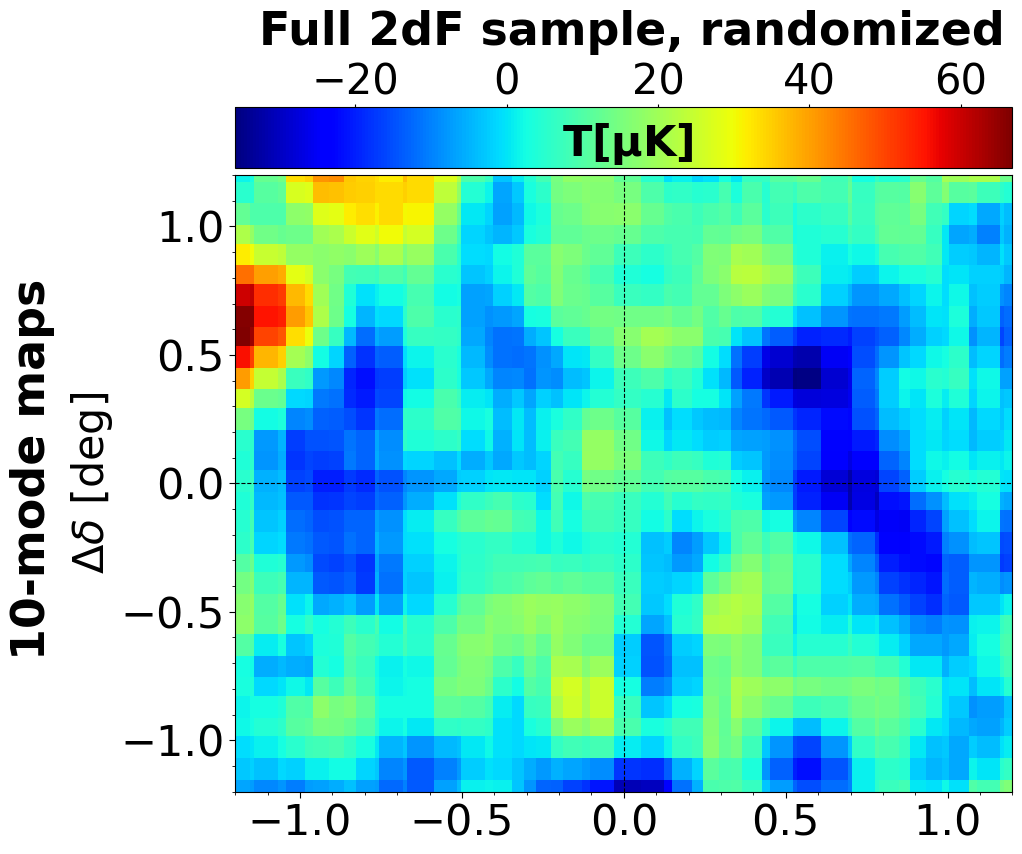}
\includegraphics[trim= 0mm 0mm 0mm 0mm, scale=0.19]{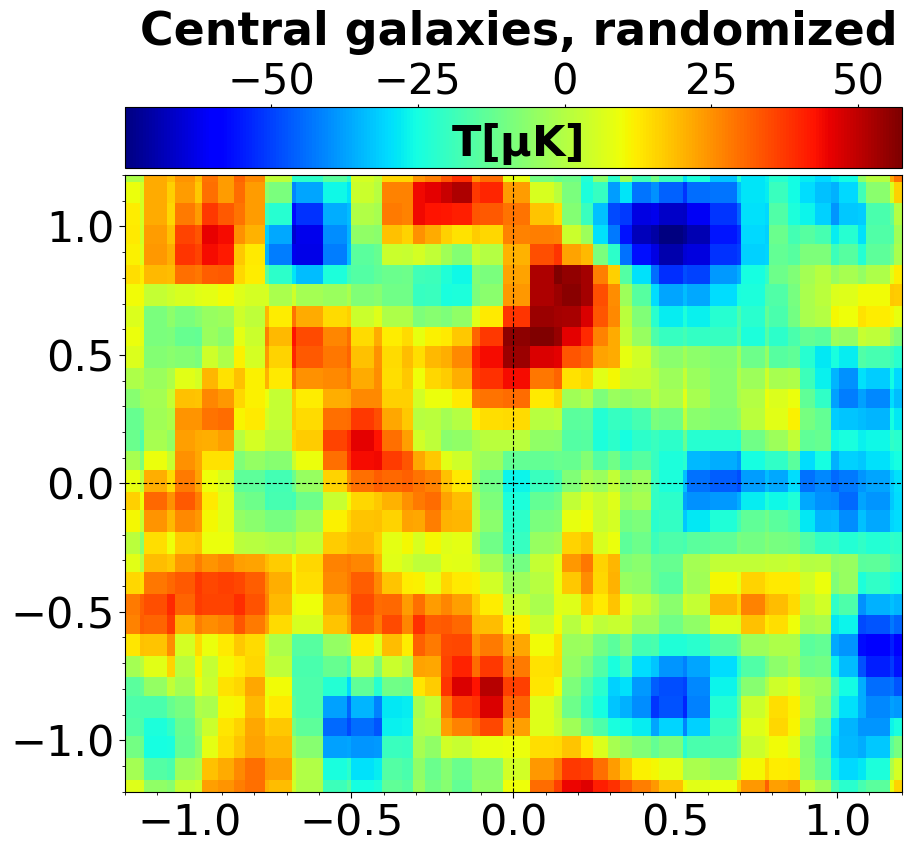}
\includegraphics[trim= 0mm 0mm 0mm 0mm, scale=0.19]{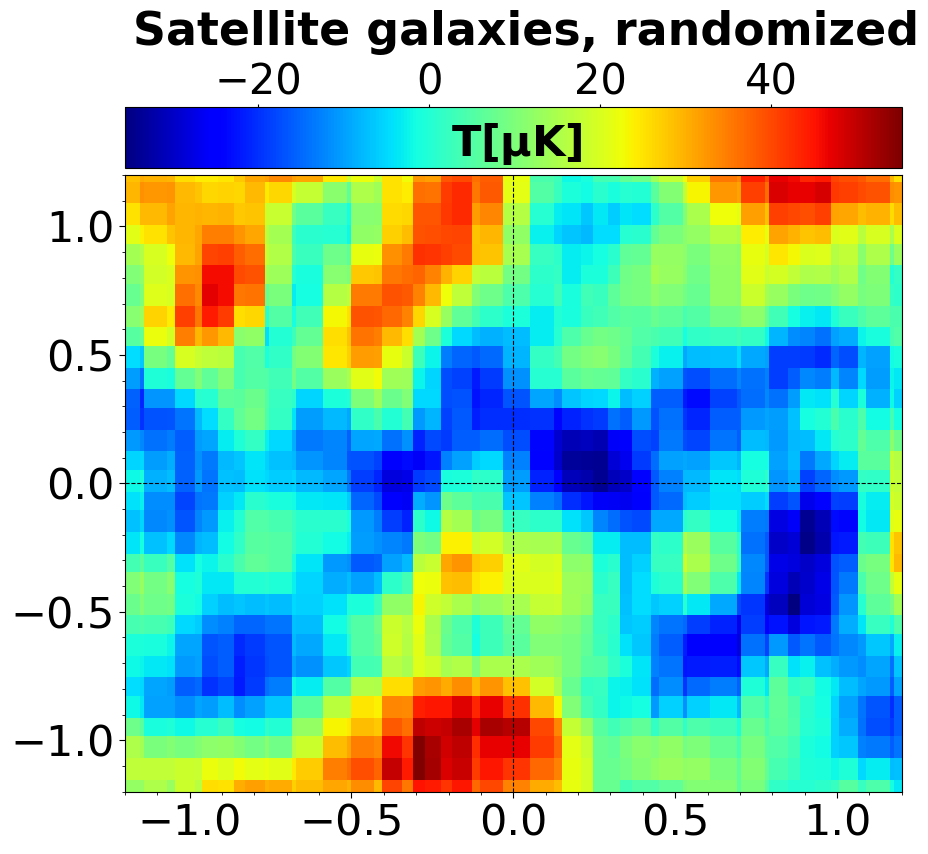}
\newline
\includegraphics[trim= 20mm 0mm 0mm 0mm, scale=0.19]{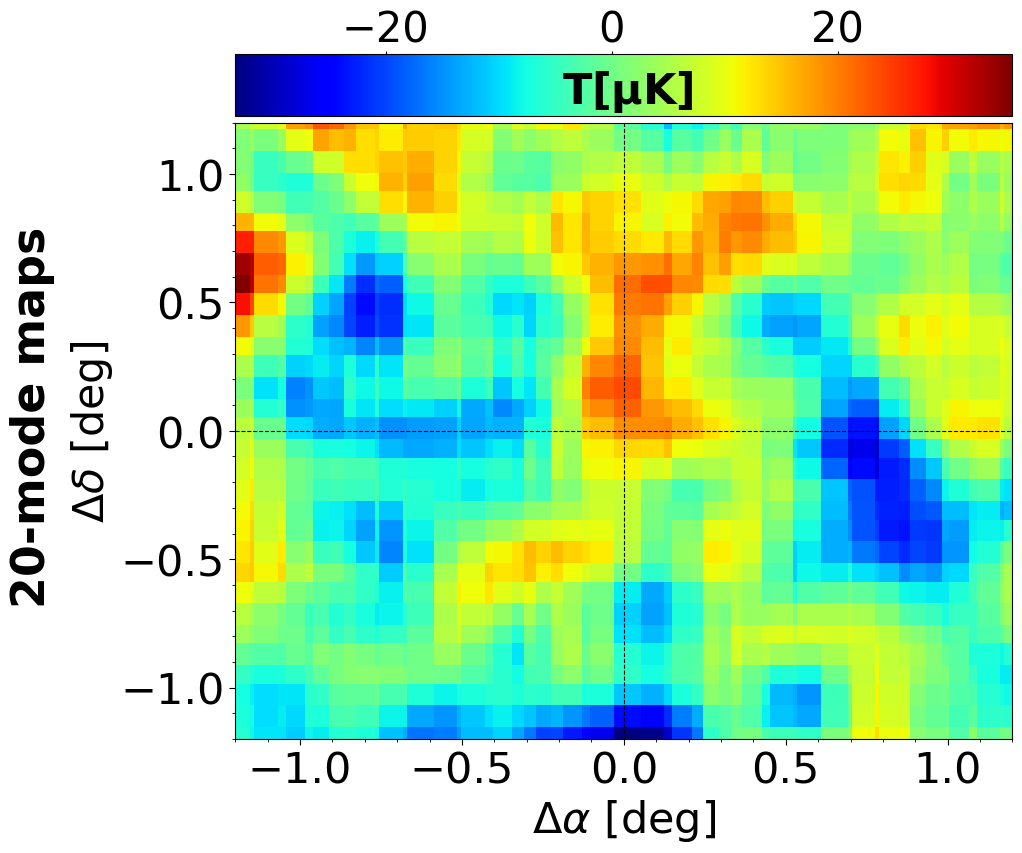}
\includegraphics[trim= 0mm 0mm 0mm 0mm, scale=0.19]{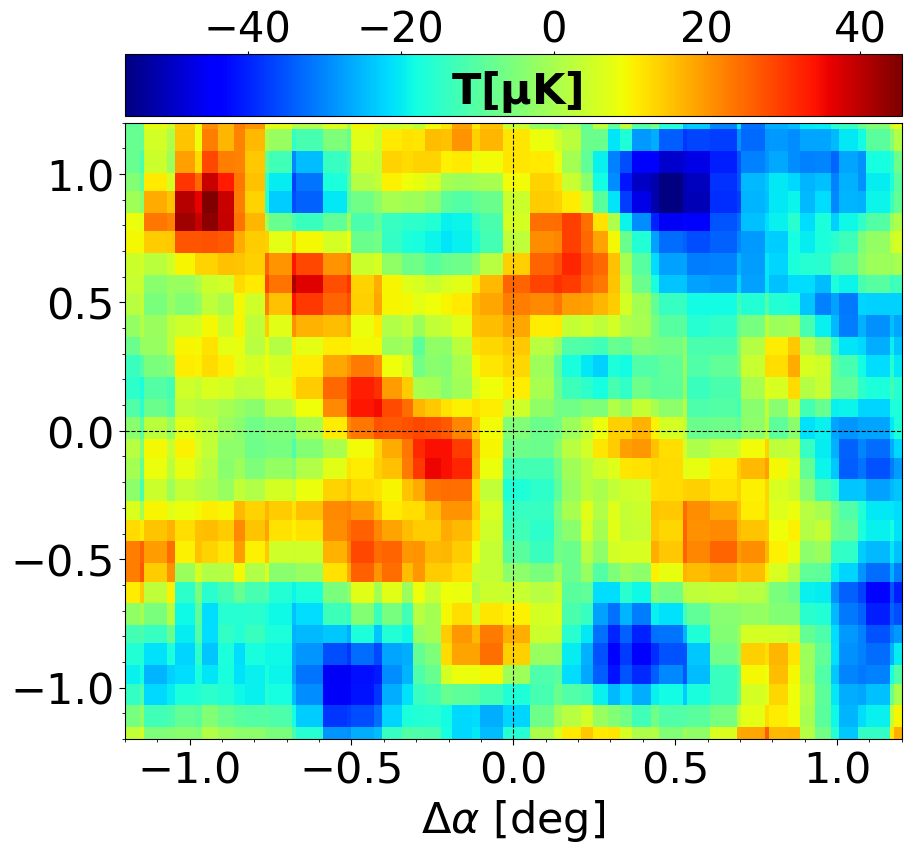}
\includegraphics[trim= 0mm 0mm 0mm 0mm, scale=0.19]{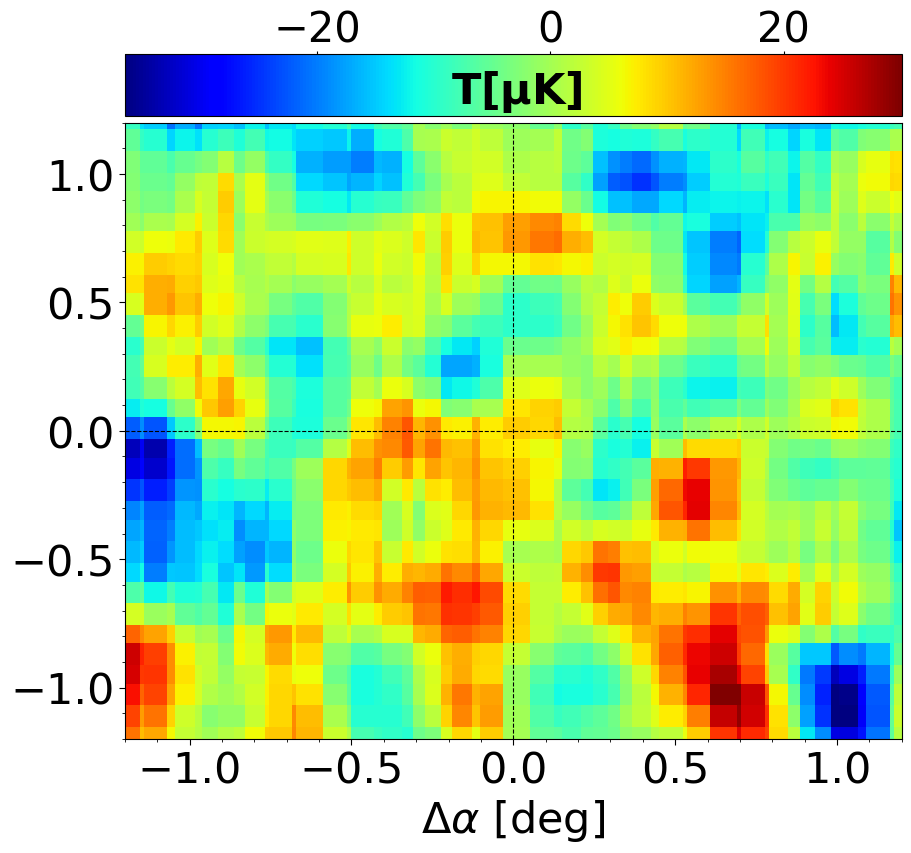}
	\caption{Example of one realisation of the randomised catalogue stacks. 
	Columns from left to right show, in order, the results from stacking the randomised full 
	galaxy sample, the randomised central galaxy sample, and the randomised satellite galaxy 
	sample on the 10-mode (top row) and 20-mode (bottom row) removed maps. The top and bottom 
	row stacks should be compared with the left column of Figs.~\ref{fig:stack10modes} 
	and~\ref{fig:stack20modes} respectively, and show the absence of the signal peak.}
\label{fig:randomstacks}	
\end{figure*}

\subsection{Error estimation}
\label{ssec:errest}

The uncertainties on the extracted profiles are derived using a bootstrap approach, consisting 
of repeating the method described so far on a set of randomised galaxy catalogues. 
More precisely, for each of the three 2dF sub-samples, we generate 500 replicas in which the 
equatorial coordinates of each galaxy are randomly chosen within the ranges allowed by the 
corresponding Parkes patch; the galactic redshift values, instead, are left unchanged. For each 
of these scrambled catalogues we repeat the same stacking procedure we described in 
Section~\ref{ssec:galstack}. In Fig.~\ref{fig:randomstacks} we show what the final stacks 
look like in one realisation chosen out of the total 500, again for both the 10 and 20-mode 
removal cases, and using the randomised versions of all three 2dF sub-samples. For simplicity, 
we show already the overall stacks obtained by combining the six Parkes patches, but the stacks 
are initially performed on individual patches in the same way as for the real catalogues.

Although in this case there is no visible halo peak, we follow the same procedure described in 
Section~\ref{ssec:profiles}, and extract the map profile over the same set of radial bins.
As a result, for each foreground removal and each sub-catalogue choice, we obtain a set of 500 
random profiles. For each bin, the dispersion of the measured brightness temperature 
across the 500 realisations yields and estimate of the 
error associated to the real profile value in that bin. If we denote by $T^{(k)}(\theta)$ the 
HI brightness temperature profile from the $k$th realisation, as a function of the angular 
separation $\theta$ from the halo centre, and
\begin{equation}
	\overline{T}(\theta) = \frac{1}{N_{\rm r}}\sum_{k=1}^{N_{\rm r}} T^{(k)}(\theta)
\end{equation}
is the mean randomised profile (with $N_{\rm r}=500$ the total number of realisations), 
then the variance for the randomised realisations is computed as:
\begin{equation}
	\label{sigmat}
	\sigma^2_{\rm T}(\theta) = \frac{1}{N_{\rm r}}\sum_{k=1}^{N_{\rm r}} \left( T^{(k)}(\theta) - \overline{T}(\theta)\right)^2.
\end{equation}
The quantity $\sigma_{\rm T}(\theta)$ represents the uncertainty associated to the real 
data profile as a function of the angular separation $\theta$, and is plotted as a shaded 
region around the observed profiles in Figs.~\ref{fig:modeprofs} and~\ref{fig:magprofs}. 
In table~\ref{tab:significance} we report for all profiles the peak temperature value with 
its uncertainty. The ratio between the two, also reported in the table, 
quantifies the significance of the detection. 

The bin-dependent uncertainty computed with Eq.~\eqref{sigmat} does not complete the statistical 
characterisation of the observed profiles. Indeed, we remind that in this case different 
bins are not statistically independent. 
A single halo profile covers an extended range of angular separations, thus introducing a 
correlation between pairs of $\theta$-bins (which becomes more important the closer the 
bins are). The proper object to summarise this correlation is
the covariance matrix. The covariance $\mathcal{C}_{\rm T}(\theta_i,\theta_j)$ between the 
angular separation bins 
centred at $\theta_i$ and $\theta_j$ is computed as: 
\begin{eqnarray}
	\label{covar}
	\mathcal{C}_{\rm T}(\theta_i,\theta_j) &=&  \frac{1}{N_{\rm r}}\sum_{k=1}^{N_{\rm r}} \left( T^{(k)}(\theta_i) - \overline{T}(\theta_j)\right) \nonumber \\
 & \times &	
	\left( T^{(k)}(\theta_j) - \overline{T}(\theta_j)\right),
\end{eqnarray}
where this time we made the discretised nature of the angular variable $\theta$ explicit. 
The individual bin variance computed with Eq.~\eqref{sigmat} corresponds to the 
covariance matrix diagonal, in the particular case of setting $\theta_i=\theta_j$ in 
Eq.~\eqref{covar}. The resultant covariance matrices are plotted in Fig.~\ref{fig:covmatrs} 
for the six combinations of maps and samples, and will 
be used in section~\ref{sec:estimation} for the parameter estimation.


\section{Discussion of data analysis results}
\label{sec:datadiscussion}

We dedicate this section to a discussion of the results obtained so far 
with the joint analysis of the Parkes maps and the 2dF galaxy sample.  

The stacks in Figs.~\ref{fig:patch10modes} and~\ref{fig:patch20modes}
show a clear detection of a 21-cm signal peaked in the map centre at a level of 
100 to 300 $\mu\text{K}$, with the 10-mode stacks having a systematically higher peak 
value than the 20-mode stacks. These stacks are showing the merged contribution 
of the 21-cm emission from the bulk HI contained in the haloes hosting the galaxies we are 
targeting. However, the background signal surrounding the central halo is not homogeneous 
and shows several structures, in some cases with an amplitude comparable to the central 
halo signal. This phenomenon is most likely due to a 
combination of the limited samples statistics and the contribution from residual foregrounds. 
Indeed, the usage of the full 2dF sample, which has the highest statistics, seems to mitigate 
this effect slightly.
However, if we look again at the last three columns of Table~\ref{tab:dataset}, it is clear 
that the number of stacked galaxies alone is not enough to account for these structures. 
Patch number 4 seems to be the most heavily affected, although it is the patch with the 
highest number of galaxies. This situation suggests that foreground residuals are also 
contributing to this effect. In this sense, it is essential to compare their amplitude 
with the amplitude scale of the 21-cm maps we employed. By looking back at 
Fig.~\ref{fig:parkes}, we see that the original Parkes maps have typical fluctuations 
at the level of $\sim 10\,\text{mK}$. Our final stacks allow us to identify structures 
at the level of $\sim 100\,\mu\text{K}$ which is two orders of magnitude lower. Any 
spurious signal initially present in the maps, such as thermal noise or foreground residuals, 
would also enter the final stacks, and its relative importance compared to the actual 
signal is determined by the available statistics. As commented in Section~\ref{ssec:parkes}, 
different patches underwent foreground removal independently, and it is possible for some 
patches to be more affected than others in this sense. Another fact corroborating 
this hypothesis is that the stacks on the 20-mode maps, which have undergone a more thorough 
foreground removal, are overall less affected by background structures. The only 
exception is patch number 6, where, although the asymmetrical structures are observed 
in both foreground removal cases, the 20-mode maps look worse. We have to take into 
account that the observed fluctuations may also proceed from actual HI emission 
distributed off-centre with respect to the selected central position for our stacking. 
If this is the case, then, using a different 2dF sub-sample should make a difference. 
In fact, for patch 6, the central galaxy catalogue and the satellite galaxy catalogue 
provide the best and worst results respectively (with the full 2dF catalogue in between), 
thus supporting the idea that the selected central 2dF galaxies are more HI-rich. 
Unfortunately, this conclusion cannot be generalised because it does not hold for 
all the six patches (for example, in patches 2 and 4, we observe the opposite
situation). 

To summarise, we can list three main properties emerging from the stacks at the individual 
patch level. First, the comparison between the 10 and 20-mode removal cases shows that 
the 20-mode removal maps provide cleaner stacks (with the only exception of patch 6). 
Second, for the same removal case, neither the central nor the satellite galaxy samples 
seem to provide consistently a better galaxy selection than the full 2dF sample, which 
is probably favoured by its higher statistics. Third, for the same removal case and 
chosen sub-sample, there is not a clear trend in the quality of the resultant stacks 
across patches, nor a clear dependence of the observed peak and fluctuation amplitudes 
on the number of galaxies available for a particular patch. That said, it is clear that 
the limited statistics of our sample is a crucial factor in this type of analysis. Even 
the choice of the full 2dF catalogue does not provide a statistically large 
enough sample to allow the study of an HI halo profile in individual patches. 
For this reason, we shall combine the contribution from the six patches and study the 
stack of each of the three 2dF sub-samples as a whole. 
Although the stacks may look better in some particular patches, we have no 
\textit{a priori} reason to discard the other ones, so the contribution from all 
six patches is considered in their combination.

The resultant overall stacks shown in the first column of Figs.~\ref{fig:stack10modes} 
and~\ref{fig:stack20modes} show the improvement in the stack quality resulting from 
the larger stacked samples. Although the background fluctuations are still visible, 
their amplitude is no longer comparable with the peak value of the HI emission. 
These maps can now be used for the study of the halo profile. We notice that the central 
galaxy sample is now clearly the worst choice, as its final stacks still show a pronounced 
asymmetry, especially for the 10-mode case. This phenomenon may suggest that the location 
of the central galaxies does not match the loci of higher HI-density in the local Universe. 
Once again, the 20-mode maps yield better results than the 10-mode maps, as the resultant 
peaks tend to be more symmetric and the backgrounds slightly more homogeneous. 
In the same figures, the projected profile maps are not particularly meaningful; 
they provide a quick visual check that the HI profiles are representative of the stack 
maps they are extracted from. The associated residual maps show a consistent cancellation 
of the central HI peak. Their amplitude fluctuations measure to what extent the initial 
stack maps deviate from circular symmetry. In this case, a better result is achieved by 
the full 2dF sample. Indeed, it is expectable that a higher number of stacked galaxies would
increase the signal-to-noise ratio of the HI emission and produce more rounded peaks. 

\begin{figure}
\includegraphics[trim= 0mm 0mm 0mm 0mm, scale=0.17]{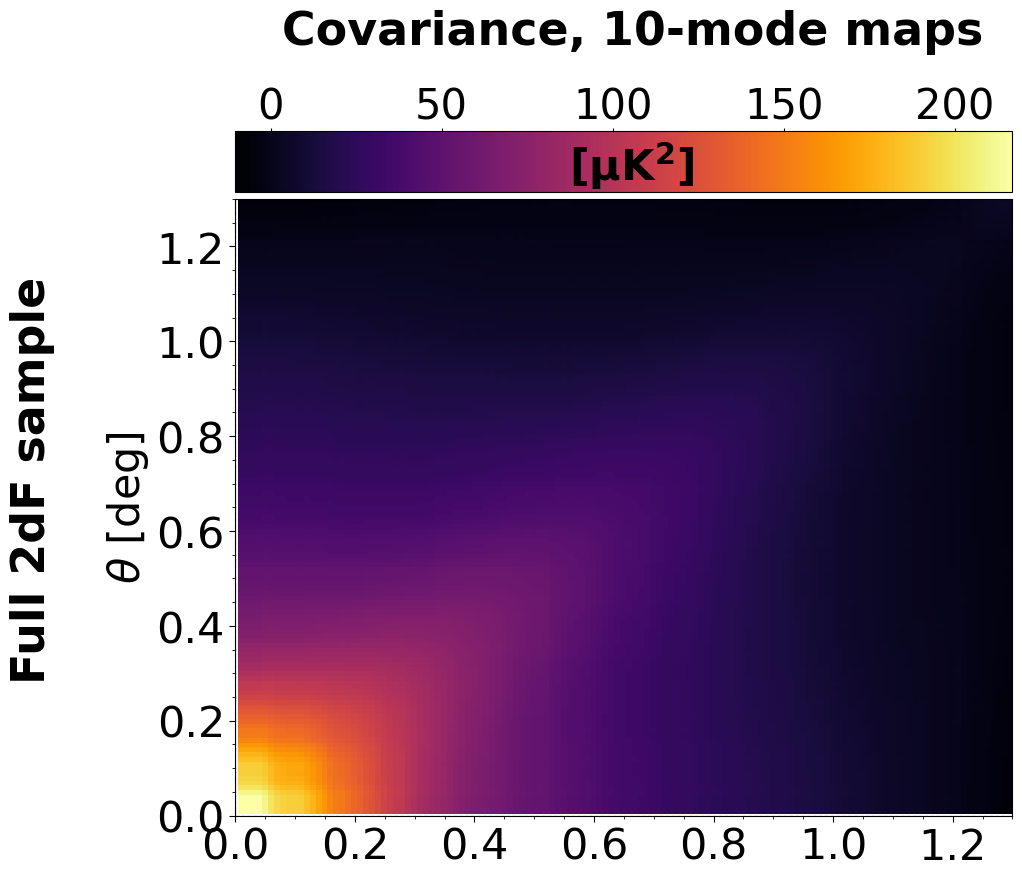}
\includegraphics[trim= 0mm 0mm 0mm 0mm, scale=0.17]{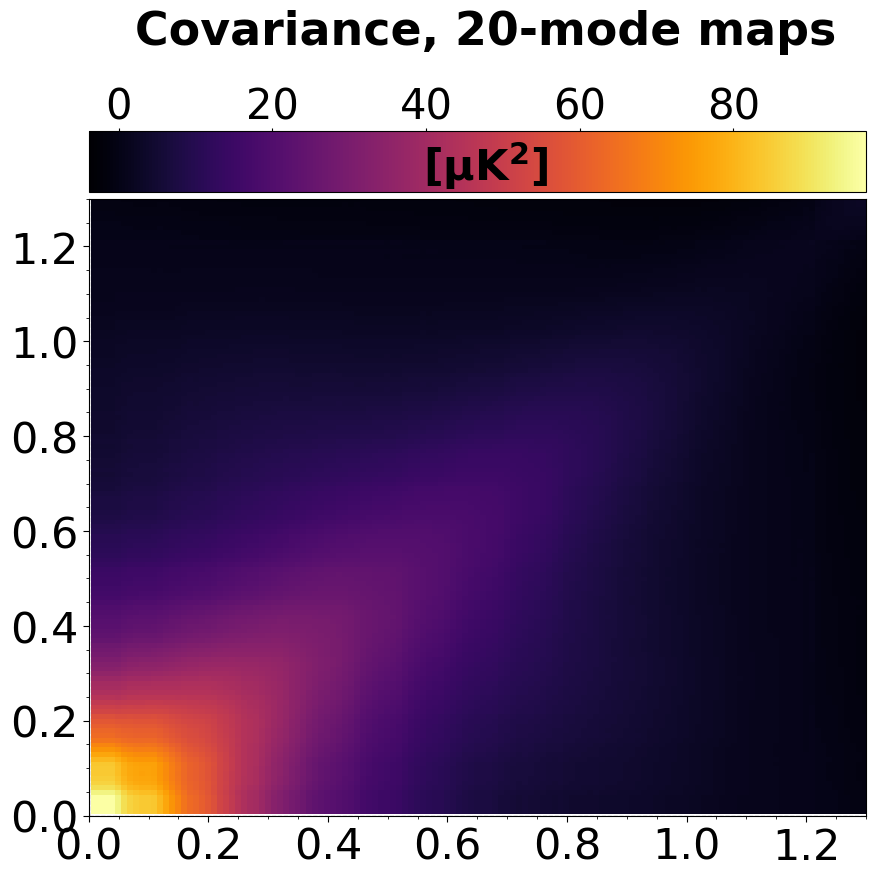}
\newline
\includegraphics[trim= 0mm 0mm 0mm 0mm, scale=0.17]{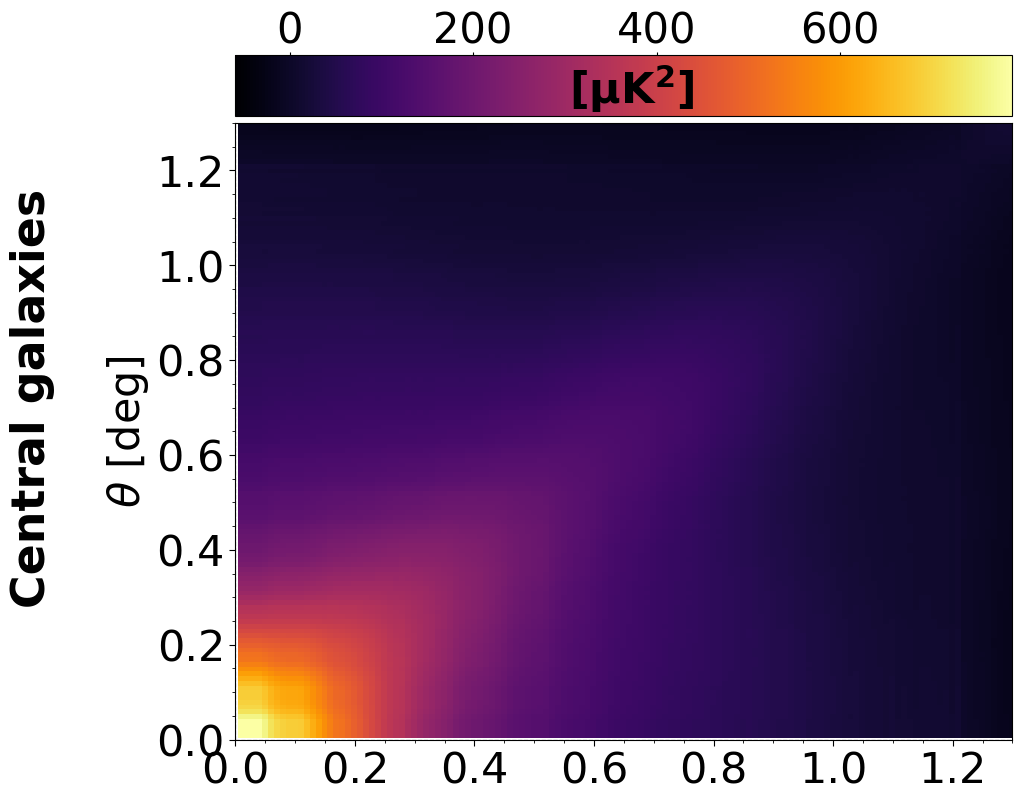}
\includegraphics[trim= 0mm 0mm 0mm 0mm, scale=0.17]{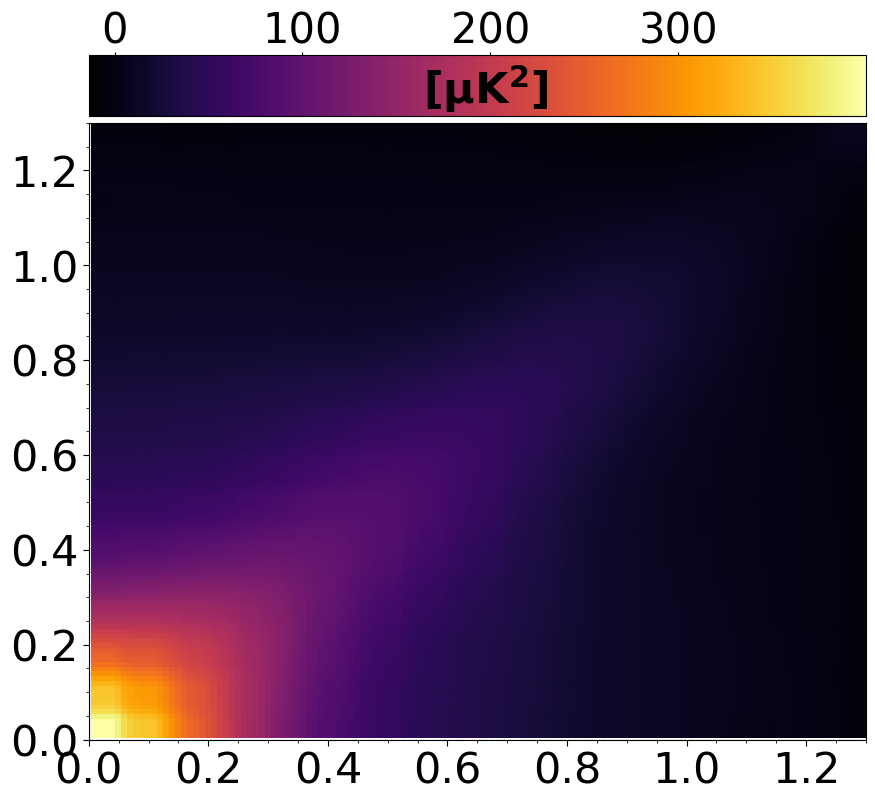}
\newline
\includegraphics[trim= 0mm 0mm 0mm 0mm, scale=0.17]{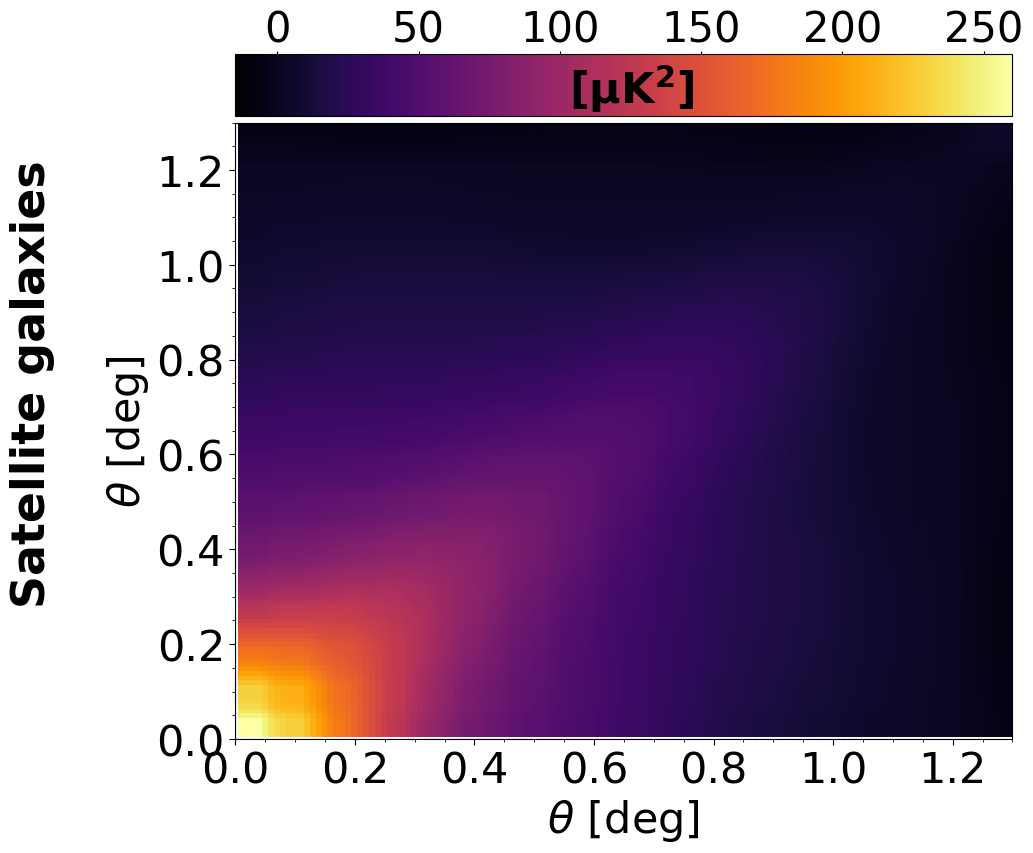}
\includegraphics[trim= 0mm 0mm 0mm 0mm, scale=0.17]{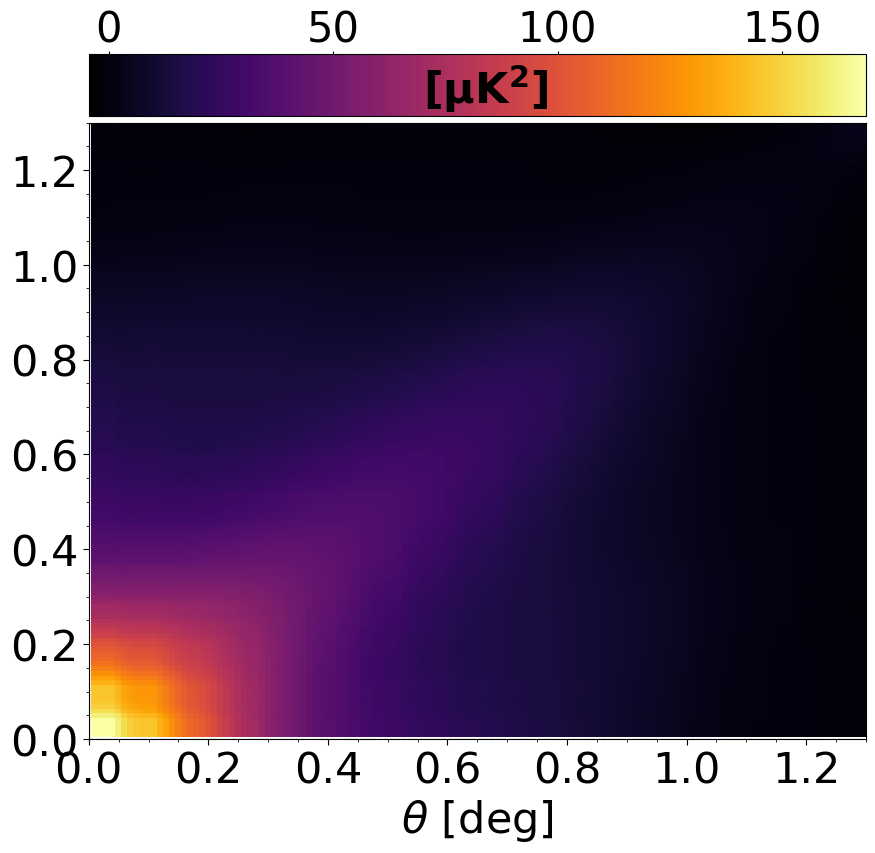}
	\caption{Covariance matrices obtained from the stacks of the 500 randomised catalogues. 
	Rows from top to bottom show, in order, the results for the full, 
	the central and the satellite galaxy samples, over the 10-mode (left) and 20-mode (right)
	removed maps. The correlation between 
	 neighbouring bins is evident, particularly for the lowest angular radii. }
\label{fig:covmatrs}	
\end{figure}

Before analysing the radial profiles, it is worth commenting on the results of the bootstrap 
method employed for estimating the statistical uncertainties of the measured signal. The 
maps reported in Fig.~\ref{fig:randomstacks} show what we obtain by stacking the randomised 
galaxy catalogues, for one particular realisation. Notice that it is still possible to compare 
stacks produced with the same sample across the two foreground removal cases, as the 
randomisation is the same. However, such a comparison is not very informative in this case. 
What is important here is to compare the final stacks with the ones obtained with the real 
data. In other words, the first and second rows of Fig.~\ref{fig:randomstacks} should be 
compared with the first column of Figs.~\ref{fig:stack10modes} and~\ref{fig:stack20modes}, 
respectively. We see that the randomisation destroys the detected halo signal, and the resultant 
stacks show absolutely no evidence of a central emission. In this case, we do not show the 
extracted halo profiles, which do not carry any useful information. However, the repetition 
of this procedure 500 times provides us with a set of profiles
that we can use to characterise the statistical properties of the observed halo emission. 
The dispersion of these profiles for individual bins provides the 1-$\sigma$ uncertainties 
associated with the halo emission, visible in Figs.~\ref{fig:modeprofs} and~\ref{fig:magprofs}. 
We decided to quote explicitly the temperature and error for the peak of all profiles 
(Table~\ref{tab:significance}) to quantify the significance associated with its measurement. 
The full 2dF sample, being the largest, clearly provides the highest significance, but even 
for the central sample the values are higher than 5, confirming the robustness of our detection. 
The correlation across pairs of bins can also be computed and is encoded in the covariance 
matrices plotted in Fig.~\ref{fig:covmatrs}. Although, as expected, the signal is peaked 
at the diagonal, we see that a significant area of the off-diagonal region has non-null 
covariance, meaning that the angular bins (particularly those closer to the peak)
cannot be considered independent. This information is to be taken into account in the 
parameter estimation described in Section~\ref{sec:estimation}.

Finally, we discuss the resultant radial brightness temperature profiles shown in 
Figs.~\ref{fig:modeprofs} and~\ref{fig:magprofs}, which are the primary goal of our data 
analysis. First of all, as already mentioned above, the fact that all profiles are considerably 
more extended than the Parkes beam implies that the emission we are observing is resolved, 
and it is then meaningful to study its shape. If the signal radial extension were comparable 
with the beam Gaussian, we would just be observing the instrumental response to sources 
that are not resolved, and it would not be possible to test any theoretical model on it. 
The finite instrumental resolution still affects our detections, as the halo HI brightness 
temperature is convolved with the telescope beam. Ideally, a smaller beam would allow to 
map smaller scales and to obtain a higher fidelity reconstruction of the physical 21-cm 
emission. In this case, however, we are observing profiles extended out to 
$\sim 0.8\,\text{deg}$, and although the 14 arcminutes are still an essential fraction of 
the signal extension, they allow to resolve it and to map the decrease of the emission 
towards larger angular separations. 

We focus now on the reconstructed halo profiles in Fig.~\ref{fig:modeprofs}, which allows 
a direct comparison of the different 2dF samples for a chosen foreground removal case.
The most evident feature is that the full 2dF sample typically lies in between the central 
and the satellite samples. Stacking satellite galaxies results in a higher peak, and this may 
suggest that this type of galaxies is spatially associated with a higher abundance of HI 
compared to the central galaxies. As a result, when considering the full 2dF sample, the 
inclusion of central galaxies lowers the mean HI content, and consequently, the full sample 
profile has a lower peak amplitude than the satellite sample profile. 
Although the separation between centrals and satellites is not neat, as the 
contribution from one type can be picked up by the beam when centering the stack on the other, 
it still shows that different environments affect the galactic HI abundance, and such 
a difference can be detected with 21-cm intensity maps.
This situation is more visible for the 10-mode removal case; the 20-mode maps, instead, 
yield remarkably consistent profiles across the different 2dF sub-samples. We can interpret 
this observation by stating that either the differences in the profiles for the 10-mode 
case are enhanced by a higher contamination of foreground 
removal, or that the 20-mode removal excises most of the physical HI emission, thus 
cancelling the differences coming from the choice of a different galaxy sample. Unfortunately, 
it is not possible to extract more information in this sense with the available data set. 
However, we have to stress that also for the 10-mode removal case we are talking about a 
separation between profiles which is still within 1-$\sigma$.
The selection of a different galaxy sample determines only a marginal variation across 
different profiles.
A final observation that can be extracted from these plots is that the profile obtained 
stacking the central galaxies is less concentrated than the one obtained by stacking 
the satellite galaxies. This effect was already visible in the final stack maps in the 
first column of Figs.~\ref{fig:stack10modes} and~\ref{fig:stack20modes}, where the central 
galaxy stacks extended over a broader region. This could reinforce the hypothesis that 
the central galaxies do not trace locations with the higher abundance of HI in the volume 
spanned by the Parkes patches. 

We conclude this section by considering the profiles plotted in Fig.~\ref{fig:magprofs}, where 
the same 2dF sub-sample is shown for both foreground removal cases. 
As it was already clear from the stack maps, the 10-mode profile is always systematically 
higher than the 20-mode one. However, it is interesting to notice that the two profiles are not
compatible when stacking the full or the satellite samples, while a much better consistency is 
achieved with the central 2dF sample. This situation is a result of the 10-mode profile being 
substantially lower for the central sample compared to the other ones. As already mentioned, 
it is likely that in this case there is an essential mismatch between the bulk of the 21-cm 
signal (whether it is actual halo HI emission or residual radio foreground) and the positions 
of the central galaxies, which has the effect 
of distributing the signal over a larger area, thus artificially broadening the profile and 
diluting its peak value.


\section{Theoretical modelling}
\label{sec:theomodel}

The last part of this work is dedicated to the theoretical prediction of the 
observed halo HI emission. Our goal is to provide a recipe to compute the mean 
expected radial brightness temperature profile for a generic halo of virial mass 
$\mvir$ at redshift $z$. This procedure will enable us to use the profiles 
extracted from the Parkes maps as a benchmark to test this theoretical model 
and to fit for the main parameters controlling the abundance and distribution 
of HI in haloes. 

Things, in this case, are complicated by the fact that our stacks merge the contributions 
from halos with different masses and radial extensions. In this case, the formally most 
correct way to reproduce the observed profiles is to cross-correlate the galaxy sample and 
the HI signal. Such a strategy has been employed in~\citet{gong19} and~\citet{tanimura20} 
to model the observed profile of the thermal Sunyaev-Zel'dovich in galaxy clusters. 
In our case, however, we do not have prior information on the stacked halo mass distribution, 
which is a crucial quantity involved in this computation.  

For this reason we limit this section to the computation of the expected 
profile observed for a single halo with chosen mass and redshift.
This prediction will be compared to the observed profiles from Section~\ref{sec:estimation},
 regarded as being produced by a single type of haloes whose properties are the mean of the 
 corresponding stacked sample.
Although this is an approximation, it is the best type of modelling we can provide 
for the current data set.  

There are two main quantities required to predict the HI halo profile.
The first is the HI mass to halo mass (HIHM) relation, which establishes the 
total HI mass to be assigned to a generic dark matter halo. The second is 
the choice of a suitable shape for the HI profile, which dictates how the HI is distributed 
inside its host halo. Different options have been explored in the literature. We 
discuss them in the following and justify our choices, before detailing the steps we took 
to theoretically predict the observed HI brightness temperature profile.

\subsection{The HIHM relation}
\label{ssec:hihm}

The relationship between the HI mass and the virial mass of a halo cannot be derived 
analytically. A universal recipe linking the halo mass to its HI content is still lacking,
most likely a result of the observed scatter in this relation~\citep{papastergis13}. 
Several functional forms have been proposed in the literature, justified by 
numerical simulation results or analytic approaches,
with varying numbers of free parameters to be estimated by fit over observations. 

The most straightforward way to model the HIHM is to take the HI mass as a fraction of the 
halo mass. Results from numerical simulations show that the HI fraction is expected to be 
maximum for halo masses of order $10^{11}\,\msun$, while disfavouring the assignment of HI 
to much lower or higher masses~\citep{pontzen08}.
Taking this information into account,~\citet{bagla10} proposed a constant fraction for the 
HI mass within a bounded range, while~\citet{barnes10} and~\citet{barnes14} adopted a 
proportionality relation between the HI mass and the halo mass with a cut-off term for 
low halo masses. This functional form was further explored 
in~\citet{villaescusa_navarro14} and~\citet{padmanabhan17}, the latter reference also 
including a high-mass cut-off and including the dependence on the virial halo mass with a 
generic power. The cut-offs are usually expressed in terms of the virial halo velocity. 
The resultant relations are expressed in a parametric form, leaving the cut-off velocities, 
the overall normalisation and possibly the logarithmic relation slope as free parameters.

A different approach is proposed in~\citet{padmanabhan17b}, where the HIHM relation is 
built on the basis of the observed HI mass function (HIMF) via abundance matching. This 
technique consists in matching the cumulative abundance of dark matter halos 
with the corresponding abundance of HI galaxies. The former is obtained by integrating 
the halo mass function, for which they chose the parametrisation from~\citet{sheth02}. 
The latter is obtained by integrating the HIMF, for which they employed a combination of the 
HIMFs fitted on the HIPASS and ALFALFA catalogues using a Schechter functional form.
The relationship between the HI mass and the halo mass is implicitly contained in this 
matching, which the authors solved in the range $10^6\,\msun < \mhi < 10^{11}\,\msun$ 
(this being the range allowed by the HI surveys results). The resultant HIHM shows that 
the HI mass increases over the full range as a function of the halo mass. However, the 
growth is slower for masses above $\mhi\sim4\times10^{11}\,\msun$. This behaviour is 
better shown in terms of the HI to halo mass fraction, which reaches a peak of $\sim1$\% 
in correspondence to the turning point, and decreases for lower or higher masses. 
On average, its value is around $10^{-3}$, showing that this relation can be applied 
for haloes of masses between $10^9\,\msun$ and $10^{14}\,\msun$. The authors fitted the 
HIHM with the following functional form
\begin{equation}
	\label{hihm}
	\mhi = 2N_{10} \mvir\left[  \left(\dfrac{\mvir}{M_{10}}\right)^{-b_{10}}  +  \left(\dfrac{\mvir}{M_{10}}\right)^{y_{10}}   \right]^{-1},
\end{equation}
where $N_{10}$, $M_{10}$, $b_{10}$ and $y_{10}$ are model-dependent parameters whose 
best-fit values are
\begin{eqnarray}
	\label{hihmpars}
N_{10} &=&  (9.89 \pm 4.89) \times 10^{-3}, \nonumber \\
M_{10} &=& (4.58\pm0.19)\times10^{11} {\rm M}_{\odot}, \nonumber \\
b_{10} &=&  0.90 \pm 0.39 \nonumber \\
y_{10} &=& 0.74 \pm 0.03,
\end{eqnarray}
as reported in the reference. As it was remarked by the authors, it is interesting to 
remind that this behaviour reminds closely the stellar mass to halo mass (SHM) relation. 
In fact, the functional form Eq.~\eqref{hihm} had been originally introduced 
in~\citet{moster10} and later employed by~\citet{moster13} to fit for the 
SHM relation, again built via abundance matching. This similarity stresses the strict 
connection between HI gas and star-forming gas in the galactic baryon cycle.

In this work we are going to use the functional form from Eq.~\eqref{hihm} with the 
parameters quoted in Eq.~\eqref{hihmpars} as the recipe for computing the HIHM. The 
reason is twofold. First of all, the stronghold of the abundance matching method is that 
the HIHM is built directly based on observational data, and as such is to some extent less 
affected by the assumptions implied by other parametrisations. The only assumption required 
in this case is that there is a monotonic relation between the HI mass and the halo mass 
and that one dark matter halo hosts one and one only HI-selected galaxy. Secondly, 
this fitting function has been built for the local Universe, which is also the target 
of our study: the HIPASS survey covers redshift values below $z\sim0.04$, while the ALFALFA 
survey below $z\sim0.06$. Our galaxy sample spans the redshift range from 0.06 to 0.1; to a 
first approximation, we can, therefore, extrapolate the results from HIPASS and ALFALFA over 
our redshift range. The analysis in~\citet{padmanabhan17b} extends the previous results by 
generalising the HIHM relation to higher redshift, by fitting for a redshift dependence of 
the controlling parameters. However, the final fit is done in combination with other parameters 
(controlling, e.g., the HI concentration ) and over observables ranging up to the much 
higher redshift limit of $z\sim 4$ (mostly DLA related quantities). Therefore, we found it 
more convenient to adopt the HIHM evaluated for $z\sim 0$, which is much closer to our data 
set and is fitted directly on its data-built counterpart.

At this point it is legit to argue that this HIHM recipe was expressely fitted 
over galactic objects, whereas the haloes detected in our work correspond to scales of groups 
and clusters of galaxies. However, the more generic halo-oriented parametrisation explored 
in~\citet{padmanabhan17} was found to be compatible with our choice, especially at $z=0$ 
and at the high-mass end.
A direct comparison between the two recipes is actually shown in~\citet{padmanabhan17}, 
up to haloes of virial masses $\sim 10^{13}\,\msun$. The aforementioned halo model includes 
a cut-off for haloes of virial velocities greater than 
$\sim 10^{4.39}\,\text{km}\,\text{s}^{-1}$; as for
very massive haloes of $\mvir=10^{16}\,\msun$ at $z=0.08$ (the mean redshift of our sample)
the virial velocity is $v_{\rm vir}\simeq 2840\,\text{km}\,\text{s}^{-1}$, we are well below that threshold
and the HIHM is still monotonically increasing with halo mass at these scales. Actually, the same 
authors in~\citet{padmanabhan17c} abandoned that upper limit cut-off, and once again showed 
the consistency across the halo-motivated HIHM and the one obtained directly from data with 
abundance matching. In this latter reference the HIHM is also plotted up to masses 
of $10^{15}\msun$, which are obviously super-galactic scales. 
The main conclusion to be drawn from this discussion is that the HIHM at very high masses is 
still largely unconstrained. It is then a fair choice 
for the current work to adopt an HIHM fitted directly on low-redshift galactic objects, which 
proved to be consistent with halo-motivated parametrisation at the highest scales sampled 
in those studies.

\subsection{The HI density profile}
\label{ssec:hiprofile}
Once the HIHM relation has assigned a specific amount of HI to a dark matter halo, the
following step is to determine how this HI is distributed inside the halo. Under the usual 
assumption of spherical symmetry, this information is encoded in a radial density profile 
for the HI gas. Again, different models have been proposed in the literature.

Some work has adopted an exponential decay for the density of neutral hydrogen as a 
function of the separation from the halo centre, which has proven effective in reproducing 
the HI content in low-redshift galaxies from numerical simulations~\citep{obreschkow09} 
and observations~\citep{bigiel12, wang14}. According to the results of those works, the most 
accredited model is an exponential radial decrease for the surface density of the gas, with 
the necessary inclusion of a central flattening to provide a better fit. We stress that this 
model is based on data from a reduced number of resolved galaxies, and mostly refers to 
HI-rich, spiral galaxies. As in our stacks we do not resolve individual 2dF galaxies, 
 using this modelling may not be appropriate.

A perhaps more general form is given by the choice of a modified 
Navarro-Frenk-White~\citep[NFW,][]{navarro96} profile of the form
\begin{equation}
	\label{mnfw}
\rhohi(r) = \rho_0 \left[ \left(\frac{3}{4} +\frac{r}{\rs} \right)  \left(1+\frac{r}{\rs} \right)^2 \right]^{-1},
\end{equation}
where the density $\rho_0$ is a normalisation factor and $\rs$ is the profile scale radius.
This profile continues the analogy with the dark matter halo framework; it was proposed 
initially by~\citet{maller04} in this form to model the distribution of the hot gas, 
which according to numerical simulations is expected to be less concentrated than dark 
matter~\citep{cole00}. The resultant model traces the original NFW profile at large radii 
but is characterised by a thermal core in the central region (at $r\sim3/4\rs$).
This form proved more suitable to fit simulation results, particularly in the case of large 
haloes (cluster scales and above). This model was then reprised in several following
HI studies~\citep{barnes10, barnes14, padmanabhan17}, and we shall adopt it in the present 
work as well. 
	
The profile in Eq.~(\ref{mnfw}) is controlled by the parameters $\rho_0$ and $\rs$.
The normalisation density $\rho_0$ is not a free parameter, because the total amount of HI 
is set by the HIHM: the integral of the HI density profile over the halo extension should 
recover the full HI mass. The scale radius $\rs$ can be related to the halo virial radius 
$\Rvir$ via a concentration parameter $\ch$ as:
\begin{equation}
	\label{rs}
	\rs = \frac{\Rvir}{\ch}.
\end{equation}
Notice that this equation is formally identical to the relation adopted in the study of 
the modified NFW profile in dark matter halos; instead of the dark matter concentration 
parameter, in this case, we are talking about
an HI concentration parameter. The latter is further decomposed into a mass and redshift 
dependent factors, in 
the same style as for dark matter haloes, as: 
\begin{equation}
	\label{cnorm}
	\ch = \czhi \left(\frac{\mvir}{10^{11}\text{M}_{\odot}}\right)^{-0.109}\frac{4}{1+z}.
\end{equation}
This model was proposed in~\citet{bullock01} and later confirmed by~\citet{maccio07} to 
fit for numerical simulation results. Eventually, the determination of the halo scale 
radius reduces to the choice of a proper normalisation for the HI concentration parameter. 
Dark matter profiles require a normalisation parameter $c_0\simeq 3.5$~\citep{maccio07}, 
while the studies as mentioned earlier of the HI distribution in haloes
suggest a higher value is required for the neutral gas.
In Section~\ref{sec:estimation} we will fit for the value of $\czhi$ on the profiles 
extracted from the Parkes maps.

\subsection{Predicting the observed halo HI profile}

We now put together the information from the halo HI abundance and distribution provided above, 
and detail the steps we follow to predict the observed halo brightness temperature profile. 
Let us consider a halo of virial mass $\mvir$ at redshift $z$. Its virial radius is computed 
as:
\begin{equation}
	\Rvir(\mvir,z) = \left( \dfrac{3 \mvir}{4\pi \rho_{\rm c}(z)\Delta_{\rm c}(z)} \right)^{1/3},
\end{equation}
with $\rho_{\rm c}(z)$ the critical density of the Universe at redshift $z$ and $\Delta_{\rm c}$ 
the overdensity as parametrised in~\citet{bryan98}.
The corresponding HI mass $\mhi$ can be computed using the HIHM relation (Eq.~\eqref{hihm}), 
and the scale radius $\rs$ using Eq.~\eqref{rs}, with the concentration parameter 
$\ch$ from Eq.~\eqref{cnorm}. For the moment, we suppose the normalisation for the HI 
concentration parameter is known \textit{a priori}. To compute the density profile 
$\rho_{\rm HI}$, the normalisation density $\rho_0$ has to be determined. This is done 
by imposing that the
volume integral of the HI density up to the virial radius yields the HI halo mass
\begin{equation}
	\label{rhonormalise}
	\mhi(\mvir,z) = 4\pi \int_{0}^{\Rvir}\,\de r\, r^2 \rhohi(r;\mvir,z),
\end{equation}
where we made it explicit that the resultant HI density profile $\rhohi(r;\mvir,z)$ 
inherits the dependence on the halo mass and redshift. We see that by plugging the 
profile expression (Eq.~\eqref{mnfw}) into Eq.~\eqref{rhonormalise} the normalisation density 
factorises out and can be easily solved for. 

At this point, we have to remember that we do not observe directly the HI density profile, 
but the profile integrated along our line-of-sight, that is, the column density. The HI number 
density profile is obtained as $\nhi=\rhohi/m_{\rm HI}$, with $m_{\rm HI}$ the mass of the 
hydrogen atom, and the column density at an observed angular separation $\theta$ from the 
halo centre yields:
\begin{equation}
	\Nhi(\theta) = 2\int_{0}^{R_{\rm max}(\theta)}\,\de r\, \nhi\left(\sqrt{\DA^2(z)\theta^2+r^2};\mvir,z\right),
\end{equation}
where $\DA(z)$ is the angular diameter distance to redshift $z$ and 
$R_{\rm max}(\theta)=\sqrt{R_{\rm out}^2-\DA^2(z)\theta^2}$ 
(the integral exploits the assumed halo circular symmetry). For our computation 
we set $R_{\rm out}=2\,\Rvir$ as the upper limit for the radial separation, and checked 
that by extending this limit the integral value does not change appreciably. 
The resultant $\Nhi(\theta)$ provides the HI column density for the considered halo, 
observed at an impact parameter
$\DA(z)\theta$ from its centre ($\theta$ is assumed to be expressed in radians 
for this computation).

Now, the radio maps obtained in Section~\ref{sec:stacking} are expressed in terms of the
HI line brightness temperature, which is the real observable in intensity mapping. 
The column density can be easily translated into the corresponding brightness temperature
with the assumption of optically thin gas (which is usually satisfied by HI observed 
in the post-reionisation Universe). The HI line absorption coefficient is
proportional to the HI number density via~\footnote{See for example 
\url{https://www.cv.nrao.edu/\~sransom/web/Ch7.html}.}:
\begin{equation}
	\label{kappa}
	\kappa_{\nu} = \frac{3c^2}{32\pi} \frac{A_{10}}{\nu_{10}} \nhi \frac{h}{\kb\ts}\varphi(\nu) \equiv \xi \nhi \frac{1}{\ts}\varphi(\nu),
\end{equation}
with $A_{10}\simeq 2.85\times 10^{-15}\,{\rm s}^{-1}$ the HI line emission coefficient 
and $\nu_{10}\simeq 1420.4\,{\rm MHz}$ the rest-frame frequency of radiation between 
the upper and lower spin states~\citep{condon16}. $\ts$ is the spin temperature, 
$\varphi(\nu)$ is the line frequency profile, and $h$ and $\kb$ are the Planck 
and Boltzmann constants respectively. In the last equality we simplified the
expression by defining the constant $\xi =2.6\times 10^{-19}\,\text{K}\,\text{m}^2\,\text{s}^{-1} $. 
We can now integrate both members of Eq.~\eqref{kappa} along the line of sight: the 
left-hand side yields, by 
definition, the line opacity $\tau_{\nu}$, and for the right-hand side it is common to 
assume the HI spin temperature uniform along the line of sight. We obtain:
\begin{equation}
	\label{tau}
	\tau(\nu)\ts = \xi \Nhi\varphi(\nu),
\end{equation}
where we moved the spin temperature to the left-hand side. 
The observed brightness temperature $\tb$ can be computed using 
the radiative transfer equation, which in the optically thin regime (with no background
sources) simplifies as $\tb\simeq \tau\ts$. We can then integrate Eq.~\eqref{tau}
both sides over frequency and obtain:
\begin{equation}
	\label{tau2}
	\int\,\de\nu\,\tb(\nu) = \xi \Nhi 
\end{equation}
(the line frequency profile is normalised to unit integral). At this point we can safely assume that the instrument bandwidth $\Delta\nu=1\,\text{MHz}$ is 
much larger than the HI line profile~\footnote{This is certainly true for the intrinsic line 
profile. But even considering Doppler broadening with a kinetic temperature of $8000\,\text{K}$ 
(the extreme case of a HII region), we obtain an additional $\Delta\nu\simeq0.09\,\text{MHz}$, 
which is still below 10\% of the instrumental bandwidth. The regions where we are observing 
HI are most likely colder. }. The integral in Eq.~\eqref{tau2} can then be approximated with 
the product of the instrument bandwidth and the brightness temperature at the central observing
frequency. The brightness temperature angular distribution for our halo is then given by:
\begin{equation}
	\label{tb}
	\tb(\theta;\mvir,z) = \frac{\xi}{\Delta\nu}\Nhi(\theta;\mvir,z).
\end{equation}

As a final step, in order to compare this quantity with the profiles extracted from the Parkes map, 
we need to convolve the halo brightness temperature with the telescope beam, which we model  
 as a Gaussian with $\theta_{\rm FWHM}=14'$. The convolved distribution can then be obtained as:
\begin{equation}
	\label{conv}
	\tbtheo(\theta;\mvir,z) = \int \,\de\theta' \tb(\theta';\mvir,z)\,B(\theta-\theta'),
\end{equation}
where the beam function is
\begin{equation}
	\label{beam}
	B(\theta) = \frac{1}{\sigma\sqrt{2\pi}}\exp{\left[-\dfrac{\theta^2}{2\sigma^2}\right]},
\end{equation}
and $\sigma=\theta_{\textrm{FWHM}}/\sqrt{(8\ln{2})}$. Notice that in order to avoid border 
effects and be consistent with the normalisation 
expressed in Eq.~\eqref{beam}, the convolution in Eq.~\eqref{conv} should be evaluated with
the profile function symmetrised around $\theta=0$.

\begin{figure*}
\includegraphics[trim= 40mm 0mm 0mm 0mm, scale=0.21]{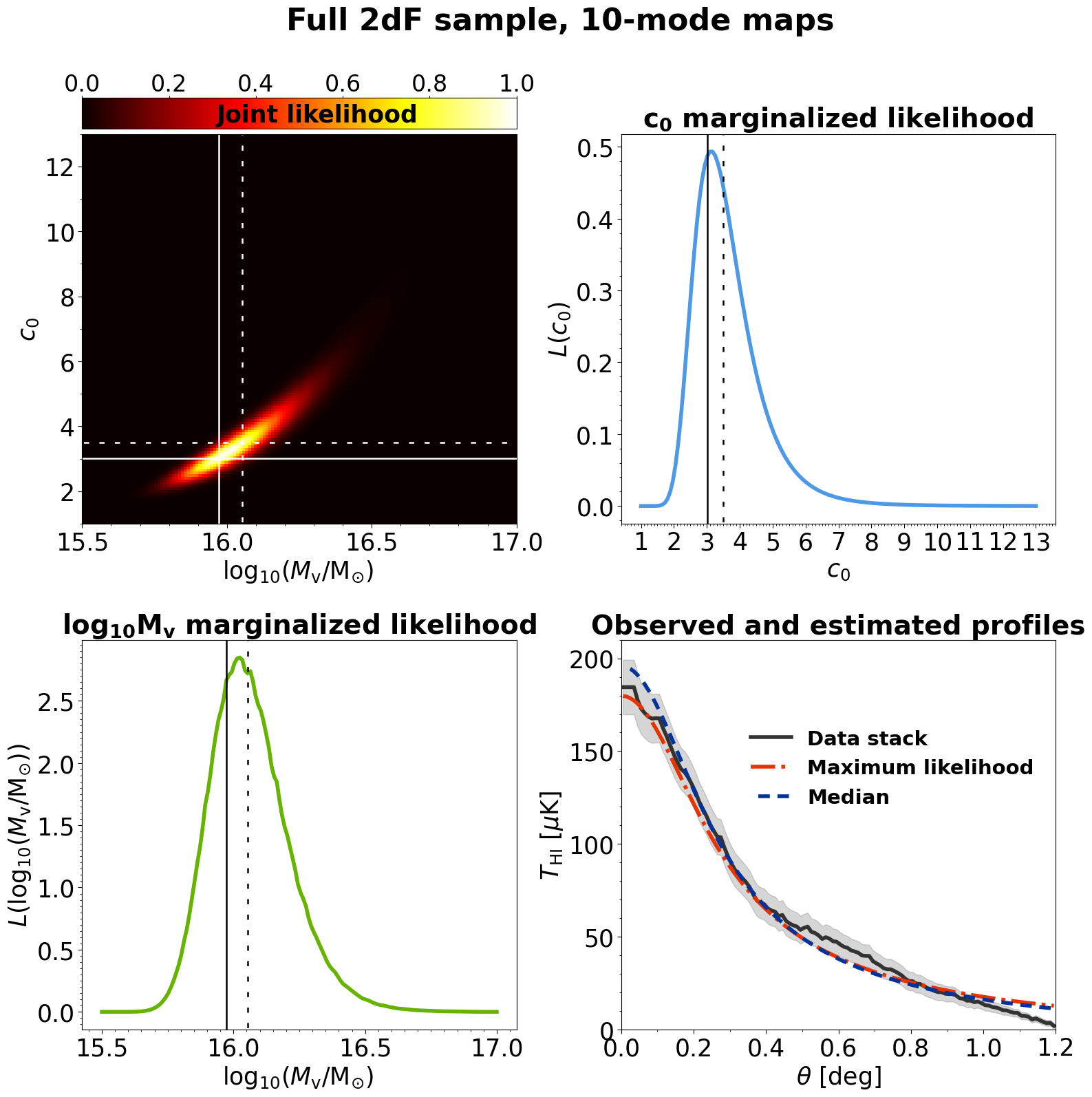}\qquad
\includegraphics[trim= 0mm 0mm 0mm 0mm, scale=0.21]{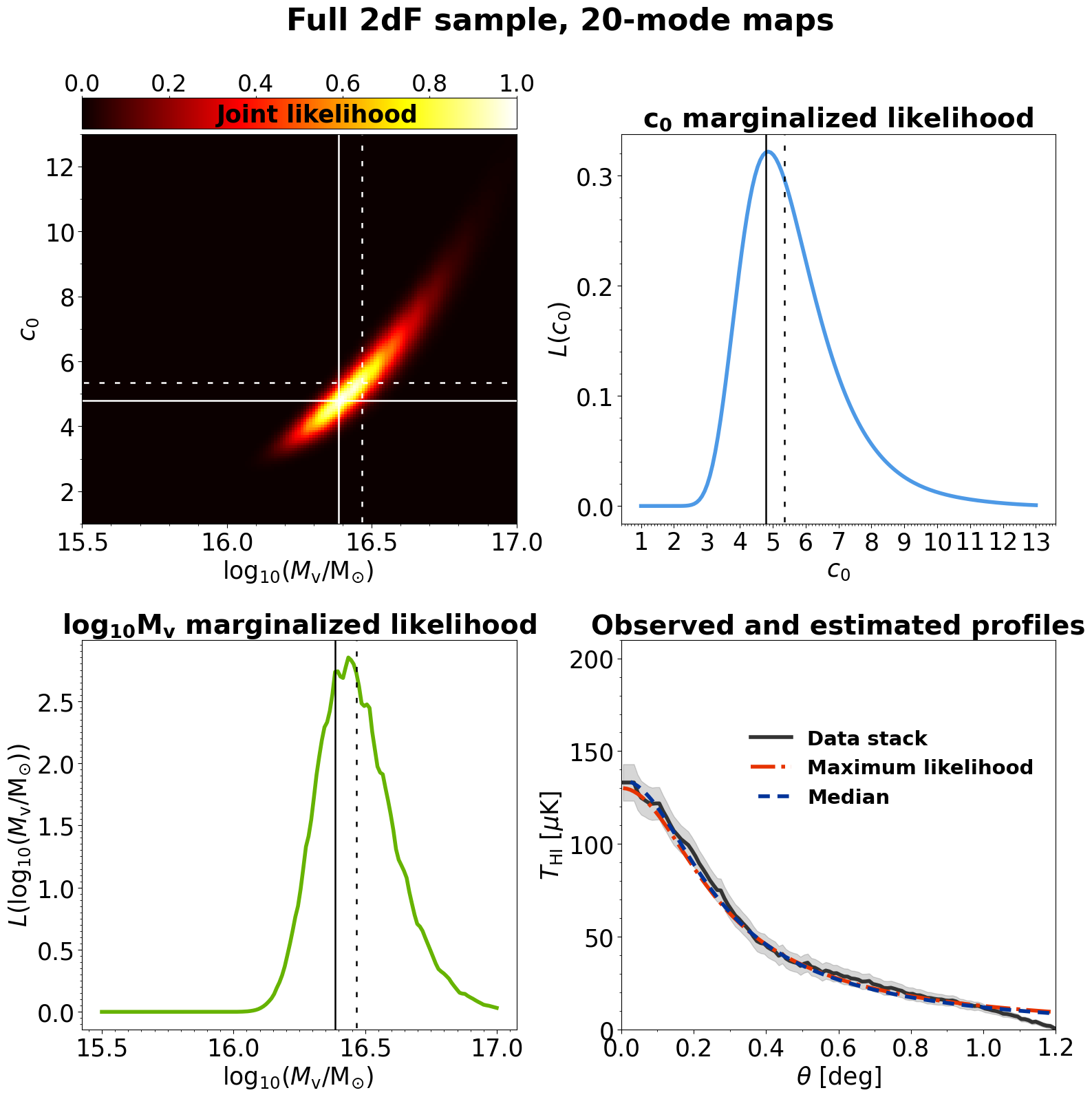}
\caption{Results of the parameter estimation described in section~\ref{ssec:estmethod}, 
	for the case of the full 2dF sample stacked on the 10-mode maps (left panel) and the 
	20-mode maps (right panel). In each case, we show the likelihood contours as a function 
	of the two parameters $\logmvir$ and $\czhi$ (top-left corner), the marginalized 
	likelihood distribution as a function of $\czhi$ alone (top-right corner) and 
	$\mvir$ alone (bottom-left corner), and the comparison between the observed and 
	estimated profiles (bottom-right). In the first three plots, the maximum likelihood 
	estimates $\tilde{\Theta}$ are marked with solid lines, the median estimates 
	$\hat{\Theta}$ with dashed lines.  In the bottom-right corner, the observed profile 
	(with associated uncertainty shown as a shaded region) is compared with the predictions 
	computed using both sets of best-fit parameters.}
\label{fig:esttot}	
\end{figure*}

\begin{figure*}
\includegraphics[trim= 40mm 0mm 0mm 0mm, scale=0.21]{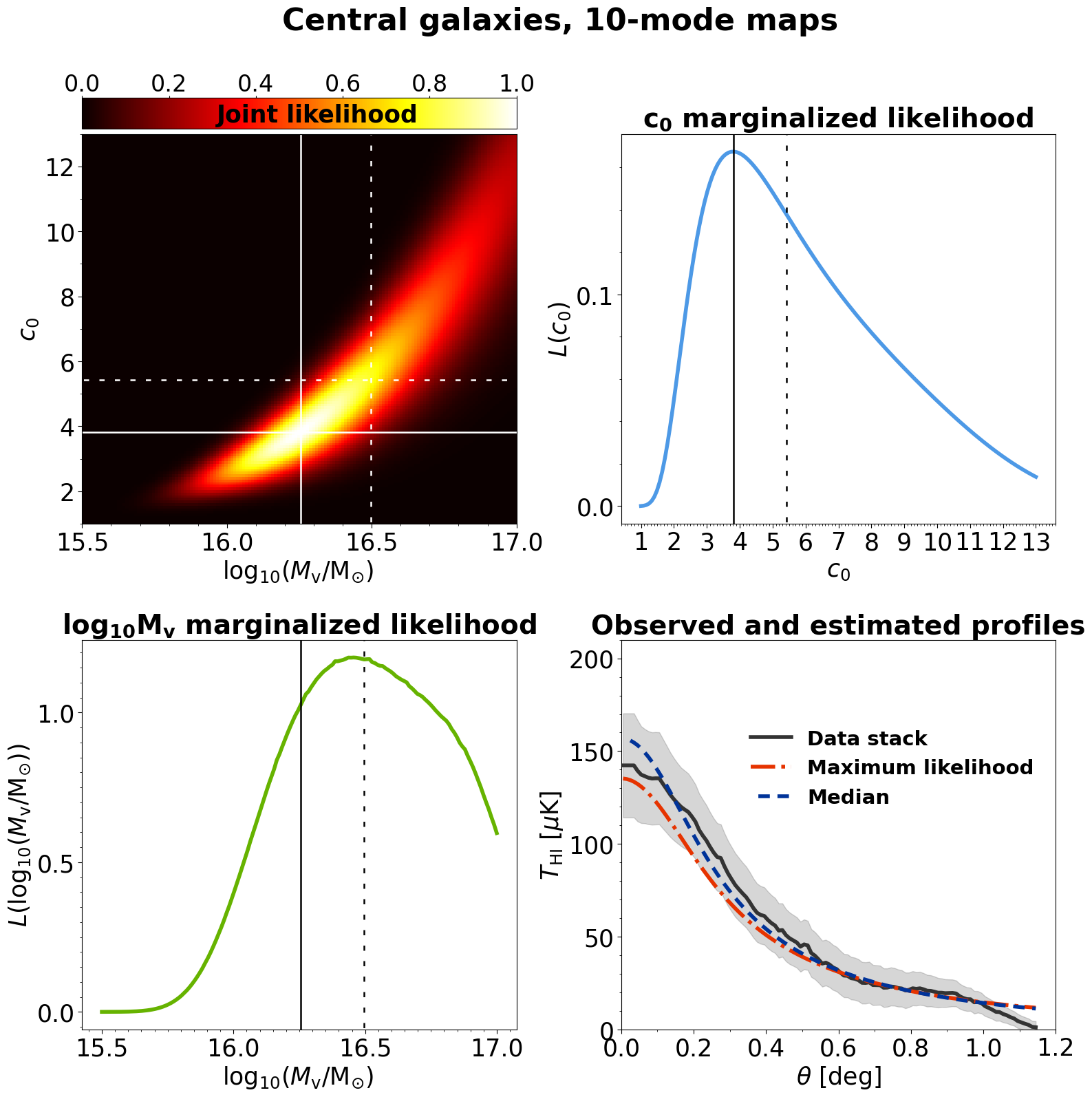}\qquad
\includegraphics[trim= 0mm 0mm 0mm 0mm, scale=0.21]{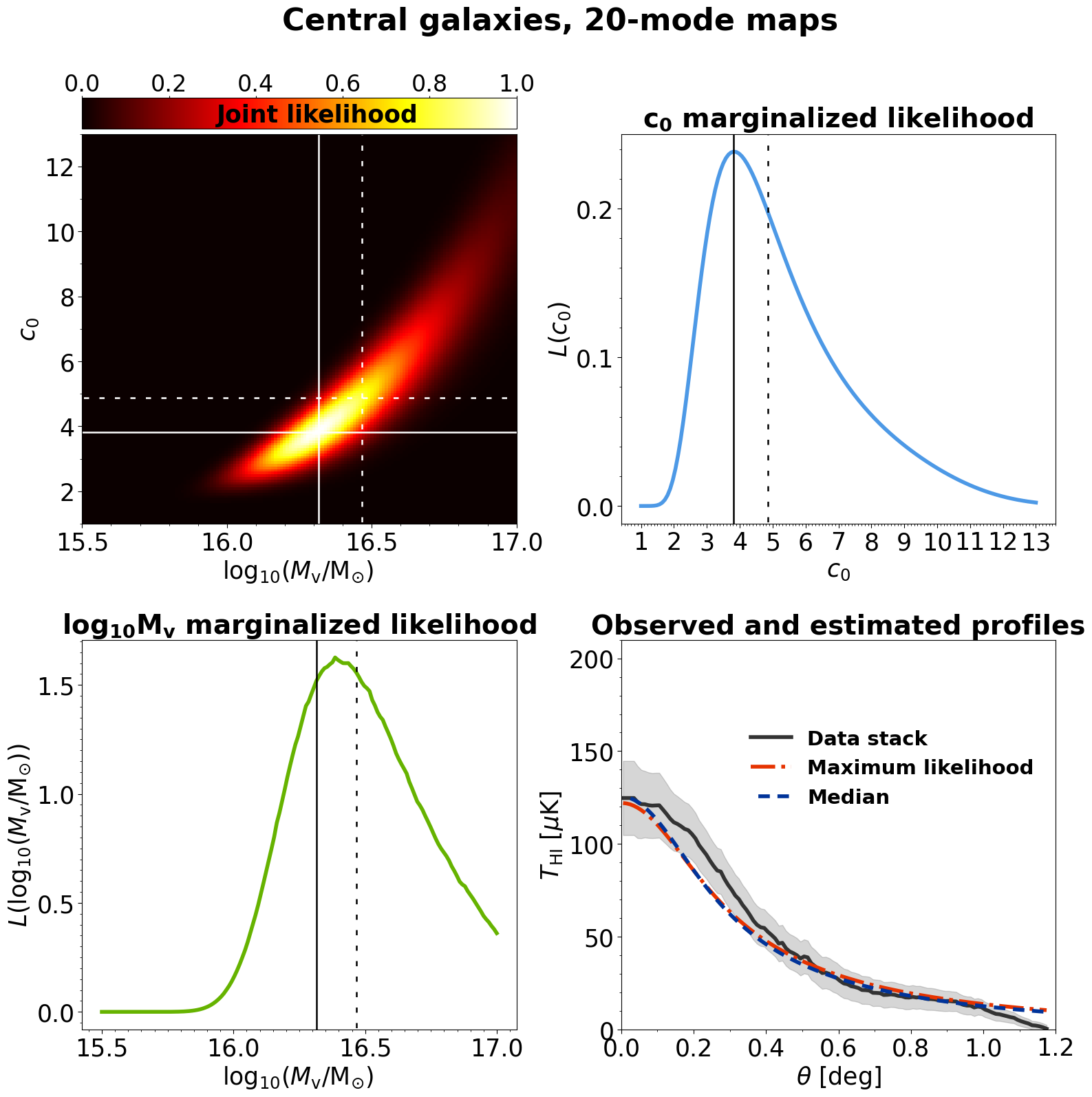}
	\caption{Same as in Fig.~\ref{fig:esttot}, but for the parameter estimation of the central galaxy sample.}
\label{fig:estloc}	
\end{figure*}

\begin{figure*}
\includegraphics[trim= 40mm 0mm 0mm 0mm, scale=0.21]{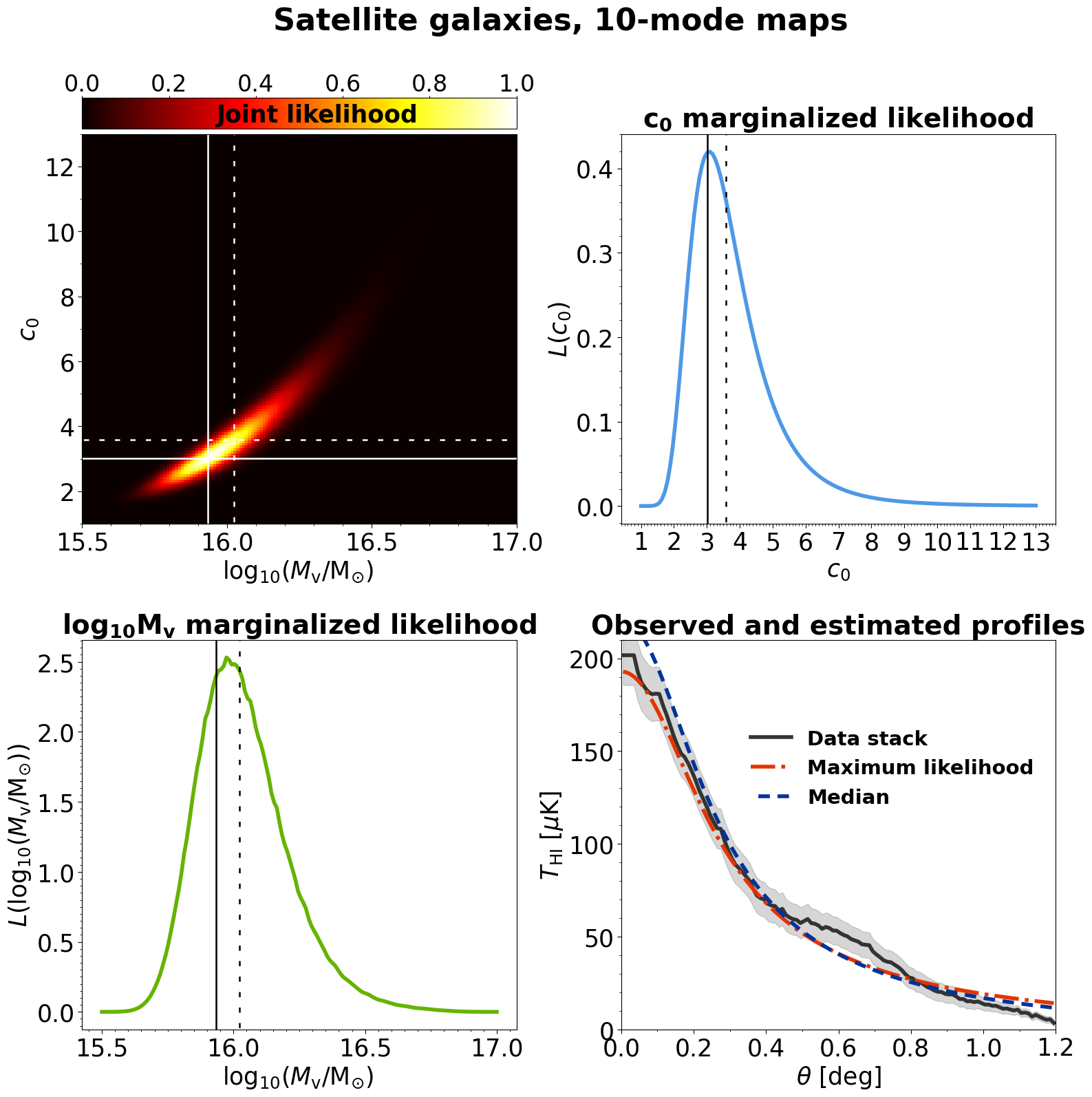}\qquad
\includegraphics[trim= 0mm 0mm 0mm 0mm, scale=0.21]{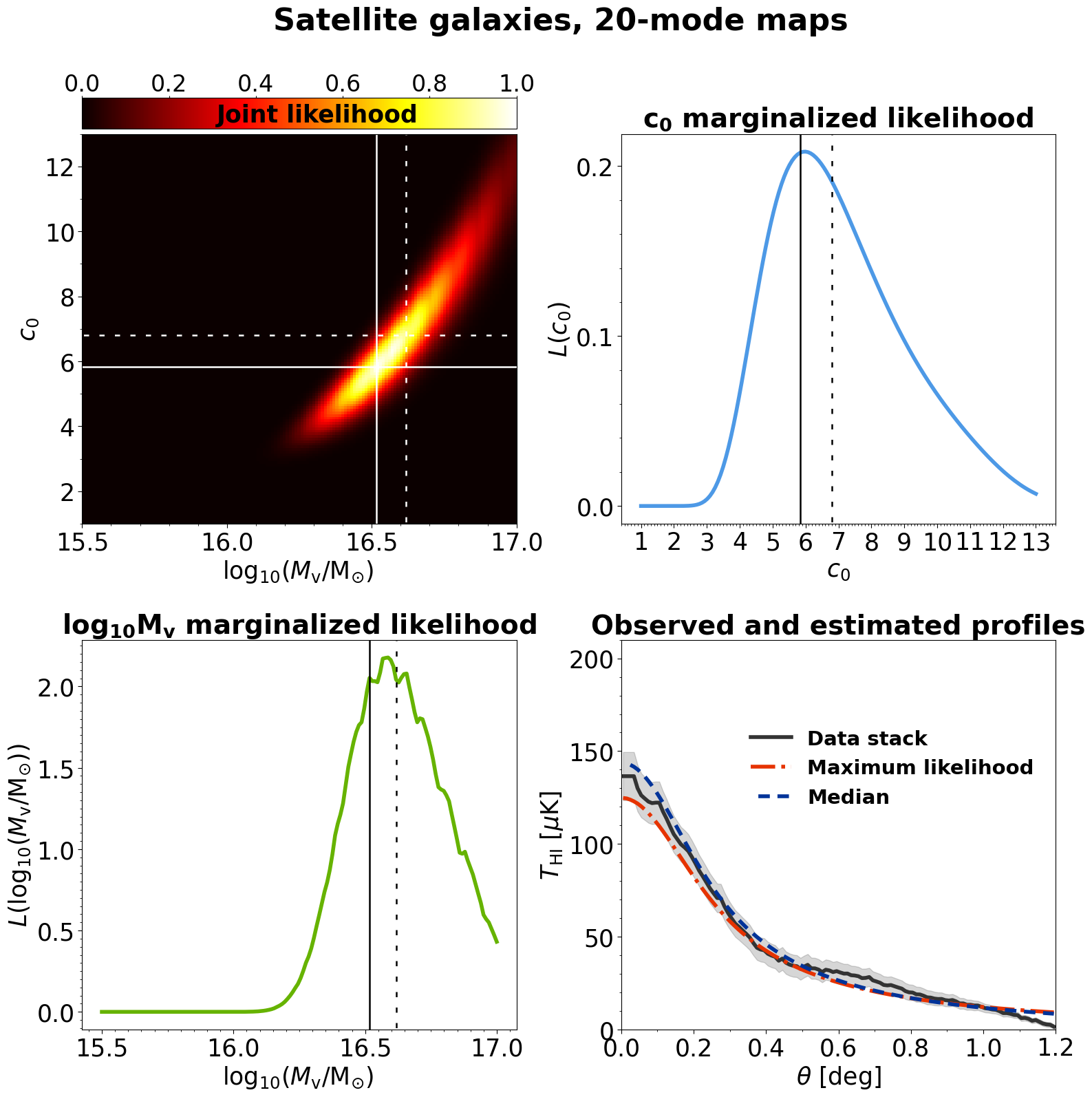}
	\caption{Same as in Figs.~\ref{fig:esttot} and~\ref{fig:estloc}, but for the parameter estimation of the satellite galaxy sample. }
\label{fig:estsat}	
\end{figure*}


\section{Parameter estimation}
\label{sec:estimation}

The formalism described in Section~\ref{sec:theomodel} allows computing the expected observed brightness temperature profile for a generic halo. In this section, we consider the inverse problem of employing the observed profiles to estimate 
the corresponding halo properties. 

\subsection{Methodology and results}
\label{ssec:estmethod}

The methodology we employ has already been exploited in the literature to estimate the 
parameters governing the HIHM and the HI density profile; these parameters are typically 
fitted over DLA observables~\citep{barnes14,padmanabhan17} or numerical 
simulations~\citep{villaescusa_navarro18}. 
Here, for the first time, we apply this type of estimation directly against the 
reconstructed halo profiles. A shortcoming of our data-set, as already mentioned, 
is that it lacks a detailed characterisation of the mass distribution of the 
objects we are stacking. Given this uncertainty, 
we cannot fit for the parameters governing the HIHM relation. In contrast, since 
the size of the haloes we detect is unknown, we want to fix the HIHM relation and 
find the best-fit values for the halo mass to accommodate our results. As anticipated, 
although haloes with different sizes are merged in the stacking, 
this methodology would only provide us with a mean value for the mass representative of 
the full samples that we are stacking. For this aim we employ the HIHM functional form 
from Eq.~\eqref{hihm} with the parameters (Eq.~\eqref{hihmpars}) obtained 
by~\citet{padmanabhan17b} with a fit over low redshift observables. 
As per the halo density profile, we adopt the modified NFW profile from Eq.~\eqref{mnfw} 
and take the normalisation $\czhi$ for the HI concentration parameter
from Eq.~\eqref{cnorm} as our second free parameter. 
Another critical variable entering the computation of the observed halo profile 
(Eq.~\eqref{conv}) is the halo redshift. Since we are modelling the observed 
profile as produced by a single halo, we are also dealing with a single value 
of redshift. This time, we can fix it to the mean value from the redshifts of 
the stacked galaxies, i.e. $z=0.08$.

Our parameter space is then two-dimensional, and can be denoted as 
$\Theta=\{\logmvir,\czhi\}$ (given the high 
sensitivity of the profile amplitude on the virial mass, we found it more 
convenient to fit for its logarithm). For each 
point in this parameter space, Eq.~\eqref{conv} provides us with the corresponding 
profile $\tbtheo(\theta;\Theta)$, which can be compared to any of the profiles plotted 
in Fig.~\ref{fig:magprofs}. The goodness of the predicted profile to reproduce the 
observed ones can be assessed by a proper likelihood function; the point $\tilde{\Theta}$ 
that maximizes the likelihood value is taken as our best fit estimate
for the combination of the two parameters. Notice that the correlation between bins that 
emerges from the covariance matrices plotted in Fig.~\ref{fig:covmatrs} means that a 
simple chi-square estimation cannot be applied. We revert then to a more general 
multi-variate Gaussian likelihood function of the form:
\begin{eqnarray}
	\label{like}
	\mathcal{L}(\Theta) &=&  \dfrac{1}{\left[(2\pi)^{N}{\rm det}\, \mathcal{C}_{\rm T} \right]^{1/2}} \exp\left[-\dfrac{1}{2}\left(\tbtheo(\Theta)-\tb\right) \right. \nonumber \\
	& \times & \left. \mathcal{C}_{\rm T}^{-1} \left(\tbtheo(\Theta)-\tb\right) \right],
\end{eqnarray}
where $N$ is the number of bins in the angular separation from the halo centre, 
$\tbtheo$ is the theoretical profile, $T_{\rm HI}$ is the reconstructed profile, 
and $\mathcal{C}_{\rm T}$ is the covariance matrix computed with~Eq.~\eqref{covar}. 
The factor multiplying the exponential is required to normalise the likelihood 
integral to unity. 

The likelihood (Eq.~\eqref{like}) computed for different combinations of the halo 
virial mass and the concentration parameter normalisation is plotted in the top-left 
corners of both panels in Figs.~\ref{fig:esttot},~\ref{fig:estloc} 
and~\ref{fig:estsat} for the case of the full, central and satellite 2dF samples, 
respectively, each panel showing a different foreground removal case. The best fit 
parameter values $\tilde{\Theta}$ are overplotted with solid lines in all cases, and 
their intersection marks the point of maximum likelihood. These plots, however, are 
also essential in showing the resultant likelihood contours that can be used to extract 
the uncertainty associated with our estimates. For this aim, we marginalise the likelihood 
for each one of the two parameters and re-normalise to unit integral; the resultant 
function can then be regarded as the associated probability distribution.  
In the same figures, the top-right corner of each panel shows the marginalised likelihood 
as a function of $\czhi$, and the bottom-left corner shows the marginalised likelihood 
as a function of  $\logmvir$. Each of these marginalised functions is integrated to obtain 
the corresponding cumulative distribution function $F(\eta)$ (where $\eta=\logmvir$ or 
$\eta=\czhi$). The percentiles of this distribution are used to estimate the uncertainty 
associated with our estimates: by defining the variables $\eta_-$, 
$\hat{\eta}$ and $\eta_+$ such as $F(\eta_-)=0.16$, $F(\hat{\eta})=0.50$ and 
$F(\eta_+)=0.84$, we compute the lower and upper uncertainties as 
$\sigma_-=\hat{\eta}- \eta_-$ and $\sigma_+=\eta_+- \hat{\eta}$. The quantity $\hat{\eta}$
is the median of the distribution and is a best-fit estimate alternative to the 
maximum-likelihood one.

In Table~\ref{tab:estimates} we quote, for each chosen galaxy sample and foreground 
removal case, the maximum-likelihood estimates $\tilde{\Theta}=\{\logmvirt,\czhit\}$ 
and the median estimates $\hat{\Theta}=\{\logmvirc,\czhic\}$ with their uncertainties. 
Both the maximum likelihood $\tilde{\Theta}$ and median $\hat{\Theta}$ estimates are 
marked in the plots mentioned above with pairs of solid and dashed lines, respectively. 
Finally, the bottom-right panel of each figure shows the profile extracted from the stacks
in comparison with the theoretical profiles computed using both sets of best-fit parameters 
$\tilde{\Theta}$ and $\hat{\Theta}$.

\begin{table*}
\centering
\caption{Results of our parameter estimation analysis. For each 2dF sub-sample and 
	foreground removal case, we quote the maximum likelihood estimates 
	$\tilde{\Theta}$ and the median estimates $\hat{\Theta}$, as defined in 
	Section~\ref{ssec:estmethod}.}
\label{tab:estimates}
\setlength{\tabcolsep}{1.2em}
\begin{tabular}{c c c c c c c}
\hline
	&	&	\multicolumn{2}{c}{10-mode maps} &  & \multicolumn{2}{c}{20-mode maps} \\
\hline
	\multirow{2}{4em}{Full 2dF sample} & & $\logmvirt = 16.0 $ & $\logmvirc = 16.1^{+0.1}_{-0.2} $ & & $\logmvirt = 16.4 $ & $\logmvirc = 16.5^{+0.1}_{-0.2} $ \\
	& & $\czhit = 3.0 $ & $\czhic = 3.5^{+0.7}_{-1.0} $ & & $\czhit = 4.8 $ & $\czhic = 5.3^{+1.1}_{-1.7} $  \\
\hline
	\multirow{2}{4em}{Central galaxies} & &  $\logmvirt = 16.3 $ & $\logmvirc = 16.5^{+0.3}_{-0.3} $ & &   $\logmvirt = 16.3 $ & $\logmvirc = 16.5^{+0.2}_{-0.3} $ \\
	& & $\czhit = 3.8 $ & $\czhic = 5.4^{+2.1}_{-3.5} $ & & $\czhit = 3.8 $ & $\czhic = 4.9^{+1.5}_{-2.6} $  \\
\hline
	\multirow{2}{4em}{Satellite galaxies} & & $\logmvirt = 15.9 $ & $\logmvirc = 16.0^{+0.1}_{-0.2} $ & & $\logmvirt = 16.5 $ & $\logmvirc = 16.6^{+0.2}_{-0.2} $  \\
	& & $\czhit = 3.0 $ & $\czhic = 3.6^{+0.9}_{-1.4} $ & & $\czhit = 5.8 $ & $\czhic = 6.8^{+1.7}_{-2.4} $  \\
\hline
\end{tabular}
\end{table*}

\subsection{Discussion}

The likelihood contour plots from Figs.~\ref{fig:esttot} to~\ref{fig:estsat} show the 
most likely combinations of the halo mass and concentration parameter that reproduce the 
reconstructed profiles. The size of the contours is representative of the uncertainty 
associated with the estimation of each parameter. We see that the contours obtained by 
stacking the full 2dF sample are narrower, a result of the higher 
statistics of the considered sample, while the contours for the central galaxy sample are 
the broadest. However, the main feature emerging from these plots is the tight correlation 
between the two parameters we fit for, that produces an evident elongation of the contours 
along the same direction in all cases. Qualitatively, such a correlation can be understood: 
the HIHM relation (Eq.~\eqref{hihm}) implies that for virial masses above the turning point 
of $\sim3\times10^{11}\,\msun$ the HI halo mass grows slowly, 
approximately as $\mvir^{0.25}$. The virial radius, however, keeps growing as $\mvir^{1/3}$, 
and as a result, the density normalisation $\rho_0$ computed using Eq.~\eqref{rhonormalise} 
decreases for higher masses. In other words, with this model, higher virial masses determine 
shallower, more extended HI density profiles; a higher concentration parameter is then required 
to shrink the profile towards the centre and increase its peak amplitude. 

The bottom-right panels of Figs.~\ref{fig:esttot} to~\ref{fig:estsat} show that the best-fit 
parameters associated with the maximum of the likelihood distribution are indeed effective 
in recovering the observed halo brightness temperature profile. The corresponding theoretical 
predictions, indeed, capture both the amplitude and shape of the reconstructed profiles. To 
assign uncertainties to the best-fit estimates, we computed the median of the contours 
following the procedure described above. The observed elongation of the likelihood profile is 
particularly visible in the marginalised likelihood plots, as it produces a clear positive 
skewness. As a result, the median estimates are always higher than the maximum-likelihood ones. 
However, it is interesting to notice how the median estimates are also very effective in 
reproducing the observed profiles, as the strong degeneracy between 
$\mvir$ and $\czhi$ allows different combinations of the two parameters that yield very 
similar predictions. Both the maximum likelihood and the median estimates with their 
uncertainties are reported in table~\ref{tab:estimates}. The virial mass seems to be 
better constrained, with a relative error typically below 2\%, but the uncertainty on 
the concentration parameter can get higher than 50\%.

We can now make some considerations on the significance of our findings. Our results 
suggest very low values for the normalisation of the concentration parameter 
(typically $\lesssim 7$). In the literature, although the concentration parameter 
for neutral gas is generally found to be higher than for the case of dark matter, its actual 
value is still largely unconstrained. For example, we can quote the estimate $\czhi\simeq25$
from~\citet{barnes10} and~\citet{barnes14}, or $\czhi\simeq133$ from~\citep{padmanabhan17b}. 
However,~\citet{villaescusa_navarro14} found that much lower values for the concentration, $\czhi\sim 3$, are favoured by numerical simulations, a result compatible with our findings. All the 
estimates mentioned above are fitted over high redshift observables, typically in the range 
of $z\sim 2$ to 4; we provide here for the first time an estimate for very low redshift, 
$z\sim 0.08$. We also need to remind that a 
direct comparison with galactic-halo estimates may be misleading, as this concentration is not 
referred to diffuse gas in the haloes, but rather to the integrated contribution of the HI 
emission from different galaxies.

The other parameter we fitted for, the virial halo mass, yields very high values 
averaging at $\mvir\sim 2\times10^{16}\,\msun$, which is the scale of the most massive
haloes in the Universe.
This estimate can actually be a biased result because of the assumptions made in 
our theoretical modelling, as it is now confirmed that our chosen functional form 
for the HIHM (Eq.~\eqref{hihm}) is being applied beyond the mass range over which the 
parameters in Eq.~\eqref{hihmpars} were 
fitted~\citep[$\sim10^{11}$ to $10^{14}\,\msun$,][]{padmanabhan17b}.
Nonetheless, it can be double-checked quickly: the virial radius for a halo 
of mass $\mvir = 10^{16}\,\msun$ is approximately 5.3 Mpc. If that halo is located at $z=0.08$, 
the angular radius it subtends is approximately 1 deg. This scale is indeed the typical 
extension of the profiles we detect in the stack maps. Plus, by taking $\czhi=4$, the 
resultant angular size corresponding to the scale radius $\rs$ is $\sim 0.25\,\text{deg}$, 
and we do observe a flexion in the profiles around $\theta\sim0.3\,\text{deg}$.
This rough estimate serves to verify that our stacks do indeed reveal
the emission towards very massive objects.  
The observed HI signal likely proceeds from the HI locked 
inside the gas richest galaxies in the 2dF sample, to which we should add the contribution from 
all the optically faint, HI-rich galaxies that were not detected in the 2dF survey.
As the low resolution of the intensity mapping technique probes the HI large-scale 
fluctuations, the estimated masses weigh the locally HI-brighter regions made 
by the contribution of all those galaxies, possibly assembled in large groups or clusters. 

It is then interesting to compare this finding with the results presented 
in~\citet{papastergis13}, according to which HI-rich galaxies are the least clustered 
population type, so that cluster member galaxies are expected not to be included in 
HI-selected samples. Neverthless, other studies pointed towards satellite galaxies located 
in the outskirts of galaxy groups and clusters to host a significant amount of 
HI~\citep{lah09, baugh19, guo20}. Now, the results of our blind stacks 
do not allow to draw further conclusions about the actual origin of the detected HI signal, 
although a hint may come from the comparison between the different 2dF subsamples we used. 
We have already pointed out in section~\ref{sec:datadiscussion} how the full sample 
shows features in between the other two, reinforcing the hypothesis that the latter are 
tracing galaxies belonging to different environments. 
We note that, for the central galaxy sample, we are most likely centring our stacks over 
groups and clusters of galaxies: we find, as a result, a dilution of the resultant HI profile, 
which appears less bright in the centre, and more slowly decaying towards larger radii. The 
stack of satellite galaxies, instead, yields brighter and more concentrated profiles. This 
phenomenon can be in agreement with the finding that even cluster-member galaxies, 
when located in the cluster outskirts, can be HI-rich. However, these conclusions are more 
evident when comparing the reconstructed profiles than the estimated parameters, because 
the tight degeneracy between $\mvir$ and $\czhi$ easily overrides the marginal difference 
provided by the selected samples. Still, looking again at table~\ref{tab:estimates}, we can 
point out some differences. For the 20-mode removal case, the mass estimate is consistent 
within a 1\% variation across different samples, while the concentration parameter is the 
highest for the satellite sample, the lowest for the central sample, and intermediate for 
the full sample. This result is in agreement with our hypothesis that the satellite galaxies 
trace better the HI distribution, yielding more concentrated profiles. The same is not valid, 
however, for the 10-mode removal case, where we obtain remarkably similar results for the full 
and satellite samples, while the central galaxy stack provides higher estimates for both the 
estimated mass and concentration parameter. Notice anyway that the 10-mode central galaxy stack 
was from the beginning the most irregular (Fig.~\ref{fig:stack10modes}), the resultant 
likelihood contours are the broadest, and the corresponding uncertainties are the largest; 
this case is then probably not as clean as the other ones. We can finally observe that the 
10-mode case generally favours lower masses and concentration parameters than the 20-mode case.  
This situation may be a result of the 20-mode maps having lost HI signal in the foreground 
removal process, which results in more irregular stacks that are artificially broadened in the 
symmetrisation process. Broader profiles, in turn, require higher mass values, and, 
because of the degeneracy, higher concentrations. 

To summarise, we can state that although our selection of specific 2dF sub-samples provides 
hints of different results, in terms of the corresponding stacks and extracted profiles, 
the resultant estimates are generally compatible with each other. Therefore, we can quote 
as our final results the estimates obtained with the full 2dF sample, which benefits from 
its more significant statistics. The strong degeneracy between $\mvir$ and $\czhi$ is an 
additional factor that hinders possible differences in the final estimates arising from 
the use of a specific sub-sample. A higher variation is instead produced by choice of the 
10 or 20-mode maps, although we cannot assess whether the differences arise from an excess 
of foreground residuals in the former or removal of physical HI emission in the latter. 
This consideration shows that proper foreground removal remains a critical element in the 
exploitation of 21-cm intensity maps for constraining the HI abundance in haloes.


\section{Conclusions}
\label{sec:conclusions}

This paper tackles the search for neutral hydrogen in the cosmic
web using 21-cm maps. The goal is to measure the HI brightness temperature profile proceeding
from the merged contribution of galaxies in massive large-scale haloes. The signal 
is searched for by blindly stacking map regions centred on the fiducial positions extracted from 
a galaxy catalogue. This approach requires to use a data set with high available statistics, and 
the substantial overlap between the spatial distribution of catalogue(s) and maps. Our choice 
reverted therefore to a set of 21-cm maps obtained with the Parkes telescope over the volume 
spanned by the 2dF galaxy catalogue, covering a large sky area of $\sim 1,300\,\text{deg}^2$ over 
the redshift range $0.06 \lesssim z \lesssim 0.08$. The same data set proved effective 
in~\citet{tramonte19} for a similar search of neutral gas in cosmic web filaments. 

We considered three different galactic samples extracted from the 2dF catalogue. These are the 
full sample comprising all the galaxies sharing the map volume, a sample made only of the locally 
brightest galaxies, which more likely trace the centre of galaxy clusters and groups, and the 
associated complementary sample of satellite galaxies. This subdivision allows us to compare
the profiles obtained with galaxies belonging to different local environments. Altogether, we employed 
48,430 galaxies for the full sample, 13,979 galaxies for the central sample and 34,361 galaxies for 
the satellite sample. 
We also considered two versions of the Parkes maps, obtained by removing a different number of modes 
(10 and 20) in a PCA foreground cleaning algorithm, to assess to what extent possible foreground 
residuals affect our results. 

For all six combinations of 2dF samples and foreground removal cases we provided the resultant 
stack maps, which clearly show the detection of the halo HI emission at a level of 
$\sim 200\,\mu\text{K}$ for the 10-mode removal case and of $\sim 150\,\mu\text{K}$ for 20-mode 
removal case. This difference can be due to either the presence of radio foreground residuals in 
the 10-mode maps or the excessive removal of HI signal in the 20-mode maps. In both cases, the 
satellite galaxy sample yields more rounded and shrunk peaks, while the central galaxy sample 
produces more irregular and extended peaks, with the full 2dF sample showing intermediate features.  
The corresponding radial profiles were obtained by circularly symmetrising the stacks and considering
the mean value of pixels found within specified angular separation bins from the halo centre. 
For each 2dF sample and foreground removal choice, we repeated the same analysis using a set of 
500 replicas of the galaxy catalogue obtained by randomising the object equatorial coordinates. 
The dispersion of the corresponding profiles across different realisations 
provides an estimate of the uncertainty associated with our measurements, and of the correlation 
between pairs of radial bins. 
The detection significance at the profile peak is always above $5\sigma$ and even higher than 
$12\sigma$ for 
the full 2dF sample. The comparison of profiles obtained with different foreground removed maps or 2dF 
samples confirms the considerations already drawn by comparing the associated stack maps. 
More importantly, we quantified the extension of the profiles out to 1 deg from the halo centre. 
This is considerably higher than the 14' FWHM Parkes beam, which ensures these halo profiles are 
spatially resolved. 

The second half of this work was dedicated to the theoretical modelling of the observed HI emission. 
We first outlined how to compute the mean observed HI brightness temperature profile expected for a 
generic halo of given mass and redshift. We adopted a functional form for the HI-halo mass (HIHM)
relation directly fitted over low-redshift observations of HI-rich galaxies, and a general modified 
NFW shape to model the HI radial distribution. 
We compared then the theoretically computed brightness temperature profile with the ones 
extracted from our stacks; the latter were considered as as produced by a single type of haloes, whose 
properties (mass, redshift and HI concentration parameter) are a mean of the corresponding sample members.
We fixed the redshift to the mean of the samples ($z=0.08$) and fitted for the logarithm of the 
virial halo mass, $\logmvir$, and for the normalisation of the HI concentration parameter, $\czhi$. 
We built a two-dimensional likelihood distribution computed with a general multi-variate Gaussian 
form that takes into account the observed correlation between angular bins. The resultant contours 
showed a clear degeneracy between the halo mass and the concentration parameters, which makes it 
difficult to extract a best-fit value without the assumption of any priors on the halo mass.
We found that, although marginal variations arise from the use of different 2dF samples, the 
estimates are mostly compatible, and favour the values $\mvir\simeq 1.3\times10^{16}\,\msun$, 
$\czhi\simeq 3.5$ for the 10-mode removal case and $\mvir\simeq 3.2\times10^{16}\,\msun$, 
$\czhi\simeq 5.3$ for the 20-mode removal case. 

The recovered halo mass is very high and corresponds to the largest haloes found in the cosmic web. 
This finding is consistent with the observed angular extension of the profiles, and it is most likely 
produced by the combined contribution of neutral gas contained in different galaxies found in groups 
and clusters. The Parkes 21-cm maps we employed, indeed, record the large-scale fluctuations of HI and 
do not allow to resolve individual galactic objects. The recovered concentration is lower than the values
typically favoured by studies of HI in haloes, probably a result of the HI gas being locked 
inside galaxies and not being diffuse inside the massive haloes we probe.
However, it is still a largely unconstrained 
parameter, and our estimates provide a direct measurement at low redshift.
 
The most crucial element entering our theoretical prediction is perhaps the HIHM relation, 
which is applied to masses outside the range 
over which it was tested. This procedure is a source of possible bias for our final estimation. But 
given the intrinsic uncertainty on the mass of the objects we are stacking, we cannot fit at the same 
time for the HIHM parameters. It would be useful, in this sense, to deal with a  
halo sample with a characterised mass distribution. Such data set would allow to fully model the 
cross-correlation between the 21-cm and galaxy sample, which in real space translates into the mean 
radial profile around halo centres. The comparison of such a model with the 
observed profiles would allow fitting for the parameters governing the HIHM. 

In conclusion, we stress that the most important results obtained in this study are the feasibility 
of measuring the HI content in large-scale haloes using 21-cm maps, and the possibility of 
testing theoretical models on the reconstructed profiles. Most of the theoretical 
modelling was previously tested on galactic-scale haloes, which are not accessible with the resolution 
of our maps. Still, we applied it to the large scale haloes we detected, formally 
treating their effective profile as if it was produced by a diffuse gas component. 
The resulting high masses and low concentrations remind us that we are in fact probing super-galactic 
scales. Nonetheless, this paper presents a novel methodology that can be applied to future, more 
suitable data sets: the next generation of radio interferometric experiment as MeerKAT~\citep{jonas16}, 
ASKAP~\citep{meyer09} and the future Square Kilometer Array~\citep{braun15} is 
expected to provide large-scale 21-cm maps with higher resolution, which may enable to extend this 
study to the proper galactic-scale haloes. It would also be interesting to apply our stack on 
the combination of the ALFALFA survey legacy catalogue with the corresponding HI cubes (the latter 
are not publicly available at present), or to the combination of the BOSS-LOWZ galaxy 
catalogue~\citep{dawson13} with maps resulting from the ongoing Commensal Radio Astronomy FAST 
Survey~\citep[CRAFTS,][]{zhang19}. The latter is scheduled to map the HI distribution over 
$\sim 20,000\,\text{deg}^2$ and up to $z\sim0.35$, with a $\sim 3.24'$ resolution  
that allows to resolve smaller galaxy group scales with typical halo masses 
of $10^{13}\,\msun$.

\section*{Acknowledgements}

We thank the anonymous reviewer 
for a number of observations that improved the clarity of this manuscript.
We thank Dr. Yi-Chao Li and Prof. Lister Staveley-Smith for valuable input. 
DT acknowledges the support from the South African Claude Leon Foundation, that 
partially funded this work. 
YZM acknowledges the support of NRF with grant no. 105925, 109577, 120385, and 120378, 
and NSFC with grant no. 11828301. The Parkes radio telescope is part of the Australia Telescope 
National Facility which is funded by the Australian Government for operation as a 
National Facility managed by CSIRO. The calibrated data underlying this article were 
provided by Dr. Yi-Chao Li and Dr. Christopher Anderson by permission. 
Data will be shared on request to the corresponding author with permission of the providers.




\bibliographystyle{mnras}
\bibliography{bibliography}



\bsp	
\label{lastpage}
\end{document}